\documentclass[11pt,a4paper]{article}
\pdfoutput=1
\usepackage{jheppub}

\usepackage{float}
\usepackage{amsmath}
\usepackage{cleveref}

\crefname{equation}{eqn.}{eqns.}
\Crefname{equation}{Eqn.}{Eqns.}
\crefname{figure}{fig.}{figs.}
\Crefname{figure}{Fig.}{Figs.}

\newcommand{\ud}{\mathrm{d}}

\title{Simulations of a bubble wall interacting with an electroweak plasma}

\author[b]{Zong-Gang Mou,}
\author[a]{Paul M. Saffin,}
\author[b]{Anders Tranberg,}

\affiliation[a]{School of Physics and Astronomy, University Park, University of Nottingham,\\ Nottingham NG7 2RD, United Kingdom}
\affiliation[b]{Faculty of Science and Technology, University of Stavanger, 4036 Stavanger, Norway}

\emailAdd{zonggang.mou@uis.no}
\emailAdd{paul.saffin@nottingham.ac.uk}
\emailAdd{anders.tranberg@uis.no}

\abstract{
We perform large-scale real-time simulations of a bubble wall sweeping through an out-of-equilibrium plasma. The scenario we have in mind is the electroweak phase transition, which may be first order in extensions of the Standard Model, and  produce such bubbles. The process may be responsible for baryogenesis and can generate a background of primordial cosmological gravitational waves. We study thermodynamic features of the plasma near the advancing wall, the generation of Chern-Simons number/Higgs winding number and consider the potential for CP-violation at the wall generating a baryon asymmetry. A number of technical details necessary for a proper numerical implementation are developed.
}

\begin{document}

\maketitle

%%%%%%%%%%%%%%%%%%%%%%%%%%%%%%%%%%%%%%%%%%%%%%%%%%%%%
\section{Introduction}
\label{sec:Intro}
%%%%%%%%%%%%%%%%%%%%%%%%%%%%%%%%%%%%%%%%%%%%%%%%%%%%%

A first order phase transition in the early Universe proceeds through the nucleation of bubbles of the low temperature phase inside the background high temperature phase. These bubbles grow and coalesce while interacting with the ambient plasma, thereby generating gravitational waves (see for instance \cite{Hindmarsh:2013xza}), with a spectrum that may be detectable by the upcoming LISA mission \cite{Caprini:2019egz}. If the first order phase transition is associated with electroweak spontaneous symmetry breaking, baryogenesis may take place in the presence of CP-violating interactions near the bubble wall \cite{Kuzmin:1985mm,Cohen:1993nk}. A first order electroweak phase transition does not arise in the Minimal Standard Model, but is a fairly generic result of extending the scalar sector with additional fields \cite{firstorder1,firstorder2,firstorder3,firstorder4,firstorder5,firstorder6,firstorder7,firstorder8,firstorder9,firstorder10,firstorder11,firstorder12}.

The precise dynamics of the bubble wall is complicated by the interactions with the plasma, and the transport of any currents and energy-momentum away from or through the wall is a highly non-equilibrium phenomenon, ultimately requiring a non-perturbative real-time treatment. One hope to simplify things is that the propagating wall has enough time to reach a steady-state regime, wherein the Higgs profile of the wall and the velocity of the wall  are roughly constant. And furthermore, that this steady-state regime persists long enough to dominate any creation of a baryon asymmetry. This is simpler to treat analytically than, say, a fast evolving transient stage prior to such a steady-state or while (or even after) the walls are colliding, the latter being the primary source of gravitational waves \cite{Hindmarsh:2013xza}. 

Ideally, one would wish for a real-time numerical simulation of the bubble nucleation process, then the growth and evolution of bubbles followed by the steady-state regime, the bubble collisions, coalescing of domains  of low temperature phase and thermalization of the released latent heat into the plasma. Also, ideally, such simulations would include all the degrees of freedom of the (extension of the) Standard Model, gauge fields, scalar fields and fermions. 

The tools for this exist (see for instance \cite{bubsim1,bubsim2,bubsim3,bubsim4}), but are numerically much too expensive to simultaneously include all degrees of freedom, with the fermions inherently quantum mechanical, the very large volumes necessary for multiple bubbles to fit, and for the microscopically quite long times involved in nucleation, growth, collision and thermalization. One solution is to split up the problem into smaller, more tractable, components some of which may be treated in or near thermal equilibrium.  A huge analytic and numerical effort over decades has been put into different aspects of this problem ( for reviews, see for instance \cite{EWBGrev2,EWBGrev1}).

In this work, we make the following simplifications: We include only the bosonic part of the Standard Model, a single Higgs field and the SU(2)$\times$U(1) gauge field, and we treat their dynamics classically. We assume that the bubble has been nucleated at some earlier stage, and consider a mostly planar Higgs wall. This wall is advancing not as a result of the thermodynamic pressure in the physical transition, but through the driving force of an external current, which we insert by hand. We are then free to choose this current to enforce a wall speed and profile (more or less) of our choosing. 

We then simulate this wall sweeping through the plasma of quasi-thermal  Higgs and gauge field fluctuations, and may compute various components of the energy-momentum tensor, as well as other field observables, including topological ones (winding number, Chern-Simons number). This will allow us to monitor the development of an out-of-equilibrium envelope near the wall, which we expect is the region of most relevance to baryogenesis.  We can follow the creation of Higgs winding number and estimate the possibility of an asymmetry being created. We will also consider the introduction of CP-violation in a particular form, but although we did perform first-principles simulations including such  CP-violation, they were inconclusive and are postponed for upcoming work.

The paper is structured as follows: In section \ref{sec:model} we will describe the field theory model, and introduce notation and basic observables. In section \ref{sec:wall} we develop our numerical setup, including the initial conditions of the plasma and the driving of the wall. We will do some basic tests of the setup, wall speed and shape. In section \ref{sec:thermo} we perform simulations focusing on the thermodynamical observables, and the transport of energy-momentum near the wall. In section \ref{sec:topology} we instead emphasize topological transitions (Higgs winding), and connect to baryogenesis.  In section \ref{sec:conc} we comment on improvements, caveats and possible future avenues for such simulations, and briefly  discuss the introduction of CP-violation through a bosonic dimension 6 operator. Finally, we conclude. 

%%%%%%%%%%%%%%%%%%%%%%%%%%%%%%%%%%%%%%%%%%%%%%%%%%%%%
\section{The $SU(2)\times U(1)$-Higgs model}
\label{sec:model}
%%%%%%%%%%%%%%%%%%%%%%%%%%%%%%%%%%%%%%%%%%%%%%%%%%%%%
The simplest extension of the Standard Model viable for electroweak baryogenesis is the 2-Higgs Doublet model (2HDM) \cite{firstorder1,firstorder2,firstorder5,firstorder9,firstorder10,firstorder12}, where the Minimal Standard Model is augmented by an additional Higgs doublet field\footnote{A first order phase transition can also be realised by just adding a singlet field, but then  insufficient CP-violation is a problem (see for instance \cite{firstorder3,firstorder4,firstorder6,firstorder7,firstorder8,firstorder10,firstorder11}).}. This new field couples to SU(2) and U(1) gauge fields exactly as the first one, and different models allow for different Yukawa coupling patterns between Higgs and fermion  fields (see \cite{Celis:2013rcs,Tanabashi:2018oca} for model descriptions and experimental constraints).

Depending on the shape of the Higgs potential, the quasi-order parameter of the transition (the operator that is close to zero outside the bubble, and large inside) may be the original Higgs field, the new Higgs field or a combination of the two. The bubble wall is then a field profile interpolating between the inside value and the outside value, minimizing the (free) energy. 

In this first investigative work, we will stick to an order parameter consisting of a single Higgs field, and postpone the case when the fields mix as we move through the wall. We will ignore SU(3) gluons and to a first approximation neglect the fermion degrees of freedom. Fermions colliding with the wall give an important contribution to the force determining the wall dynamics. But because we are moving the wall by hand, the backreaction on the wall is not our primary concern. Also, the numerical methods available to simulate real-time fermions are prohibitively expensive \cite{Saffin:2011kn}.

This means that we will approximate the electroweak plasma by the SU(2) and U(1) gauge fields, coupled to a single Higgs doublet. The doublet is fully dynamical, and interacts with the gauge fields but, as we will describe, an external current forces the Higgs to have a wall-like profile and drives it through the plasma with an overall shape and velocity of our choosing.

Our SU(2)-U(1)-Higgs model is then described by the continuum action
\begin{align}
{\mathcal L} =
-\frac{1}{4} W_{\mu\nu}^aW^{\mu\nu,a}
-\frac{1}{4} B_{\mu\nu}B^{\mu\nu}
-\left( D_\mu \phi \right)^\dagger\left( D^\mu \phi \right) -\frac{m_H^2}{2v^2} \left( \phi^\dagger \phi -\frac{v^2}{2} \right)^2
+\Delta {\mathcal L_2}
,
\end{align}
where $\phi$ is the Higgs scalar doublet, $W^a_\mu$ are the SU(2) gauge fields and $B_{\mu}$ the U(1) hypercharge gauge field.
We have defined
\begin{align}
W^a_{\mu \nu} =\partial_\mu W^a_{\nu} - \partial_\nu W^a_{\mu} + g_2\epsilon^{abc}  W^b_{\mu} W^c_{\nu}
,\quad
B_{\mu\nu} = \partial_\mu  B_\nu -\partial_\nu  B_\mu,
\end{align}
and the covariant derivatives
\begin{align}
D_\mu = \partial_\mu - i\frac{g_2}{2} \sigma^a W^a_{\mu} + i\frac{g_1}{2} B_\mu,
\end{align}
where the Higgs field has hypercharge $Y=-1/2$. We have introduced the parameters
\begin{align}
g_2=\frac{e}{\sin\theta}, 
\quad
g_1=\frac{e}{\cos\theta},
\quad
m_W=\frac{g_2v}{2}
,\quad
m_Z=\frac{g_2v}{2\cos\theta},
\end{align}
and we take
\begin{align}
m_W=80 \textrm{ GeV}
,\quad
m_Z=91\textrm{ GeV}
,\quad
m_H=125 \textrm{ GeV}
,\quad
v=246 \textrm{ GeV},
\end{align}
so that
\begin{align}
g_2=0.65
,\quad
g_1=0.35
,\quad
\cos\theta=0.88
,\quad
\sin^2\theta=0.227
,\quad
e=0.31\approx\sqrt{\frac{4\pi}{137}}.
\end{align}
The term $\Delta {\mathcal L_2}$ represents a potential CP-violating term, which we would like to introduce to bias the dynamics. We will return briefly to this in the conclusion.

%%%%%%%%%%%%%%%%%%%%%%%%%%%%%%%%%%%%%%%%%%%%%%%%%%%%%
\section{The wall and initial conditions}
\label{sec:wall}
%%%%%%%%%%%%%%%%%%%%%%%%%%%%%%%%%%%%%%%%%%%%%%%%%%%%%

In a first order electroweak phase transition, the high temperature global minimum of the effective potential at $|\phi|=0$ becomes degenerate with a second minimum at $0<|\phi|<v/\sqrt{2}$ as the temperature is lowered to some critical $T_c$. Lowering the temperature further, the second minimum takes over as the global minimum and the minimum at $|\phi|=0$ becomes first a local minimum, and eventually ceases to be a minimum altogether.

If the system starts out in the first minimum $|\phi|\simeq 0$, there is a period of metastability while this minimum still exists, but is not the global minimum. This becomes more and more pronounced as the (free) energy difference between the two minima increases with decreasing temperature. At some nucleation temperature, $T_N$, random thermal fluctuations have a sufficient probability of generating a large enough bubble of the low-temperature phase. If the bubbles are large enough, they will dynamically continue to grow rather than collapse back, and the phase transition is triggered. 

The walls of the bubble interpolate between the metastable minimum (outside) and the global minimum (inside), with a profile so as to minimize the (free) energy of the wall. In vacuum this profile may be found by shooting methods, or one may find an approximate expression for its shape by invoking a thin wall approximation or a thick wall approximation. Ultimately, one may solve for the profile numerically, also in a  thermal plasma (see for instance \cite{bubsim1}). It will depend on the supercooling $T_c-T$, the Higgs potential and the thermal fluctuations through the free energy associated with a unit area of wall, the wall tension. If the bubble is not very large relative to the wall thickness, curvature effects may also play a role. 

To tune the parameters and the temperature to the exact nucleation point, and to realise this transition in individual classical simulations is difficult. We will instead force a moving wall into our system by adding a current  $R({\bf x},t)$, so that the potential for the Higgs field is time and space dependent,
\begin{eqnarray}
V[\phi] =\frac{m_H^2}{2v^2} \left( \phi^\dagger \phi -\frac{v^2}{2} \right)^2 + R({\bf x},t)\phi^\dagger\phi.
\end{eqnarray} 
Whenever and wherever $R= 0$, we recover the Standard Model potential, which at zero temperature has a global minimum at $|\phi|= v/\sqrt{2}$. At leading order, thermal fluctuations at some temperature $T$ will contribute a term analogous to $R\simeq T^2$ and raise the global minimum, and at high enough temperature $T>T_c$, the global minimum is at $|\phi|=0$. This thermal contribution is the same inside and outside the bubble, and so in a true phase transition, an ``outside'' and an``inside" is established not because $R$ varies in space, but because the temperature (and hence potential) is so finely tuned that it has two minima and  both phases are allowed simultaneously.

As we will describe below, we will operate at a lower temperature than the critical one, and instead drive the wall through a choice of $R$ 
\begin{eqnarray}
R({\bf x},t)&=&\frac{1}{2}m_H^2 \quad \textrm{outside the bubble},\\
R({\bf x},t)&=&0 \quad\qquad \textrm{inside the bubble}.
\end{eqnarray}
In this way, the potential has a minimum at $|\phi|=0$ in the region of space we want to be outside the bubble and minimum at a non-zero value of the field in the region we want to be inside. By changing $R$, we can make the bubble grow, by first having $R$ non-zero in all of space, and then flipping it to zero in an ever expanding region. We emphasize that we do not directly stipulate the wall, but rely on the field dynamics to generate the wall itself in the background of this current. In particular, the Higgs field is free to have physical fluctuations around the overall wall shape. Moreover, $R$ is a gauge singlet, and so introducing does not affect Gauss's law, which is an important consideration in the numerical simulations. It turns out that for the initial conditons we have in mind (to be described shortly) it is important that the zero-temperature mass outside  (or prior to the appearance of) the bubble is small. By choosing the current $R$ to be $\frac{1}{2}m_H^2$ outside the bubble  this mass is zero, and we avoid that the random fluctuations, defined by the Bose-Einstein distribution below, are cut off by a large mass. 

%%%%%%%%%%%%%%%
\subsection{Initialization of Higgs  and gauge fields}
\label{sec:initial}
%%%%%%%%%%%%%%%

Our simulations emulate a region of space with a thermal plasma in the high-temperature phase, where a bubble wall sweeps through as it expands. In our numerical setup, we have a finite 3-D box, of dimension $L_xL_yL_z$, where $z$ denotes the direction of motion of the wall, and $L_z\gg L_x,L_y$. The box has periodic boundaries, a point to which we shall return below.

For the situation prior to the arrival of the bubble wall, we wish to introduce fluctuations in the Higgs and  gauge fields that are quasi-thermal , and may provide an ensemble of random classical field realisation. This is in principle straightforward for a free theory, but because the fields interact, the true thermal state is non-linear and must obey Gauss's law. This may in turn be resolved by brute force, for instance by Monte Carlo methods \cite{Tranberg:2003gi}, but this is numerically heavy. Since we are not attempting to reproduce an exactly thermal state with a specific temperature, we allow ourselves some short cuts. These may later be ameliorated.

For each of the four real component of the Higgs scalar doublet we set  $\phi_a=0$, $a=1,2,3,4$. The momentum variables $\pi_a = \partial_t\phi_a$ are taken to be non-zero, and to satisfy the free-field two-point correlations 
\begin{align}
\label{eq:initpi}
\langle \pi_a({\bf p}) \pi_b({\bf p}') \rangle  &= 2\omega_p n_p(2\pi)^3\delta_{ab}\delta^3({\bf p}-{\bf p}').
\end{align}
The total energy of a free field system is 
\begin{align}
E= \frac{1}{2}\int \frac{\ud^3{\bf p}}{(2\pi)^3} \sum_a\left(\pi_a({\bf p})^\dagger\pi_a({\bf p}) +\omega_p^2\phi_a^\dagger({\bf p})\phi_a({\bf p})\right).
\end{align}
For an equipartitioned system the two terms  in the bracket give the same contribution, and if we insist on not initialising $\phi$ (the second term), we can mimic the correct initial condition by putting twice as much energy in the momentum component $\pi$  (the first term). This explains the overall factor of 2 in (\ref{eq:initpi}).

The initial state is then determined by the particle number $n_p$ and the frequency $\omega_p$, which we take to follow a Bose-Einstein distribution,
\begin{align}
n_p=\frac{1}{e^{\omega_p/T}-1},
\label{eq:BE}
\end{align}
with 
\begin{align}
\omega_{p} =\sqrt{p^2+m^2_{\rm eff}}
,\quad 
m^2_{\rm eff} = \left(\frac{3g_2^2}{2}+\frac{g_1^2}{2}\right)\frac{T^2}{12} + \frac{3m_H^2}{v^2} \int \frac{d^3{\bf p}}{(2\pi)^3}\frac{n_p}{\omega_p}
.
\label{eq:thermmass}
\end{align}
We have introduced the one-loop thermal mass, $m^2_{\rm eff}(T)$, to take into account (at leading order) that the system is interacting \cite{Carrington:1991hz,Arnold:1992rz}. We find the thermal mass self-consistently by iterating Eqs. (\ref{eq:BE}), (\ref{eq:thermmass}). Since we have chosen the zero-temperature mass to be zero outside the bubble, only the thermal mass enters in the dispersion relation. This is convenient, since it is fairly small and allows for sizable occupation numbers in the IR, while still cutting off the IR divergence. 

By choosing this initialisation, where $\phi_a=0$, we ensure that the initial local Higgs charge vanishes, rendering the Gauss law much simpler to solve. Obviously,  already one time-step further down the line, both  $\phi_a$ and $\pi_a$ will be non-zero. However, the ensemble does not start out thermal, since only the momentum variables are initialized. Rather quickly, the energy will equipartition to some near-equilibrium state, but the parameter $T$ is no longer the actual physical temperature. We should therefore treat it as a parameter fixing the energy density, which in turn is related to the actual temperature of the system.  For more on  real-time initialization of Higgs fields on the lattice, see \cite{Mou:2017zwe}.

We initialize the gauge fields in a similar way, but now we set the momentum variables, the electric fields, to zero $E^a_i=0$. The field variables $A_{i}$ are initialized as a free-field-like thermal ensemble, 
\begin{align}
\langle A^a_i ({\bf p})A^b_j({\bf p}')\rangle &= 2\frac{n_p}{\omega_p}\delta^{ab}\left(\delta_{ij}-\frac{p_ip_j}{p^2}\right)(2\pi)^3\delta^3({\bf p}-{\bf p}'),
\end{align}
again with an overall factor of $2$, so that the total initial energy is approximately correct.
This applies to the three $SU(2)$ gauge fields and single $U(1)$ gauge field. In practice, we write 
\begin{align}
A^a_i(x) &=\sqrt{\frac{2}{V}} \sum_p e^{ipx} \sqrt{\frac{n_p}{2\omega_p}} \sum_{\lambda} \epsilon_i({\bf p},\lambda)\xi^a({\bf p},\lambda),
\end{align}
where $\xi^a(p,\lambda)$ is complex random number with variance $\langle\left(\xi^a({\bf p},\lambda)\right)^*\xi^a({\bf p},\lambda)\rangle=2$ and the two transverse polarization vectors are 
\begin{align}
\label{eq:polar}
{\bf \epsilon}({\bf p},1)=\frac{{\bf r}\times {\bf p}}{|{\bf r}\times {\bf p}|} 
,\quad
{\bf \epsilon}({\bf p},2)=\frac{{\bf\epsilon}({\bf p},1)\times {\bf p}}{|{\bf \epsilon}({\bf p},1)\times {\bf p}|} 
,
\end{align}
with ${\bf r}$ a random vector.
The initial particle number again follows the Bose-Enstein distribution,
\begin{align}
n_p=\frac{1}{e^{\omega_p/T}-1}
,\quad
\omega_{p} =|p|.
\end{align}
In the symmetric phase, we take all the gauge fields to be massless, and so the zero mode is singular. We set it to zero initially. 

With this initialization, Gauss's law is explicitly obeyed everywhere on the lattice initially, and the classical equations of motion ensure that this is then also true at all later times. 

%%%%%%%%%%%%%%%
\subsection{Establishing the wall}
\label{sec:establishwall}
%%%%%%%%%%%%%%%

\begin{figure}[H]
\begin{center}
\includegraphics[width=1.0\textwidth]{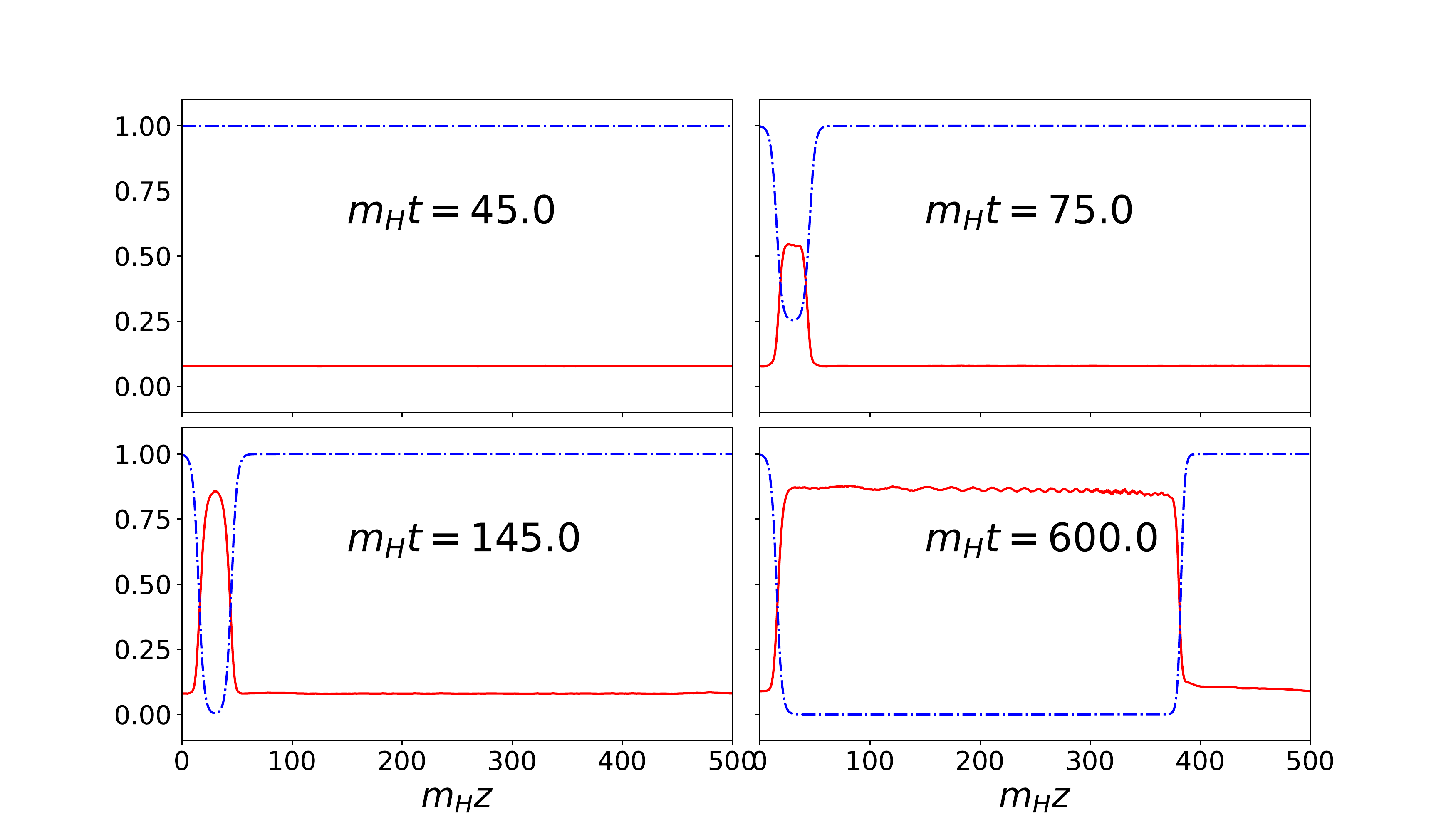}
\caption{
An approximately thermal state in the symmetric phase is created (top left) by evolving the initial data of section \ref{sec:initial} for $0<m_Ht<50$. The current, $R$ (blue dashed line), is switched off in a small region of space for $50<m_Ht<100$, creating two stationary Higgs field walls (red line) with the broken phase in the middle, and symmetric phase outside (top right). Once the current $R$ has reached its minimum value in that region we hold it fixed for $100<m_Ht<150$ to allow transients to dissipate (bottom left). The right hand boundary of the current then starts to extend in the positive $z$ direction, creating an expanding bubble of broken phase ready for the simulation (bottom right).  $v=0.75$, $T/m_H=1$, $am_H=0.5$.
}
\label{fig:wallup} 
\end{center}
\end{figure}

The procedure for preparing our initial condition is described in the following, and is depicted in Fig. \ref{fig:wallup}:
\begin{itemize}
\item We start by initialising the fields throughout the lattice at a temperature $T_{i}$ (see section \ref{sec:initial}) in the symmetric phase with  $R({\bf x},t)=\frac{1}{2}m_H^2$. We will generate a whole ensemble of such classical realisations, and observables quoted below are averages over such an ensemble of real-time simulations.
We leave the system to equilibrate for a time $m_Ht_{\rm thermal}=50$, as seen in Fig. \ref{fig:wallup}, top left.
\item
We then establish the wall by changing the current  $R({\bf x},t)$, so  that it depends on time and space, but in the $z$-direction only
\begin{align}\label{eq:R_create_bubble}
R({\bf x},t)=R(z,t)=\frac{m_H^2}{4}\left[ 2+ c(t)\tanh\left(\frac{z-z_{r}}{0.25d}\right) - c(t)\tanh\left(\frac{z-z_{l}}{0.25d}\right) \right].
\end{align}
We will think of this as smoothed-out step functions, and will establish the profile of the wall through the parameter $d$. 

We are creating two walls, one centered at $z_r$ and one at $z_l$. This is so that we may have periodic boundary conditions in the  $z$-direction throughout. The  $x$- and $y$-directions along the wall are also periodic. We may later decide whether both walls move outwards or just one of them. $z_r-z_l$ determine the size of the initial bubble, prior to expansion. Hence before we make the walls move, part of the volume is already in the broken phase. 

The time dependent parameter $c(t)$ changes from 0 to 1 and determines how fast the bubble wall is established. If we do this very fast, the wall is shocked into existence and this may bring in unwanted, and unphysical, transient effects. In practice, we find that choosing 
\begin{align} 
c(t)=\frac{1}{2}\left(1+(t-\tau_{\rm thermal})/\tau_{\rm wall}\right)
\end{align}
is convenient\footnote {We checked, that the precise prescription for changing $c(t)$ from 0 to 1 does not influence the result much.}, with $m_H\tau_{\rm wall}=m_H\tau_{\rm thermal}=50$. This is seen in Fig.~\ref{fig:wallup}, top right.
\item After the current is established ($c(t)=1$), we leave the wall to settle into its shape and for any transients to damp away. We do this for a time $m_H\tau_{\rm stable}=50$. We are then ready for the simulation proper to commence. Fig.~\ref{fig:wallup}, bottom left, and bottom right show these phases.
\end{itemize}

%%%%%%%%%%%%%%%%
\subsection{Running the wall}
\label{sec:running}
%%%%%%%%%%%%%%%%%

\begin{figure}[H]
\begin{center}
\includegraphics[width=0.8\textwidth]{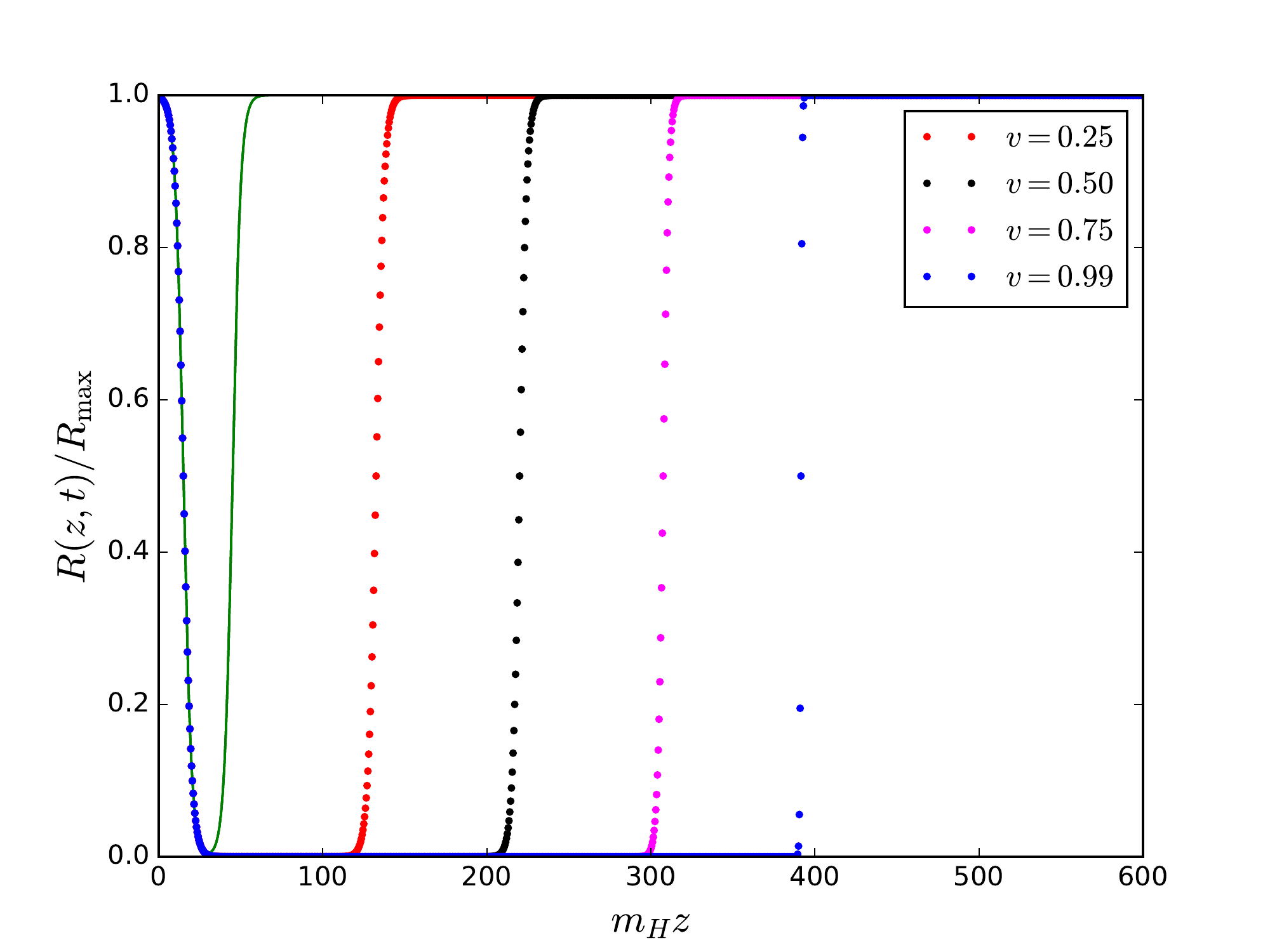} 
\caption{The bubble walls prior to moving  is the solid (green) curve. The dotted curves are snapshots at $m_Ht=500$ for  different speeds.}
\label{fig:wallspeeds}
\end{center}
\end{figure}

After the walls have been established and have settled down, we may control their motion by further changing the current $R$. We use the form
\begin{align}
R(z,t)=\frac{m_H^2}{4}\left[ 2+ \tanh\left(\gamma\frac{z-z_{r}-v(t-\tau_{\rm move})}{0.25d}\right) - \tanh\left(\frac{z-z_{l}}{0.25d}\right) \right],
\end{align}
%\comment{\ref{eq:R_create_bubble} has $m_H^2/4$, I think this should also have 1/4, right?}
with starting time, $\tau_{\rm move}=\tau_{\rm thermal}+\tau_{\rm wall}+\tau_{\rm stable}$, taken to be $=150$. $v$ is the speed of the wall, and we have included the Lorentz factor $\gamma = 1/\sqrt{1-v^2}$ for the moving (scalar) wall. Note that we may choose to move both walls, but to take maximum advantage of the available lattice volume, we move only one, keeping the other at rest. In  Fig.~\ref{fig:wallspeeds} we show the initial wall and walls after a time $m_Ht=500$ with different speeds. Dots indicate the lattice points, and give an impression of the discretization. Clearly, by $v=0.99$ our resolution of the wall is no longer satisfactory.

To summarize: On a lattice of size $N_x=N_y=64\ll N_z=1000$, physical size $L_i=N_ia$ with lattice spacing $a$, $am_H=0.5$, we initialize Higgs and gauge fields using a parameter $T$, and let it settle. We then grow a pair of walls spaced by $z_r-z_l$ with a width parameter $dm_H=15$, and let it settle. We then drive one of those walls through the plasma with a speed parameter $v$, until we run out of box. 

With this many parameters, we have had to make some choices, and although we have tried quite a few combinations to find some sensible values, a complete sweep has not been done. For the largest possible volumes $N_x,N_y$, $N_z$, we focus on the dependence on $v$ for a few values of  $T$.

%%%%%%%%%%%%%%%%%%%%%%%%%%
\section{Observables}
\label{sec:observables}
%%%%%%%%%%%%%%%%%%%%%%%%

Given the initial conditions presented above, the simulation is ready to go. We are interested in monitoring a broad set of observables, as we carry out the simulation. We monitor total Gauss law and total energy over the lattice to check our numerics. But the physically most interesting quantities are space-, and in particular $z$-dependent. We will therefore introduce a number of quantities averaged over x-y spatial slices, such as the Higgs field
\begin{eqnarray}
\label{eq:higgsav}
\phi^2(z,t) = \frac{1}{L_xL_y}\int dx\, dy\, \phi^\dagger\phi(x,y,z,t).
\end{eqnarray}
This may further be integrated over $z$, to get average Higgs field over the whole volume $\phi^2(t)$, if required. Similarly, for the observables introduced in the following, we will consider local, slice and global versions as appropriate.

%%%%%%%%%%%%%%%%
\subsection{Thermodynamical observables}
\label{sec:thermobs}
%%%%%%%%%%%%%%%%

We may compute diagonal components of the energy-momentum tensor. For the scalar field the energy density is
\begin{align}
\label{eq:higgsenergy1}
T_{\phi}^{00} = 
\pi^\dagger \pi + (D_i\phi)^\dagger D^i\phi + \left(R-\lambda v^2\right)\phi^\dagger\phi +\lambda(\phi^\dagger\phi)^2. 
\end{align}
The three diagonal space components are
\begin{align}
T_{\phi}^{zz} = 2\pi^\dagger \pi 
+2\left(D_z \phi\right)^\dagger \left(D_z \phi\right)
-T_{\phi}^{00},
\end{align}
and 
\begin{align}
T_{\phi}^{xx}=T_{\phi}^{yy} = 2\pi^\dagger\pi 
+2\left(D_{x,y} \phi\right)^\dagger \left(D_{x,y} \phi\right)
-T_{\phi}^{00},
\end{align}
where the equality of the latter two is ensured by the symmetry of the system.

Similarly, for the gauge fields we have
\begin{align}
T_{W,B}^{00} =
\sum_{i}\left[
\frac{1}{2} W_{i0}^aW_{i0}^a
+\frac{1}{2} B_{i0}B_{i0}
\right]
+\sum_{i,j>i}\left[
\frac{1}{2} W_{ij}^aW_{ij}^a
+\frac{1}{2} B_{ij}B_{ij}
\right],
\end{align}
while the three diagonal components are 
\begin{align}
T_{W,B}^{zz} =
\sum_{i\neq z}\left[
 W_{i0}^aW_{i0}^a
+ W^a_{zi}W^a_{zi}
+ B_{i0}B_{i0}
+ B_{zi}B_{zi}
\right]- T_{W,B}^{00},
\end{align}
and 
\begin{align}
T_{W,B}^{xx} =T_{W,B}^{yy} =
\sum_{i\neq z}\left[
 W_{i0}^aW_{i0}^a
+ W^a_{(x,y)i}W^a_{(x,y)i}
+ B_{i0}B_{i0}
+ B_{(x,y)i}B_{(x,y)i}
\right]- T_{W,B}^{00}.
\end{align}
For all of these quantities, we will present results averaged over a slice perpendicular to the $z$-direction (similar to (\ref{eq:higgsav})). 

It is not unreasonable, as a first approximation, to assume that the plasma acts as a perfect fluid. In that case, the energy-momentum tensor can be written as
\begin{eqnarray}\label{eq:stress:perfect_fluid}
T^{\mu\nu}=(\rho+P)U^\mu U^\nu+P\eta^{\mu\nu}.
\end{eqnarray}
Under the assumption that the fluid moves only in the $z$-direction, we can write
\begin{eqnarray}
U^\mu=\frac{1}{\sqrt{1-u^2}}\left(\begin{tabular}{c}
1\\0\\0\\u
\end{tabular}\right)
=
\left(\begin{tabular}{c}
a\\0\\0\\b
\end{tabular}\right).
\end{eqnarray}
Inserting this into (\ref{eq:stress:perfect_fluid}), we find
\begin{eqnarray}
T^{\mu\nu}=\left(
\begin{tabular}{cccc}
$a^2(\rho+P)-P$&0&0&$ab(\rho+P)$\\
0&$P$&0&0\\
0&0&$P$&0\\
$ab(\rho+P)$&0&0&$b^2(\rho+P)+P$
\end{tabular}
\right),
\end{eqnarray}
so that
\begin{eqnarray}
\label{eq:fluid}
P=T^{xx} =T^{yy},\quad\rho= T^{xx}-T^{zz}+T^{00}, \quad u^2=\frac{T^{zz}-T^{xx}}{T^{xx}+T^{00}}.
\end{eqnarray}
We see that if the space-like diagonal components are the same, i.e. $b=\frac{u}{\sqrt{1-u^2}}=0$, then the energy density is simply the 00-entry. We will see below, however, that because of the push of the wall, the $zz$-component is different from the other two, and there is a net space-dependent  fluid velocity, which we extract as indicated in Eq.~(\ref{eq:fluid}).

%%%%%%%%%%%%%
\subsection{Topological observables}
\label{sec:topobs}
%%%%%%%%%%%%%%%

\begin{figure}[H]
\begin{center}
\includegraphics[width=01.0\textwidth]{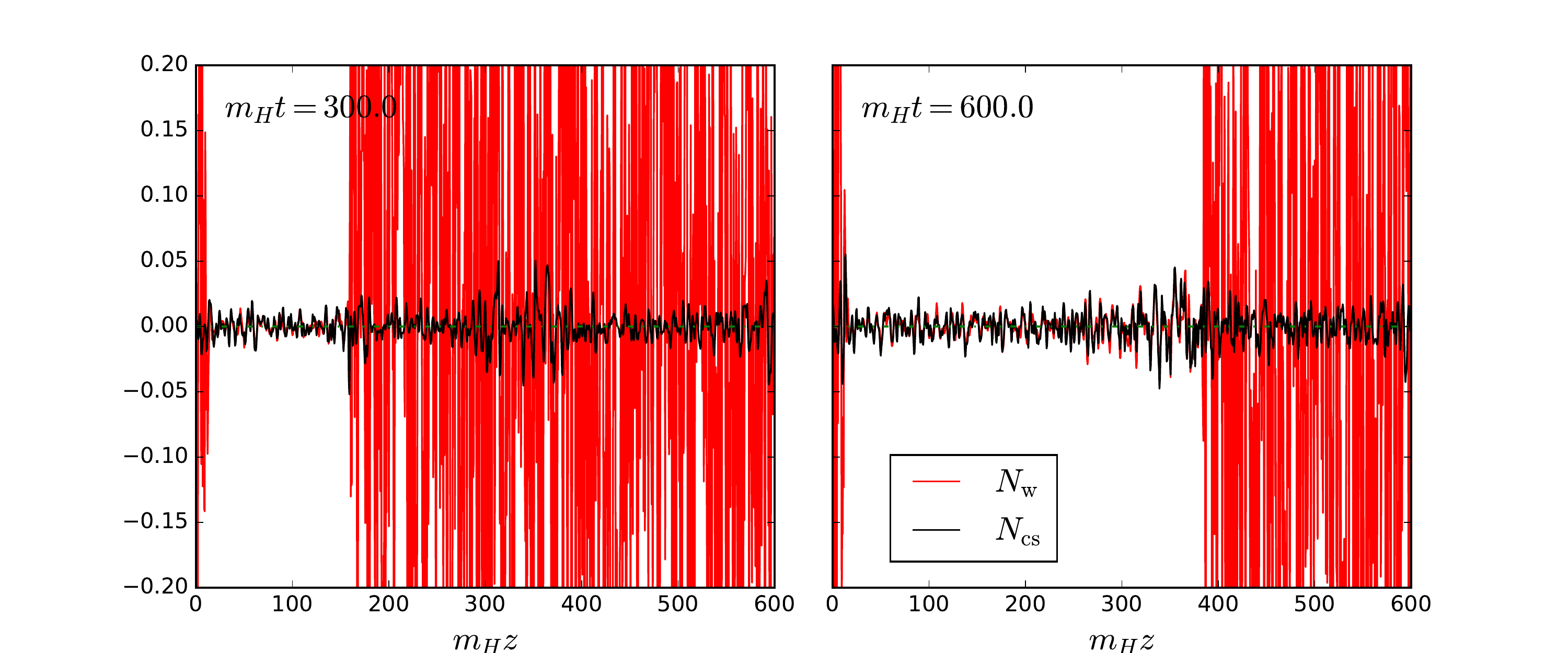} 
\caption{ The freeze-in of topological charge for two different times, for a single initial condition realisation. The Chern-Simons number density (black) is fairly well behaved throughout, while the (lattice definition of) the Higgs winding number (red) is wild in the symmetric phase (on the right and far left or the lattice). As soon as the wall passes, however, the winding number density freezes into a value close to the Chern-Simons number.
$am_H=0.5$, lattice size $36\times 36\times 1200$, $T/m_H=0.5$, $v=0.75$,  $m_Hd=10$.
}
\label{fig:freezein_nw}
\end{center}
\end{figure}

Electroweak baryogenesis is closely related to the chiral anomaly, relating baryon number to the Chern-Simons number of the underlying gauge field. 
\begin{align}
&B(t_f)-B(0) = L(t_f)-L(0) = 
3[N_{\rm cs} (t_f)-N_{\rm cs}(0)]\\
&= 
\frac{3}{64\pi^2} \int_{t_0}^{t_f} dt \int d^3x
~ \epsilon^{\mu\nu\rho\sigma}
\left[g_2^2 W^a_{\mu\nu}W^a_{\rho\sigma}
-g_1^2B_{\mu\nu}B_{\rho\sigma} \right].
\end{align}
For the degrees of freedom included in our simulations, only Chern-Simons number is available to us, one for the SU(2) field and one for the U(1) hypercharge field,
\begin{align}
N_{\rm cs,2} (t_f)-N_{\rm cs,2}(0)=& \frac{g_2^2}{32\pi^2}\int d^3x \epsilon^{ijk} 
\left(W^a_i W^a_{jk}-\frac{g_2}{3}\epsilon^{abc}W^a_i W^b_j W^c_k\right)
,
\end{align}
and
\begin{align}
N_{\rm cs,1} (t_f)-N_{\rm cs,1}(0)=& -\frac{g_1^2}{32\pi^2}\int d^3x \epsilon^{ijk} 
B_i B_{jk}
.
\end{align}

A particularly useful proxy for the baryon asymmetry turns out to be the Higgs winding number 
\begin{align}
N_{\rm w}&=-\frac{1}{24\pi^2} \int d^3x \epsilon^{ijk}{\rm Tr} \left[ \partial_i\Omega\Omega^{-1}  \partial_j\Omega\Omega^{-1}  \partial_k\Omega\Omega^{-1}  \right]
,\quad
\Omega = \frac{\Phi}{|\Phi|},
\end{align}
where the matrix form of the Higgs field $\phi$ is defined by
\begin{align}
\Phi = \left(\phi, i\sigma_2\phi^*\right).
\end{align}
Close to equilibrium and at low temperatures, $N_{\rm w}\simeq N_{\rm cs}\simeq N_{\rm cs,2}$. The first relation follows from minimization of the covariant derivative which is achieved at low energies, when the gauge fields are close to a ``pure gauge" vacuum. It is not an identity, and Chern-Simons number and winding number may be very different in a violent non-equilibrium environment. The second relation follows from the field space of the U(1) field, which has a single vacuum at $N_{\rm cs,1}= 0$. In contrast, the SU(2) field has infinitely many degenerate vacua, enumerated by integer values of $N_{\rm cs,2}$. 

We will consider Higgs  winding number ``density" in $z$, corresponding to the integrand of the expression for $N_{\rm w}$, but which is then integrated over $x$ and $y$. Outside the bubble, where $\phi\simeq 0$, the winding number density observable is very noisy, and integrating it up as described is demonstrated in Fig.~\ref{fig:freezein_nw}. We will therefore choose to show only the winding number inside the bubble for either the whole bubble volume (which grows over time as the bubble expands), or a smaller, constant volume immediately behind the wall.

%%%%%%%%%%%%%%%%%%%%%%%%%%%%%%%%%%%%%%%%%%%%%%%%%%%%%
\section{Thermodynamics of the plasma}
\label{sec:thermo}
%%%%%%%%%%%%%%%%%%%%%%%%%%%%%%%%%%%%%%%%%%%%%%%%%%%%%
%
%
%%%%%%%%%%%%%%%%%%%%%%%%%%%%%%%%%%%%%%%%%%%%%%%%%%%%%
\subsubsection*{Wall motion and energy release}
%%%%%%%%%%%%%%%%%%%%%%%%%%%%%%%%%%%%%%%%%%%%%%%%%%%%%
%
\newpage

\begin{figure}[H]
%\vspace{-3cm}
\begin{center}
\begin{tabular}{lll}
%\hspace{-2cm}
\includegraphics[width=0.27\textwidth]{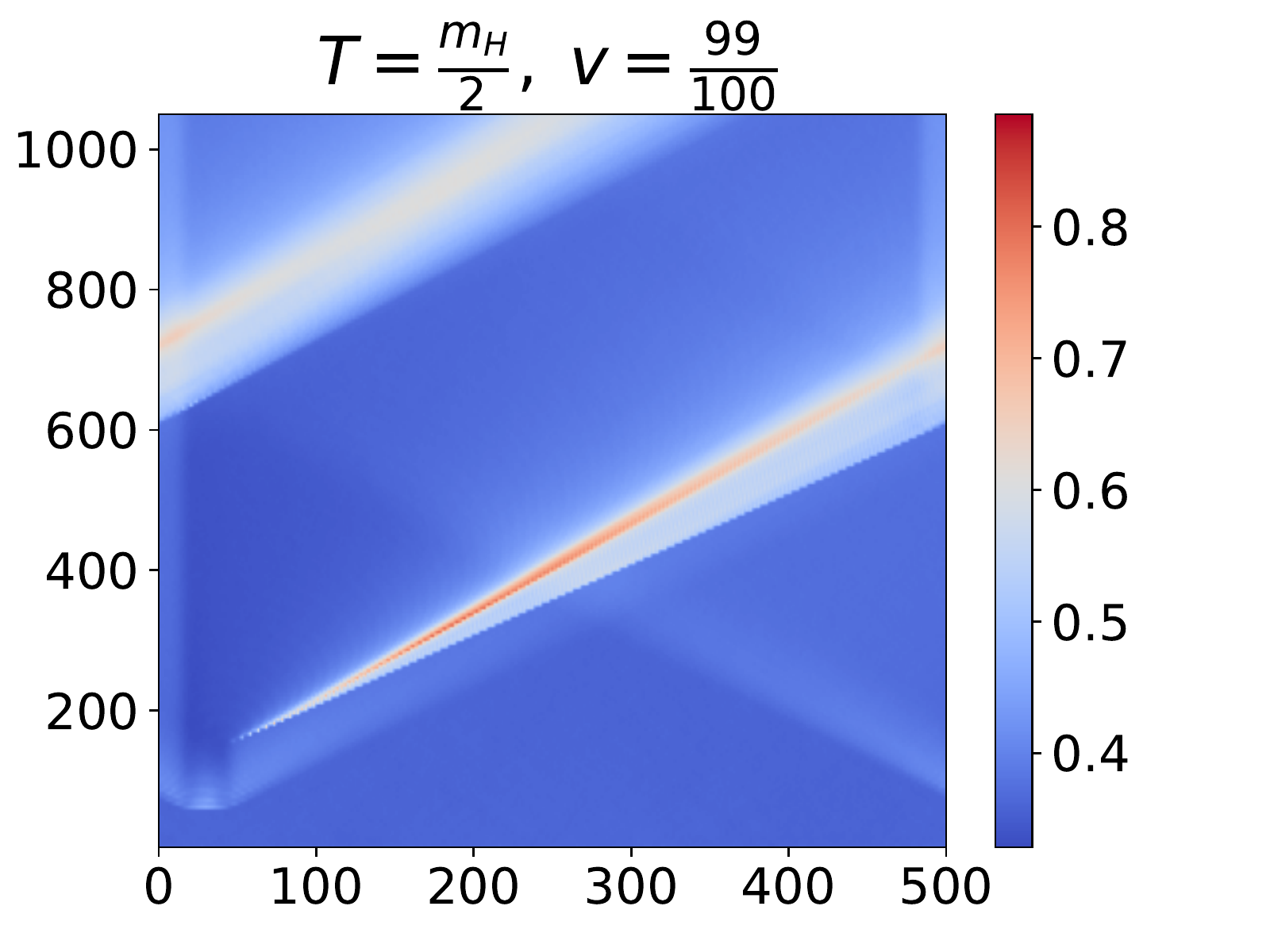} 
&
\includegraphics[width=0.27\textwidth]{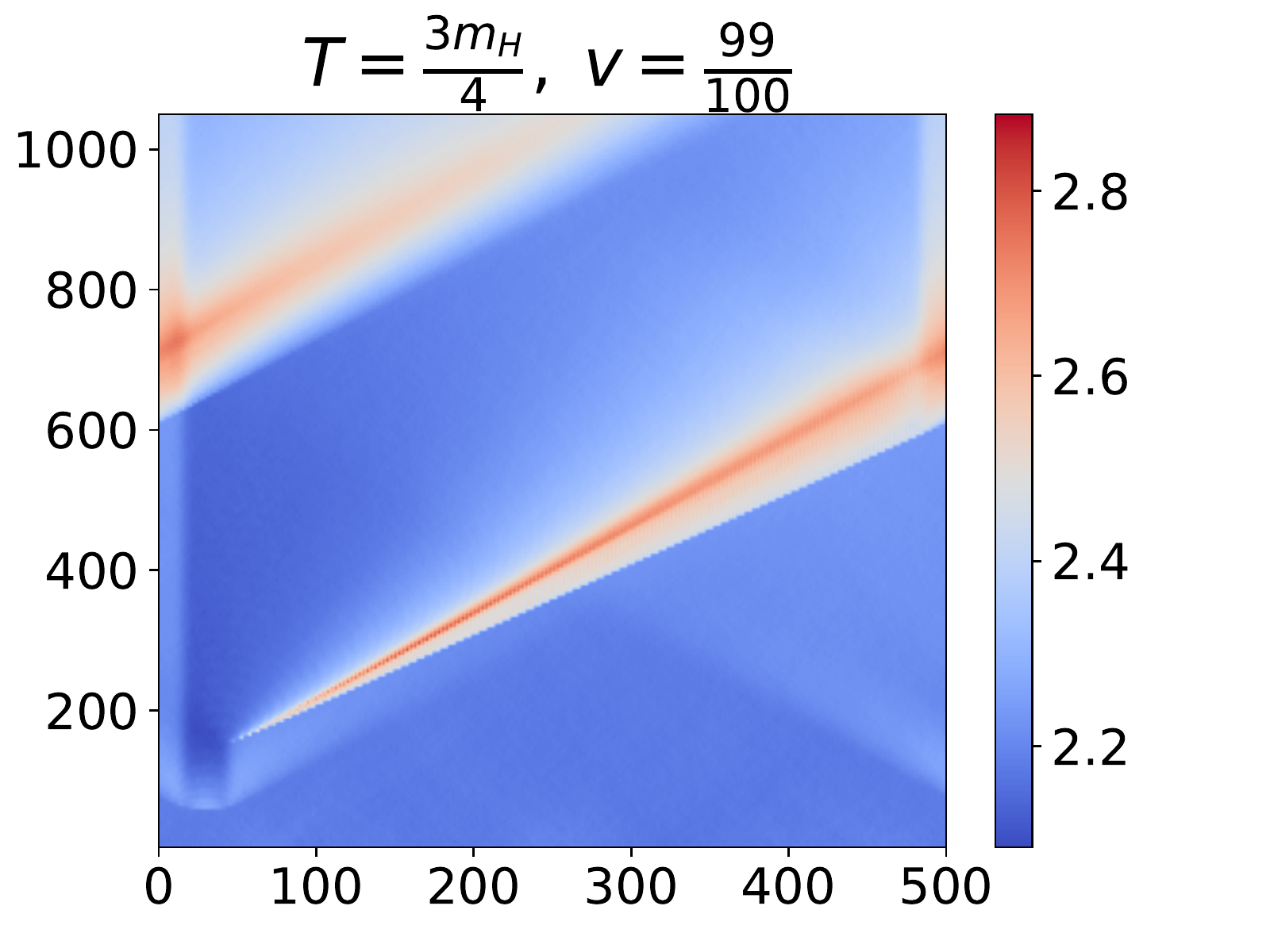} 
&
\includegraphics[width=0.27\textwidth]{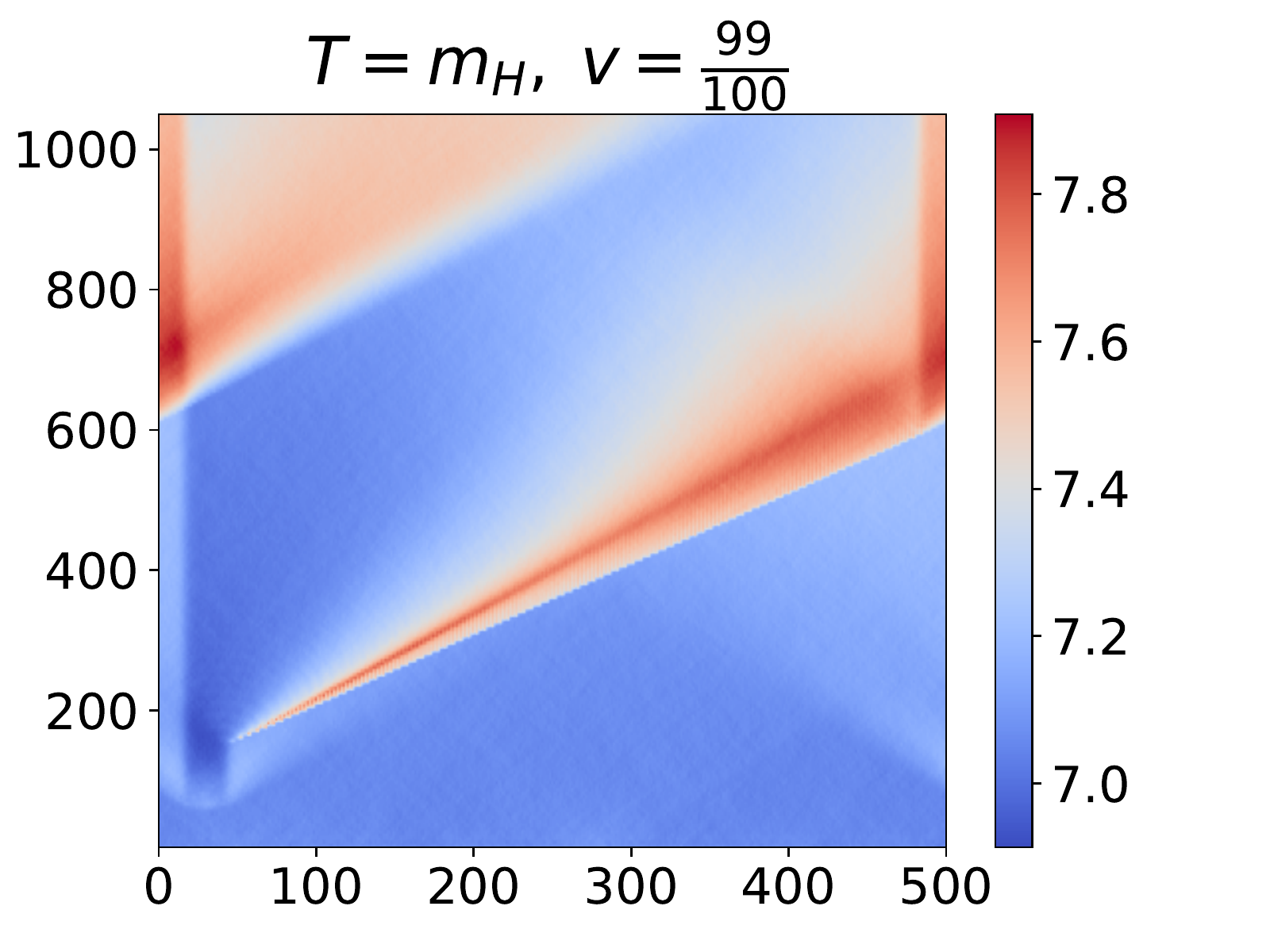} 
\\
\includegraphics[width=0.27\textwidth]{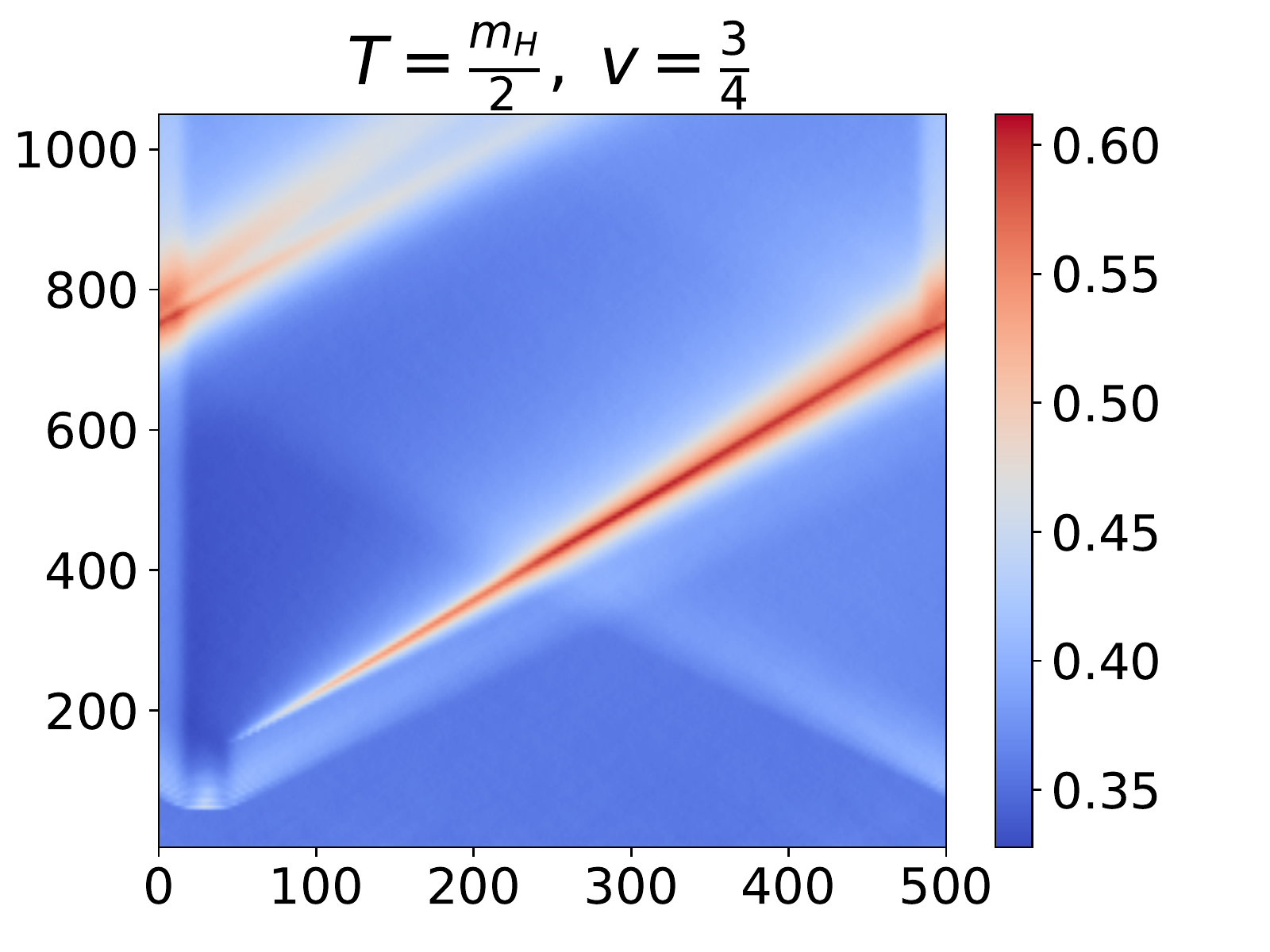} 
&
\includegraphics[width=0.27\textwidth]{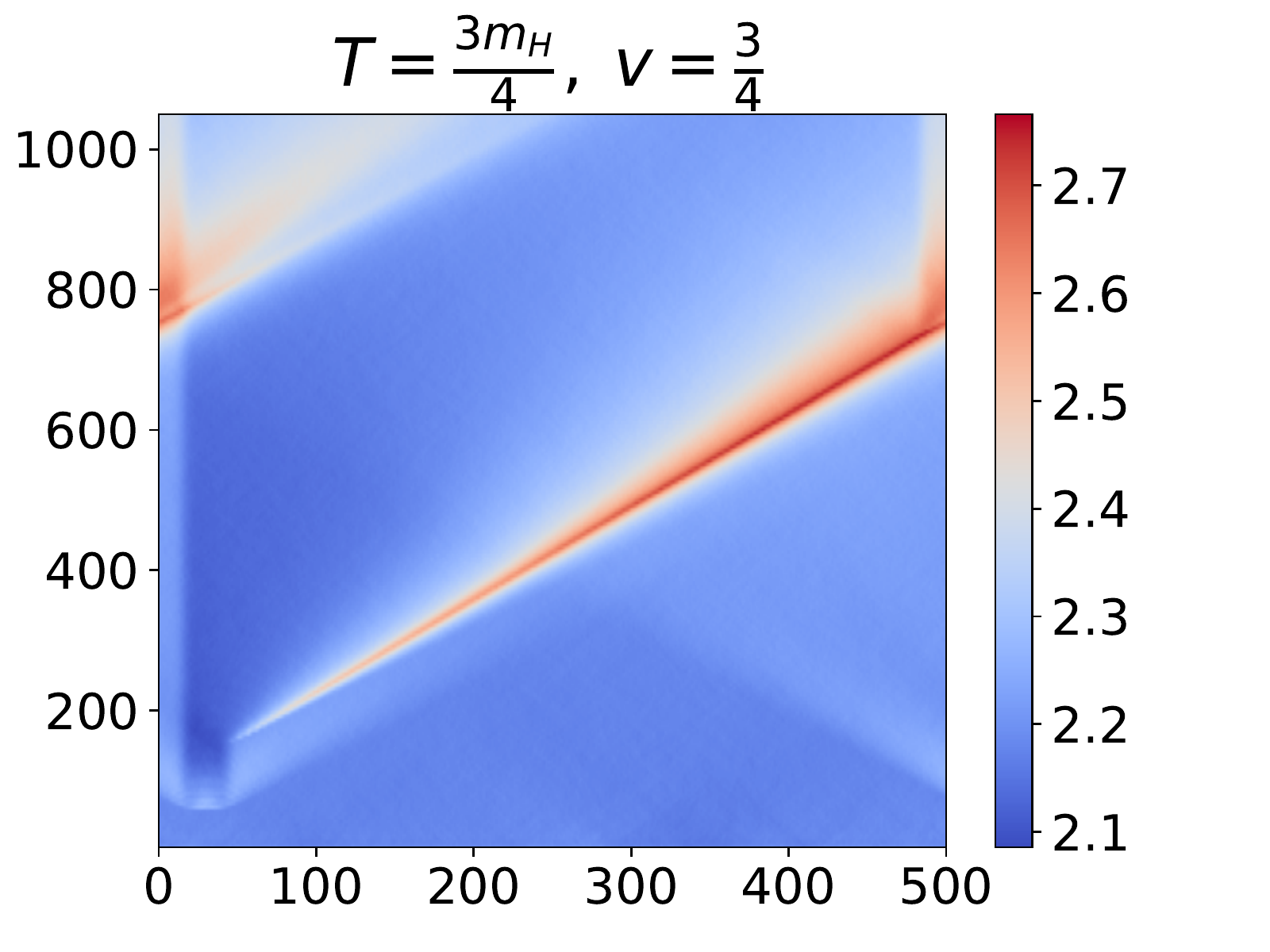} 
&
\includegraphics[width=0.27\textwidth]{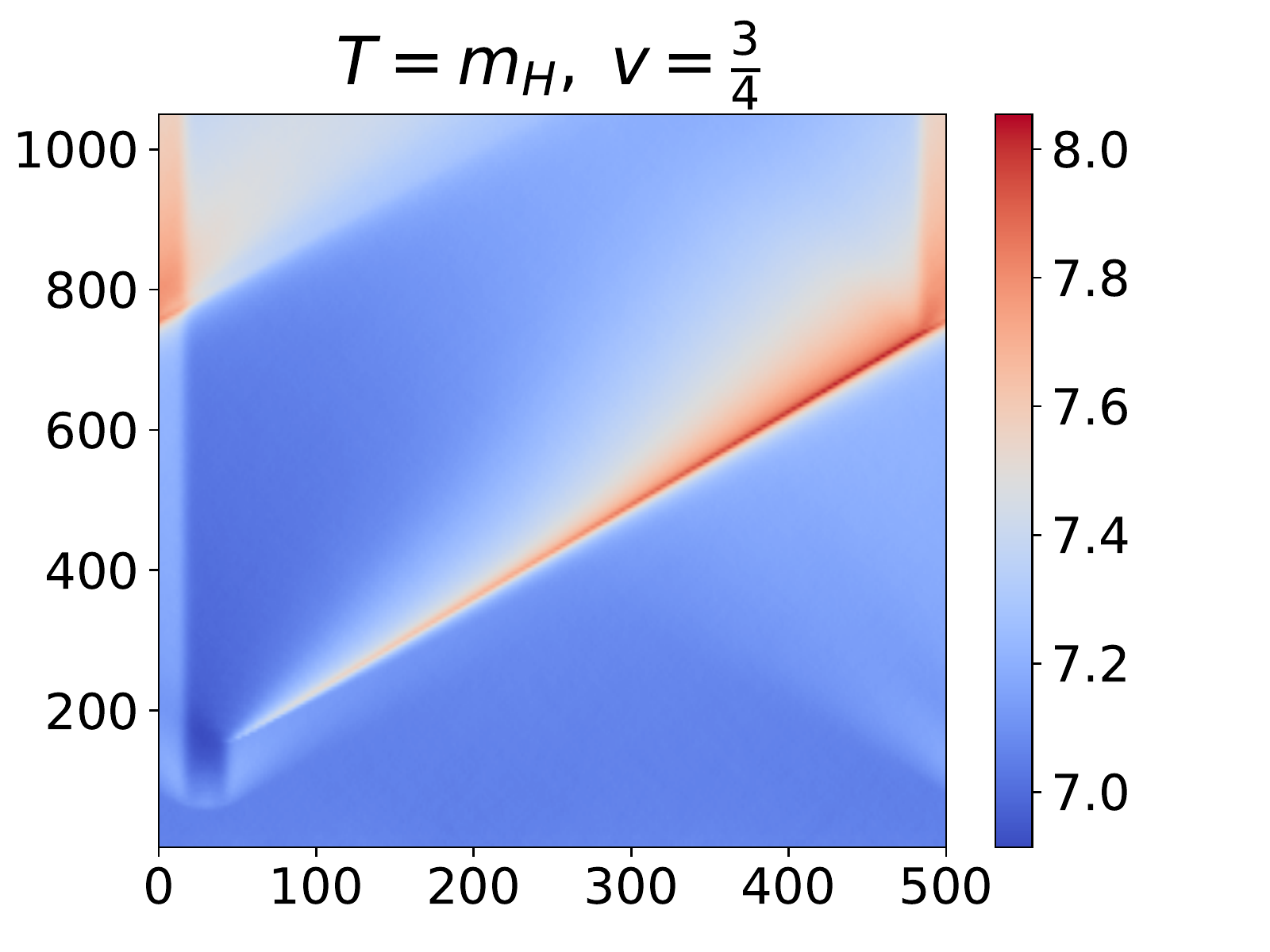} 
\\
\includegraphics[width=0.27\textwidth]{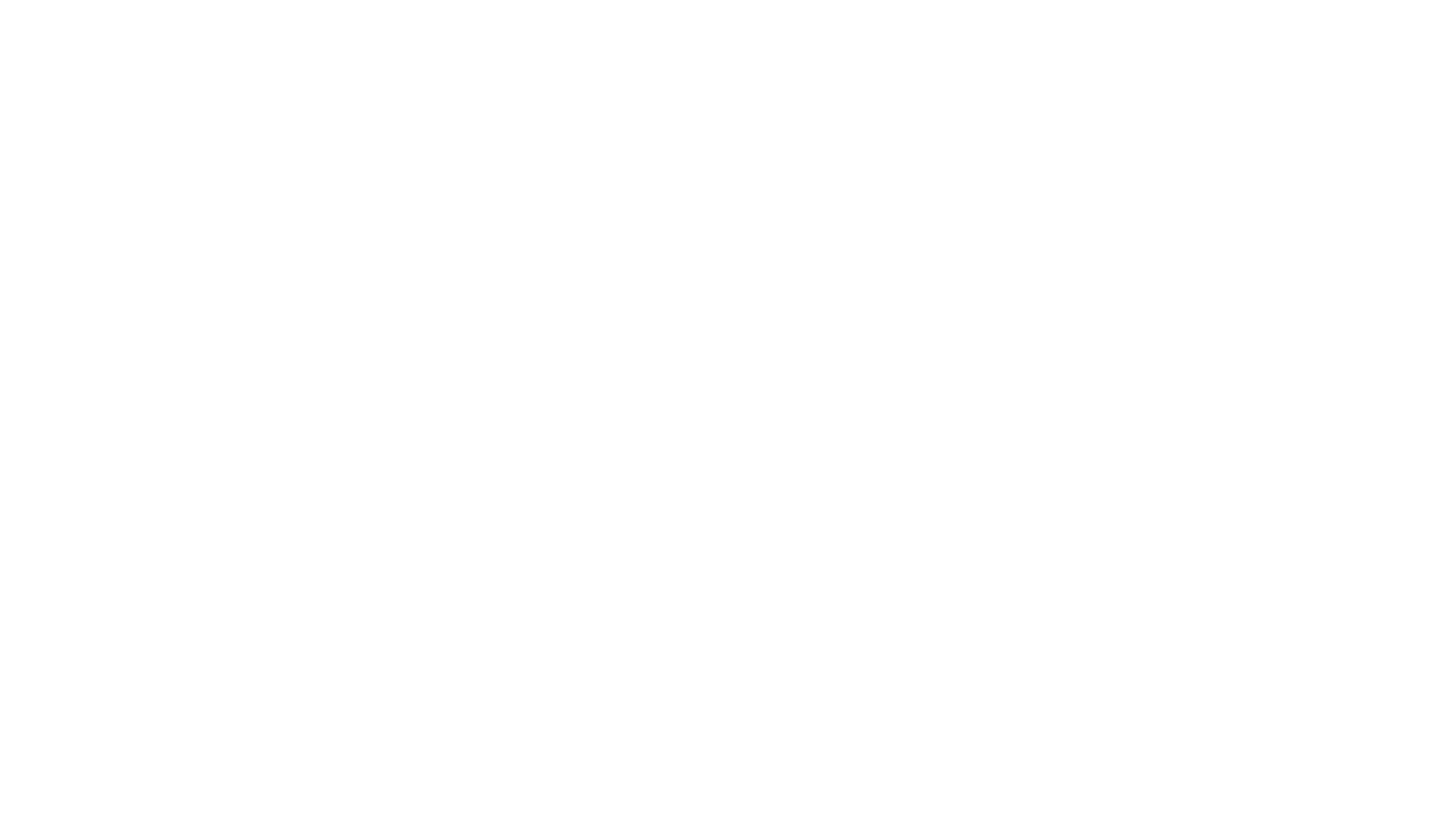} 
&
\includegraphics[width=0.27\textwidth]{Placeholder.pdf} 
&
\includegraphics[width=0.27\textwidth]{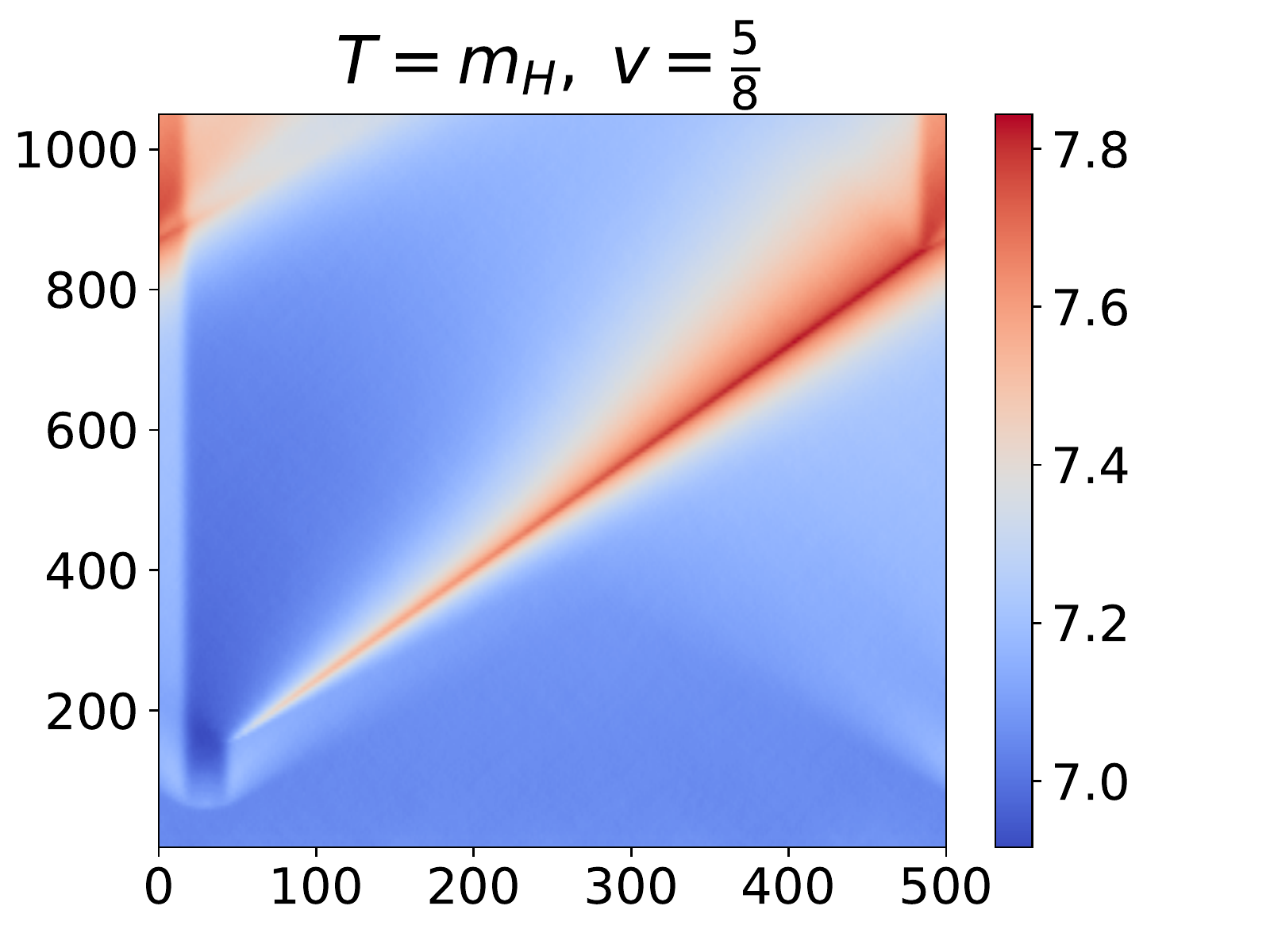} 
\\
\includegraphics[width=0.27\textwidth]{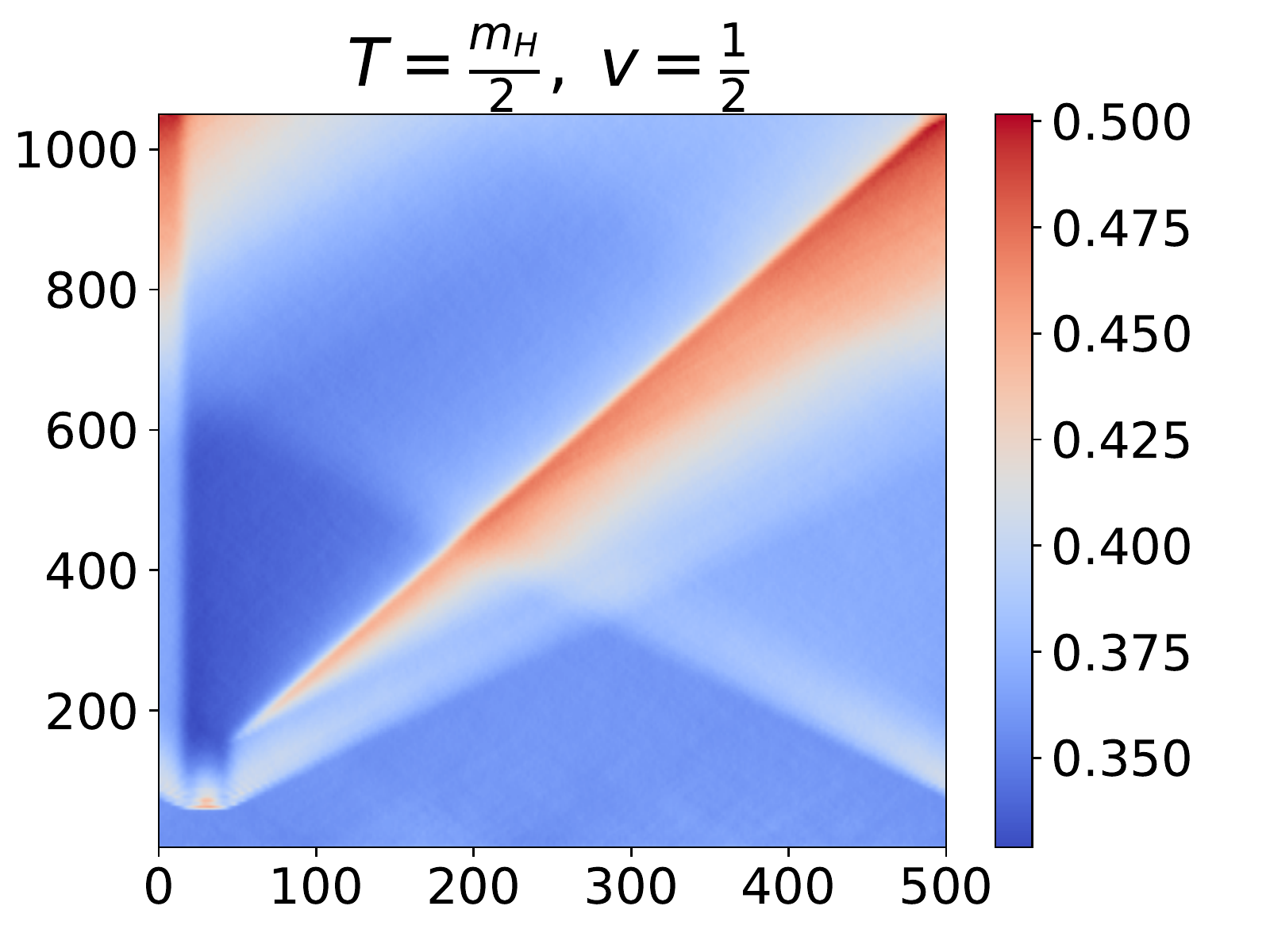} 
&
\includegraphics[width=0.27\textwidth]{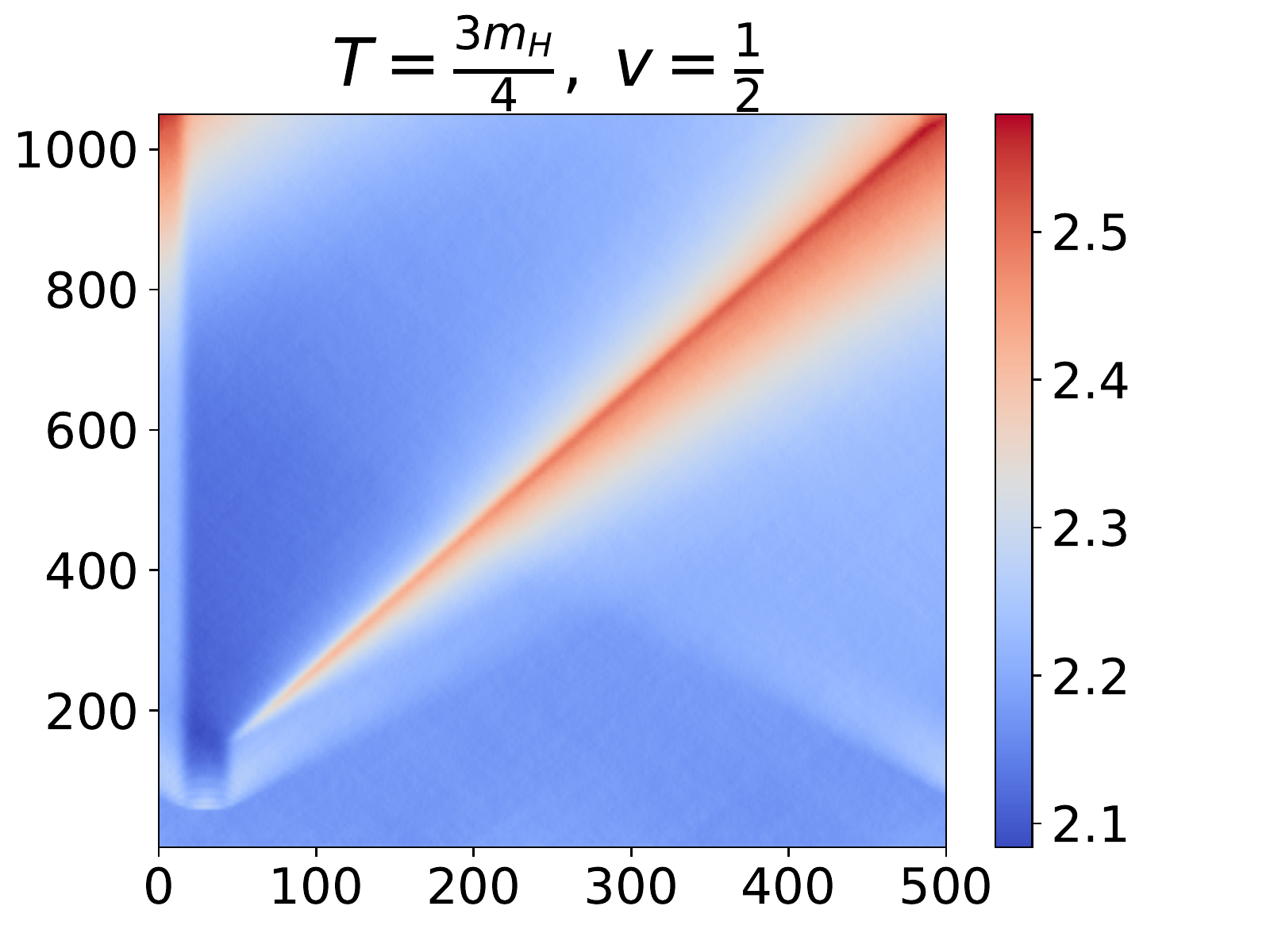} 
&
\includegraphics[width=0.27\textwidth]{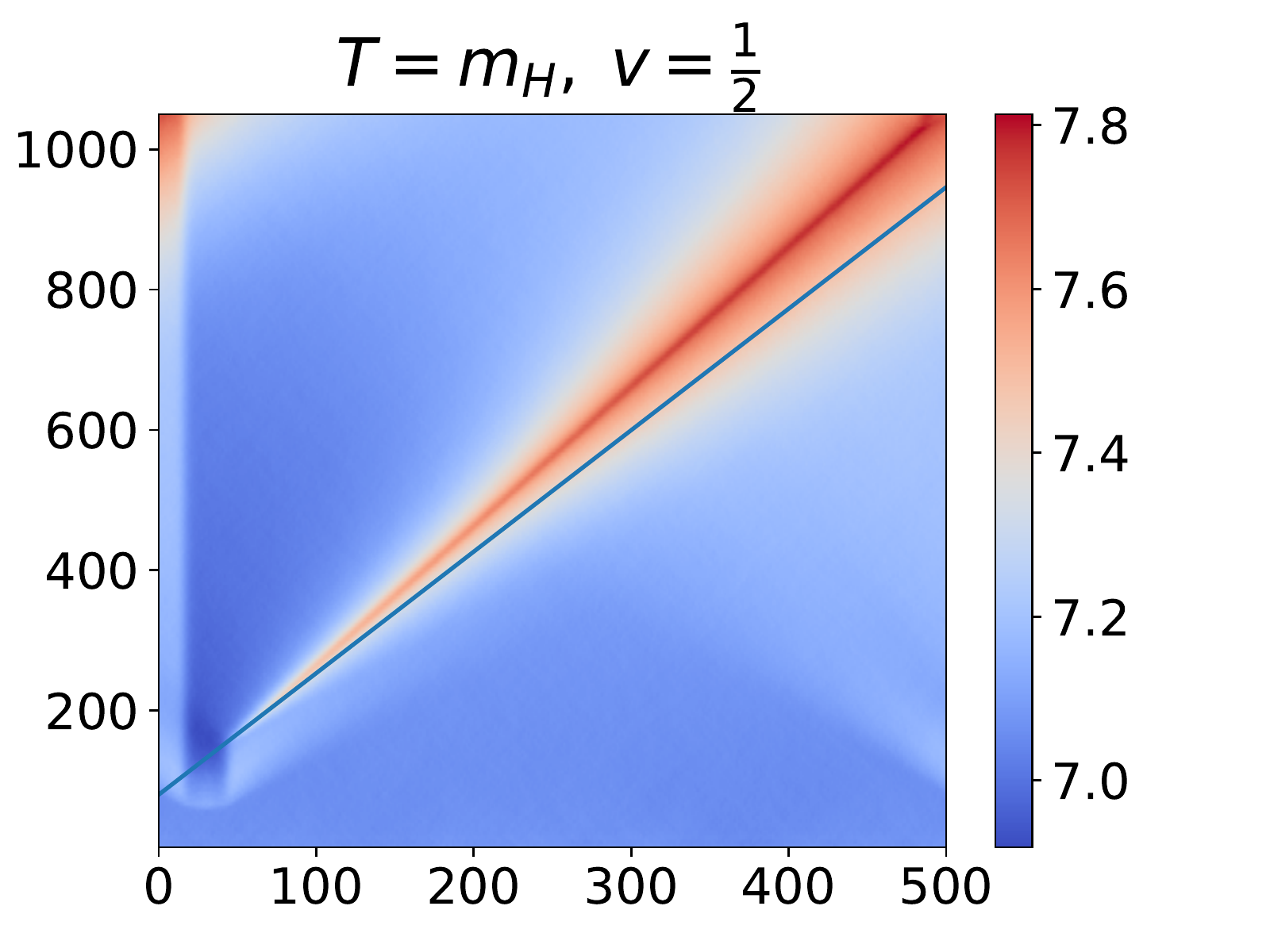} 
\\
\includegraphics[width=0.27\textwidth]{Placeholder.pdf} 
&
\includegraphics[width=0.27\textwidth]{Placeholder.pdf} 
&
\includegraphics[width=0.27\textwidth]{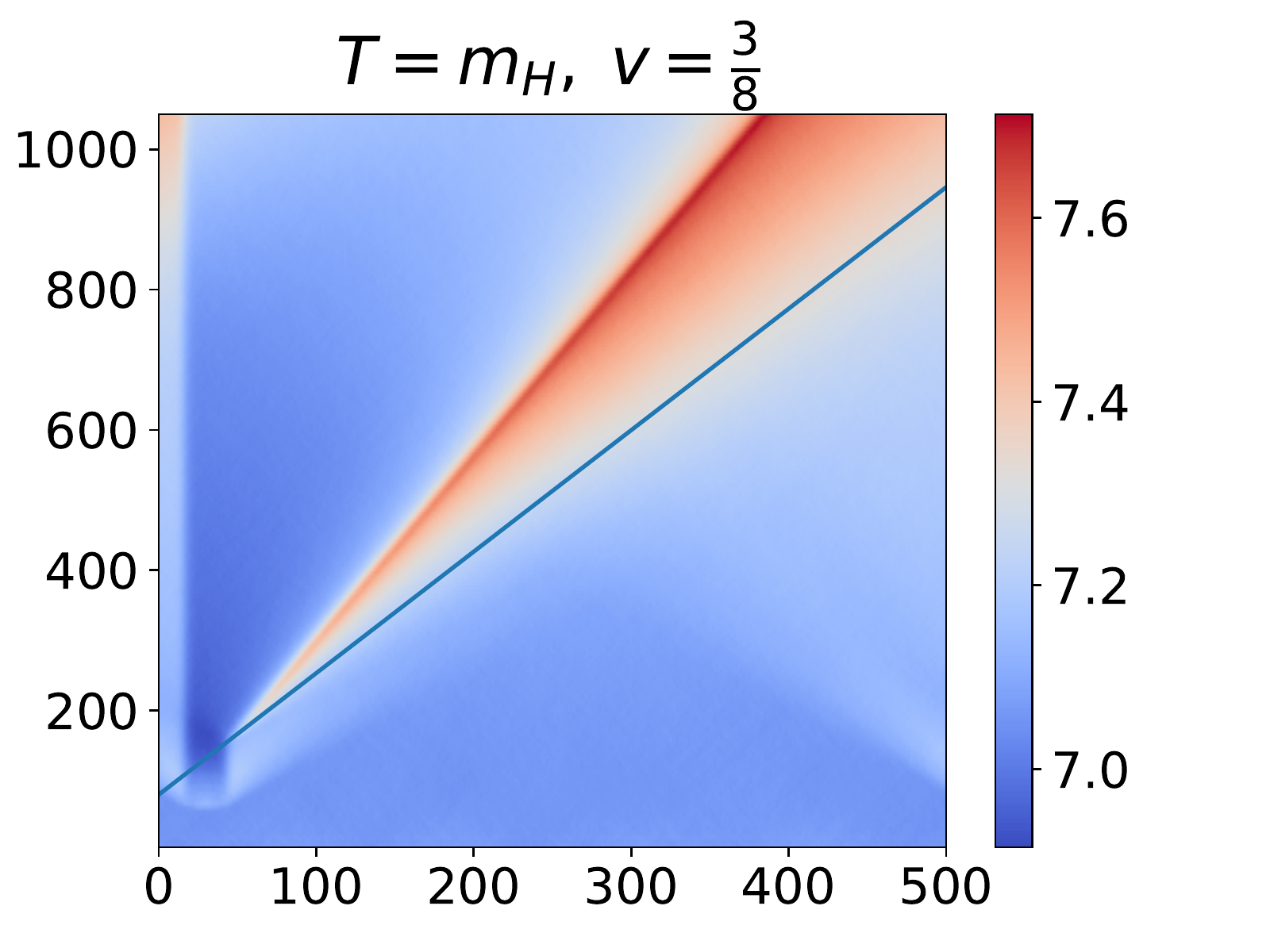} 
\\
\includegraphics[width=0.27\textwidth]{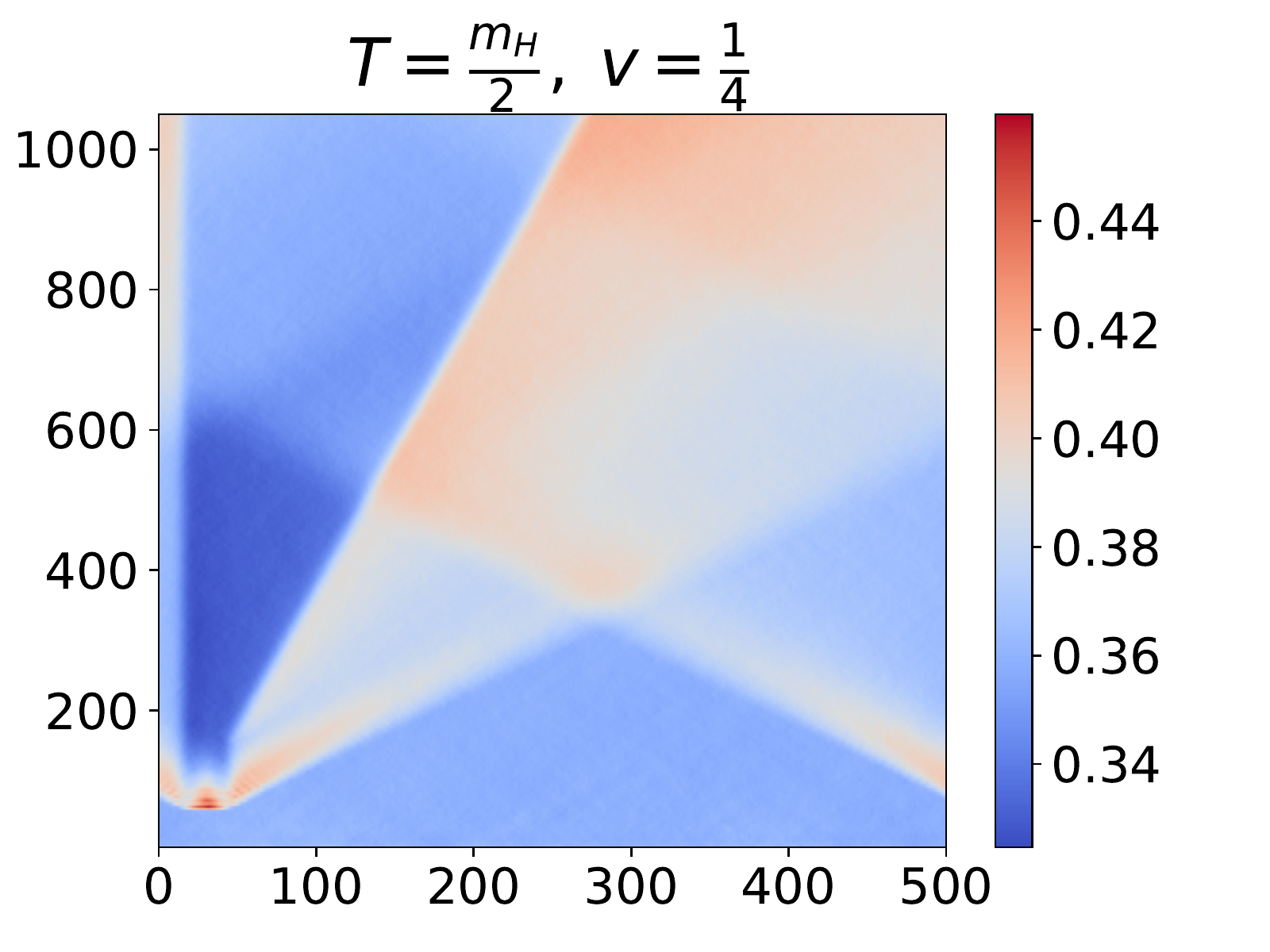} 
&
\includegraphics[width=0.27\textwidth]{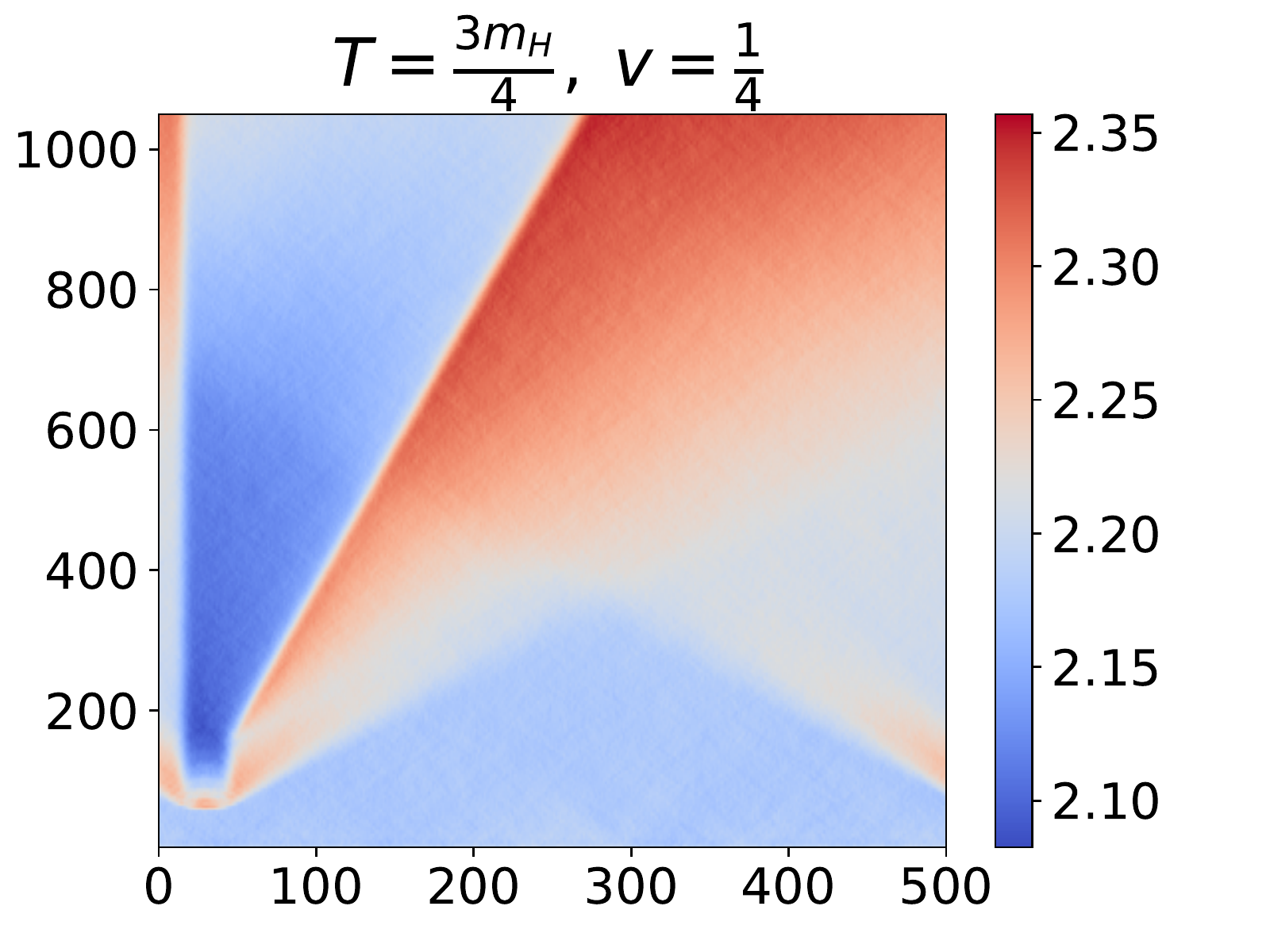} 
&
\includegraphics[width=0.27\textwidth]{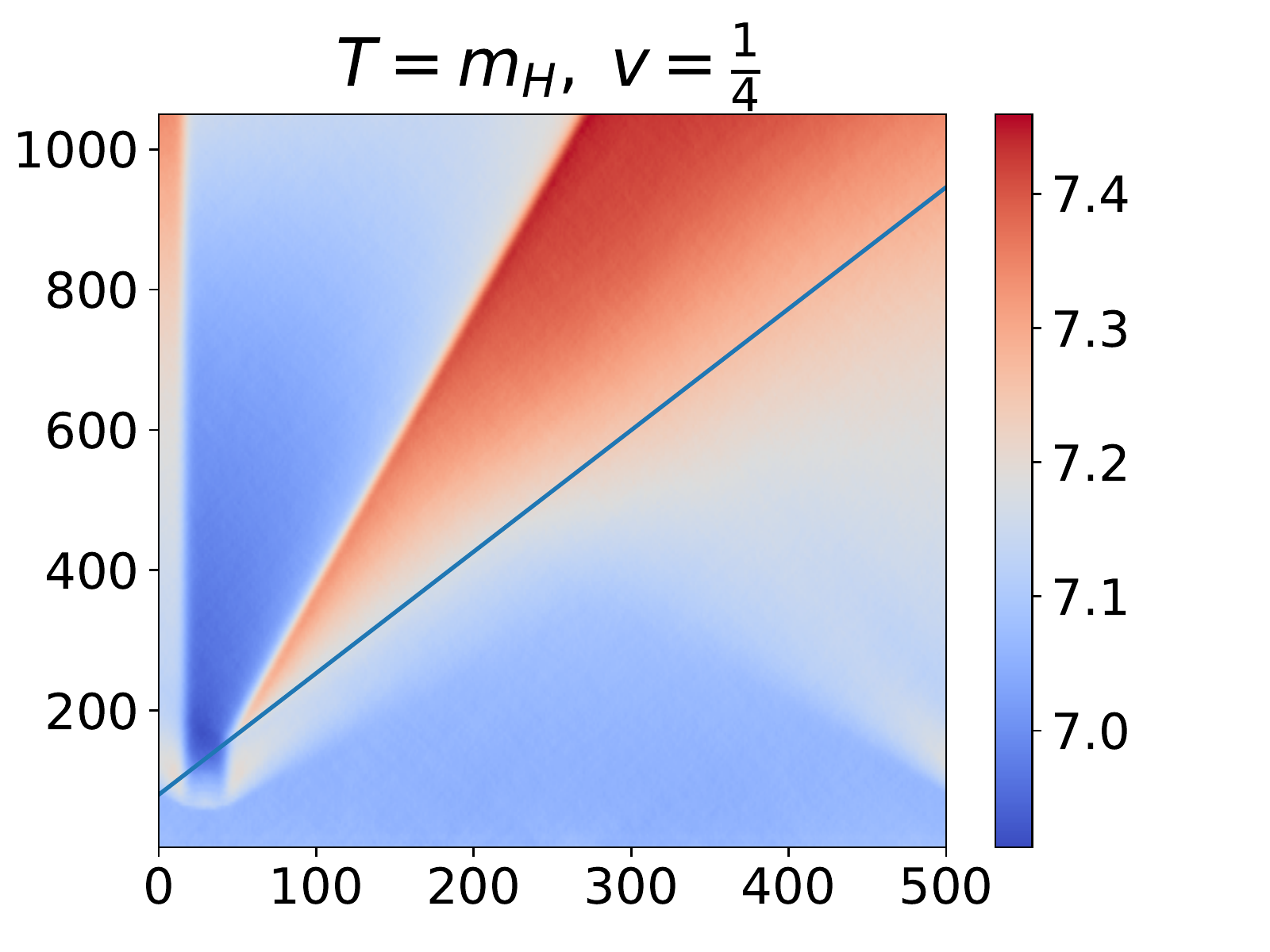} 
\\
\end{tabular}
\caption{Slice-averages of the energy density along the $z$direction (horizontal axis) for different times (vertical axis). Results are shown for different $v$ and initial temperatures, increasing $T/m_H$ moving right in the plot table, increasing wall speed $v$ moving up. $am_H=0.5$, lattice size $64\times 64\times 1000$, $m_Hd=15$. For each plot, we have averaged over 20 configurations, except for the $T/m_H=1$ ( the whole right-most column), which are averaged over 100. In the plots with $T=m_H$ and $v=0.25,\;0.5$ we also show a line corresponding to a velocity of $\frac{1}{\sqrt 3}$, the sound speed in a relativistic fluid.
}
\label{fig:thermal1}
\end{center}
\end{figure}

Let us now consider how the plasma reacts to a bubble wall sweeping through it. In Fig.~\ref{fig:thermal1}, we show the energy density integrated over the $x$-$y$ plane. The axes of the plots represent $z$ (running from 0 to 500 in units of $m_H$) and time $t$ (running from 0 to 1000, also in units of $m_H$). The different plots correspond to different initial temperatures ($T/m_H=0.5, 0.75, 1.0$) and wall speeds  ($v=0.25, 0.375, 0.5, 0.625, 0.75, 0.99$).

Looking first at the bottom left plot ($T/m_H=0.5, v=0.25$) the positions of the initial walls are around $z=30$, and as the bubbles come up in the time interval 50-100 a small spike in energy appears. Then after settling down, the right-hand wall starts moving to the right from time 150. We see that the straight line with speed 0.25 corresponds to the upper edge of the red wedge.  With excess energy density coloured red, this means that the energy  is deposited in front of the wall, spreading out far ahead of it to create a large region of  activity. From this plot we can see the effect of the periodic boundary conditions, with the energy flow from the moving bubble wrapping around the box. This will need to be borne in mind when viewing the profiles of thermodynamic quantities. Furthermore, we note that two energy ``beams" from the initial growth of the walls travel in both directions around the lattice. These are an artefact of the lattice implementation , and may be reduced by an even slower wall initialization.

The simulation ends at a time 1000, which is 850 after the wall started moving, Indeed, in this plot the wall has moved to $850v$ from the initial wall position to $m_Hz\simeq 250$.

Increasing speed (moving up in the left-most column of plots), the energy wedge becomes more pronounced, but narrower, as the wall is able to more and more keep up with the released energy. Whereas for $v=0.5$ all the energy is still outside the bubble, as we increase the speed to $v=0.75$ some of the energy is overtaken by the wall, so that energy is also deposited inside the bubble. For the fastest speed of $v=0.99$, the wall and energy profile are very localised and, as indicated earlier, we begin to be wary of discretization and resolution effects. We also remark that the current $R$, responsible for the bubble, is stopped when it reaches the edge of the simulation. For wall speeds of $v=0.75$ and $v=0.99$ this happens before the end of the simulation, as can bee seen in the top three  rows of Fig.~\ref{fig:thermal1}. 

Moving instead to the right in the array of plots, temperature increases, and we see a similar picture in each case, but with much larger numbers and contrasts (note the difference in colour scale). In particular, for faster speeds, energy density is definitely overtaken by the wall and is deposited both inside and outside the bubble. Another important feature in all these plots is the scaling, or self-similar, behaviour. We can clearly see that the energy density forms a wedge-like shape as the bubble expands, which is expected as the time since nucleation is the only relevant macroscopic variable in the simulation. This means that at late times, thermodynamic profiles are functions of $\xi=z/(t-\tau_{\rm move})$, as is expected from relativistic hydrodynamics \cite{Rezzolla-Zanotti}.

Finally, in a relativistic fluid made up of massless particles we have an equation of state $P=\frac{1}{3}\rho$, leading to a sound speed (squared) of $c_s^2=\frac{\partial P}{\partial \rho}=\frac{1}{3}$. In the high temperature simulations ($T=m_H$) of Fig. \ref{fig:thermal1}, for wall speeds of $v=0.25$ and $0.5$ , we have  included a curve corresponding to this sound speed . As we can see, the outer part of the fluid envelope is propagating at a speed consistent with this sound speed.

%%%%%%%%%%%%%%%%%%%%%%%%%%%%%%%%%%%%%%%%%%%%%%%%%%%%%
\subsubsection*{Energy, pressure and fluid flow }
%%%%%%%%%%%%%%%%%%%%%%%%%%%%%%%%%%%%%%%%%%%%%%%%%%%%%

\begin{figure}[H]
\begin{center}
%\vspace{-3cm}
\begin{tabular}{lll}
%\hspace{-2cm}
\includegraphics[width=0.27\textwidth]{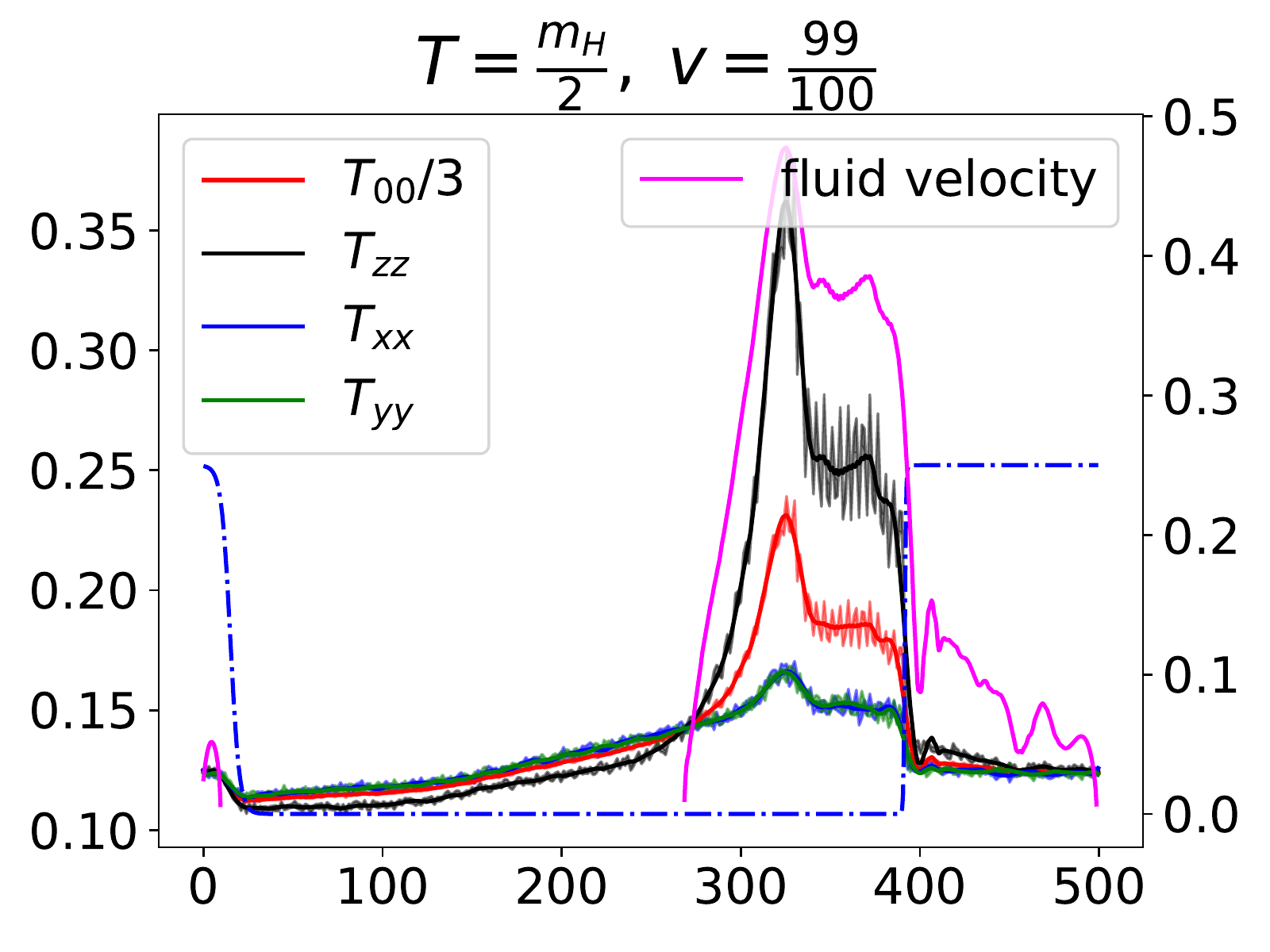} 
&
\includegraphics[width=0.27\textwidth]{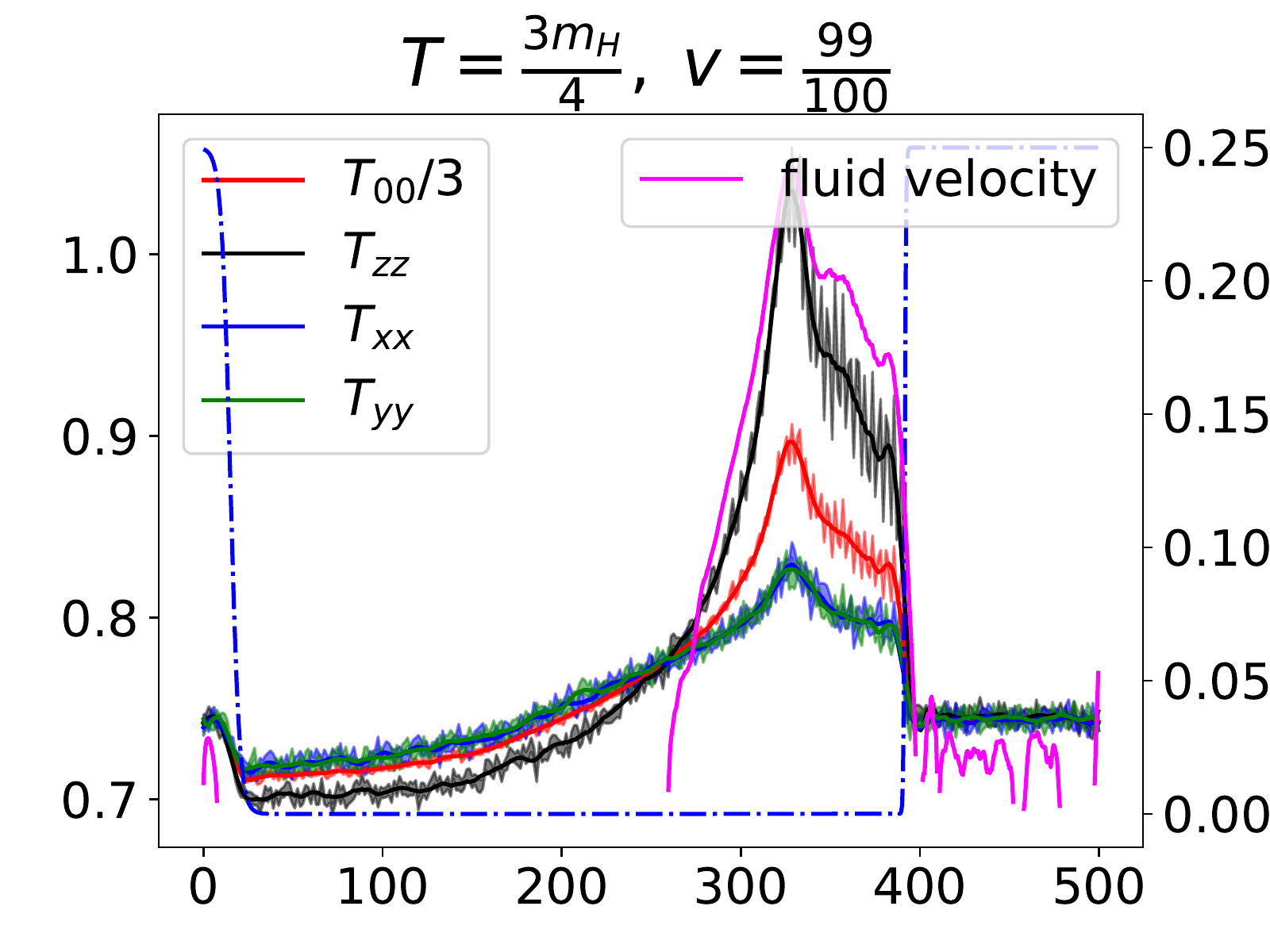} 
&
\includegraphics[width=0.27\textwidth]{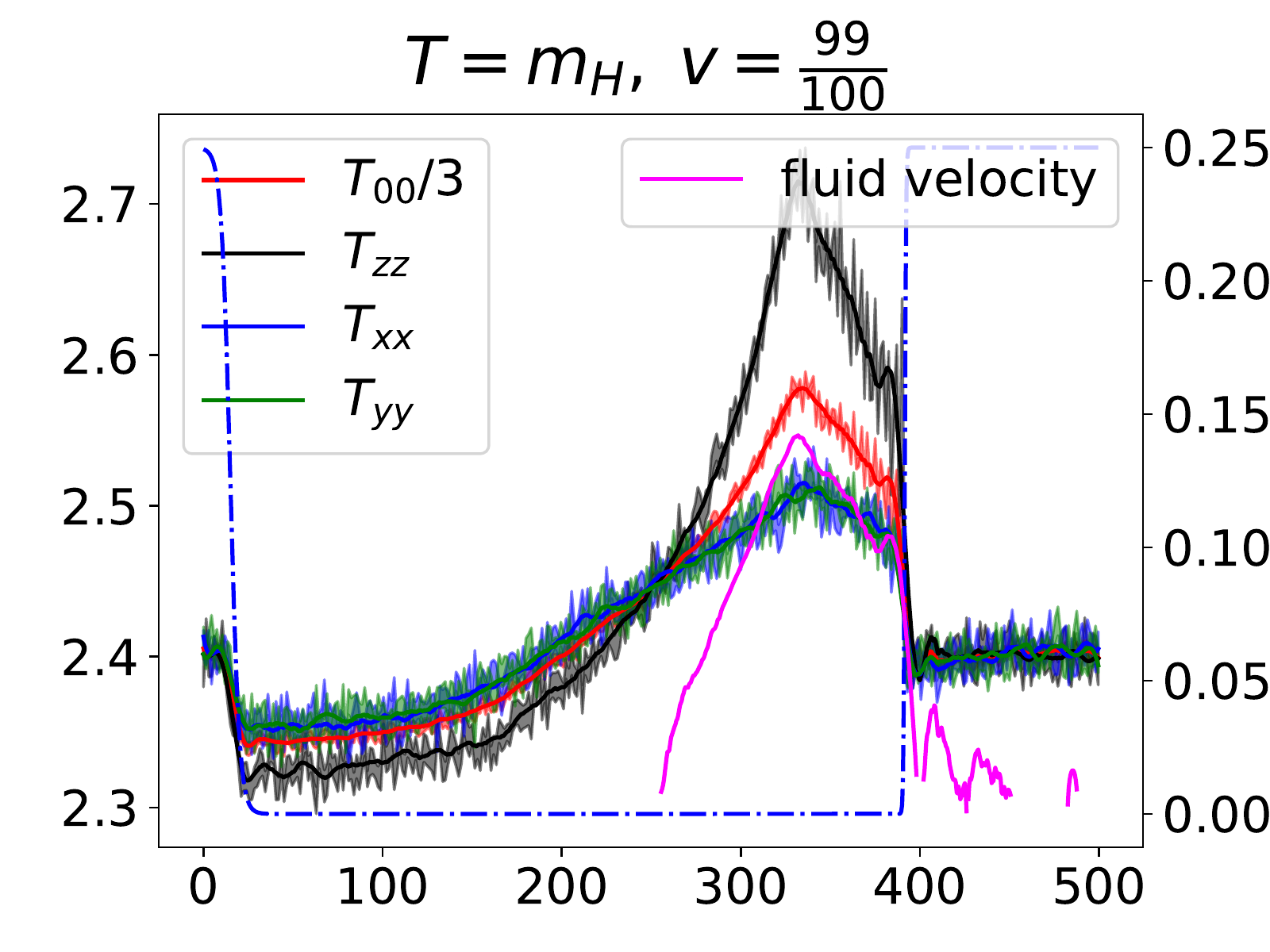} 
\\
\includegraphics[width=0.27\textwidth]{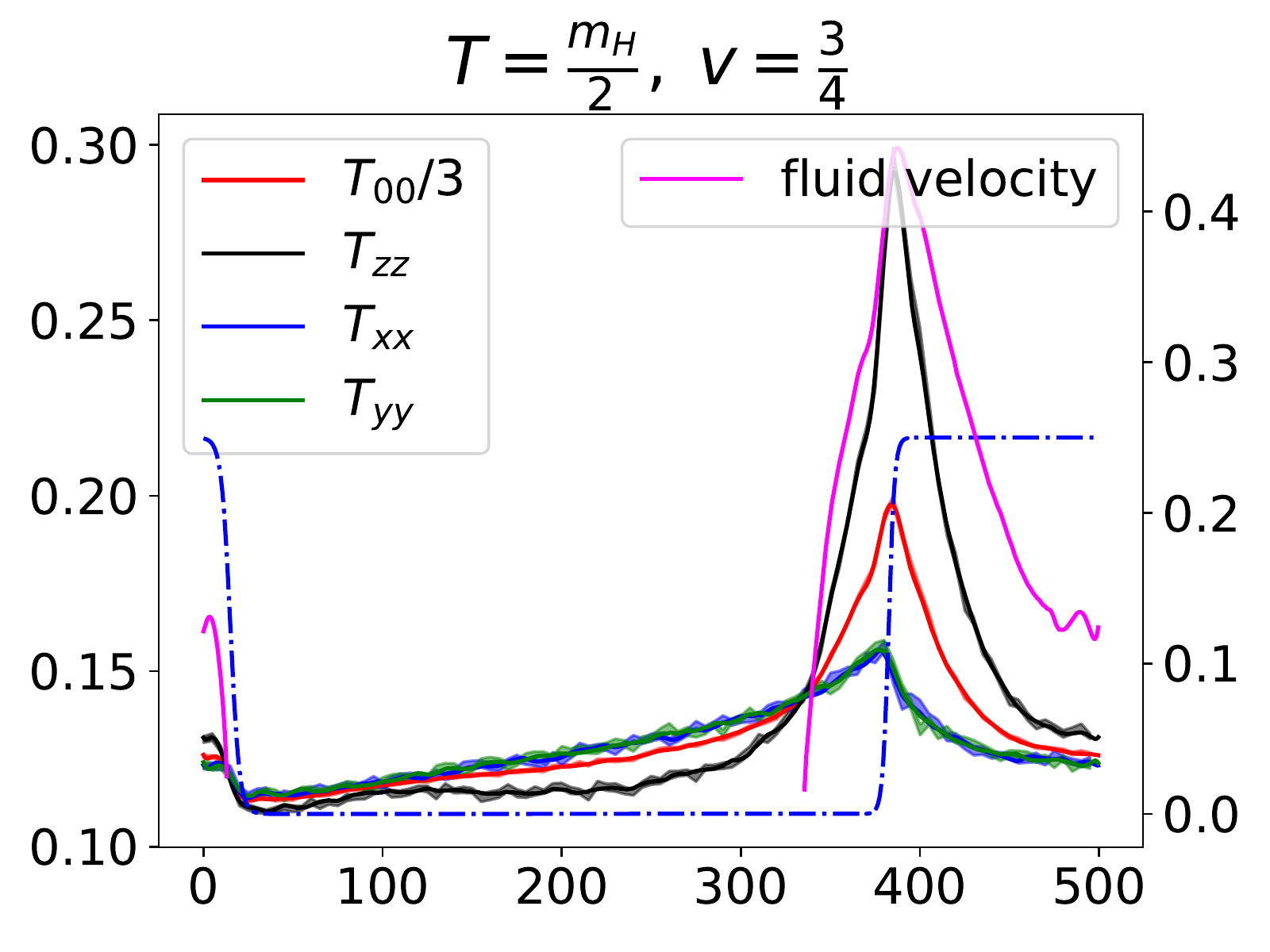} 
&
\includegraphics[width=0.27\textwidth]{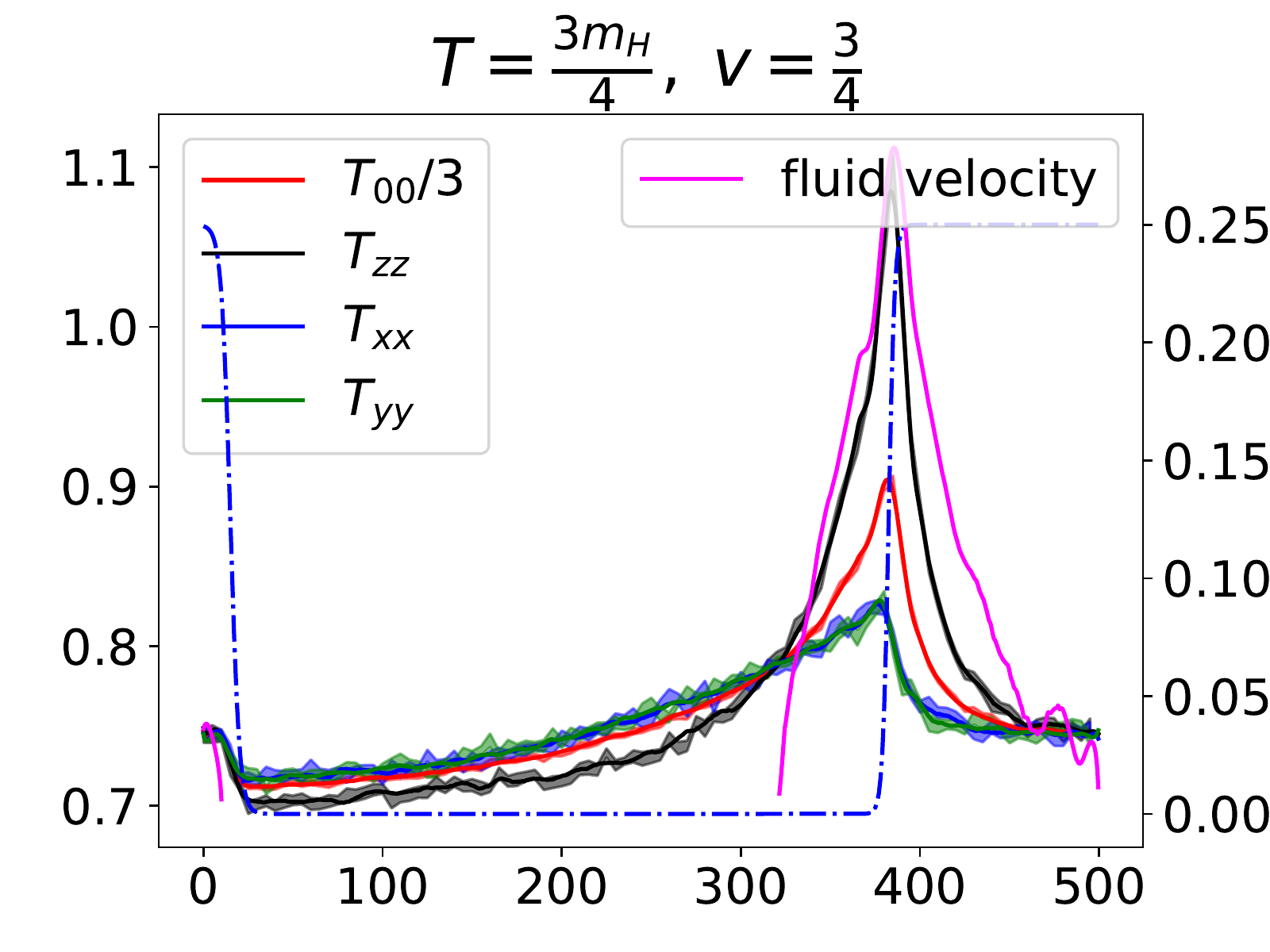} 
&
\includegraphics[width=0.27\textwidth]{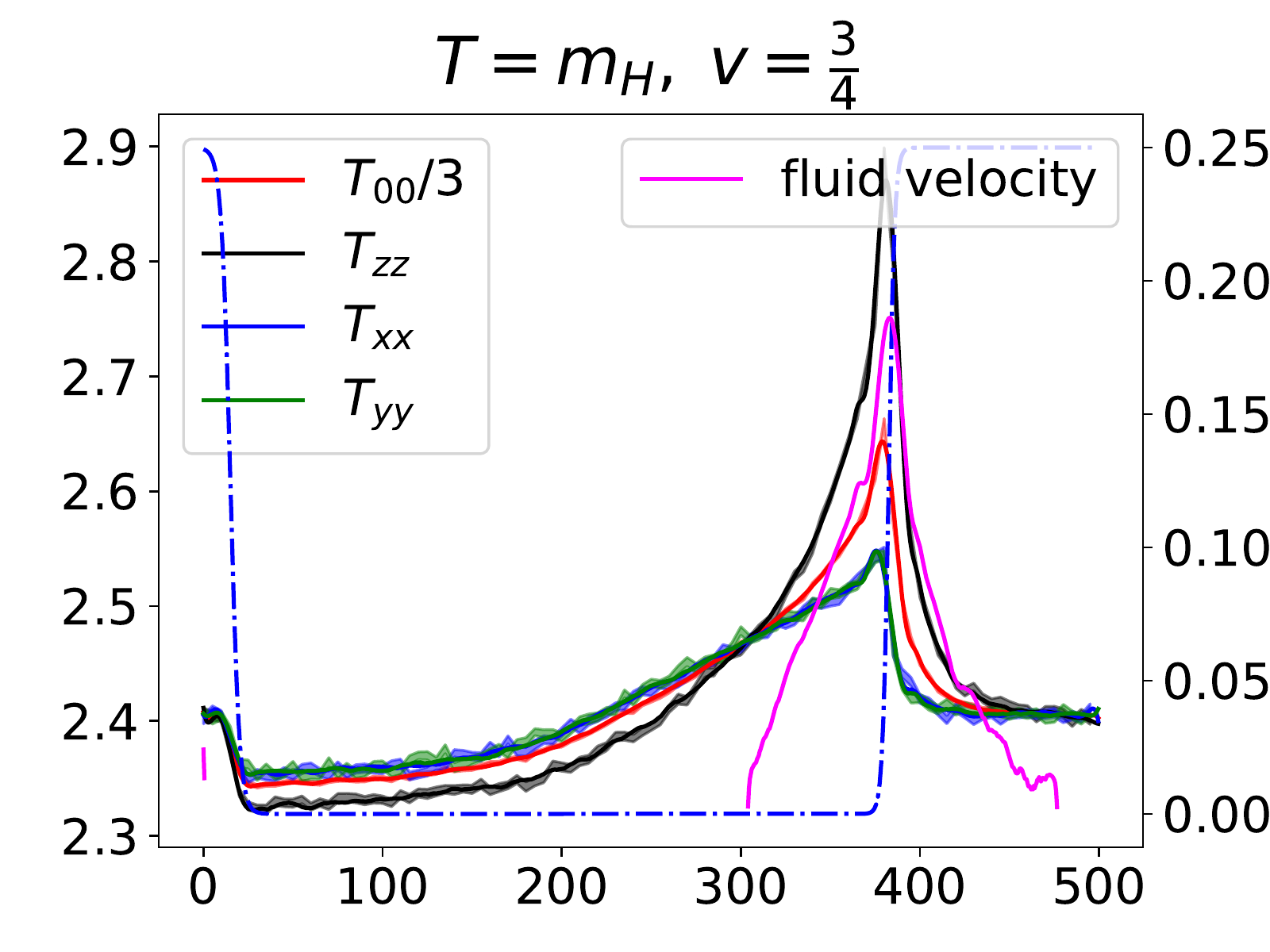} 
\\
\includegraphics[width=0.27\textwidth]{Placeholder.pdf} 
&
\includegraphics[width=0.27\textwidth]{Placeholder.pdf} 
&
\includegraphics[width=0.27\textwidth]{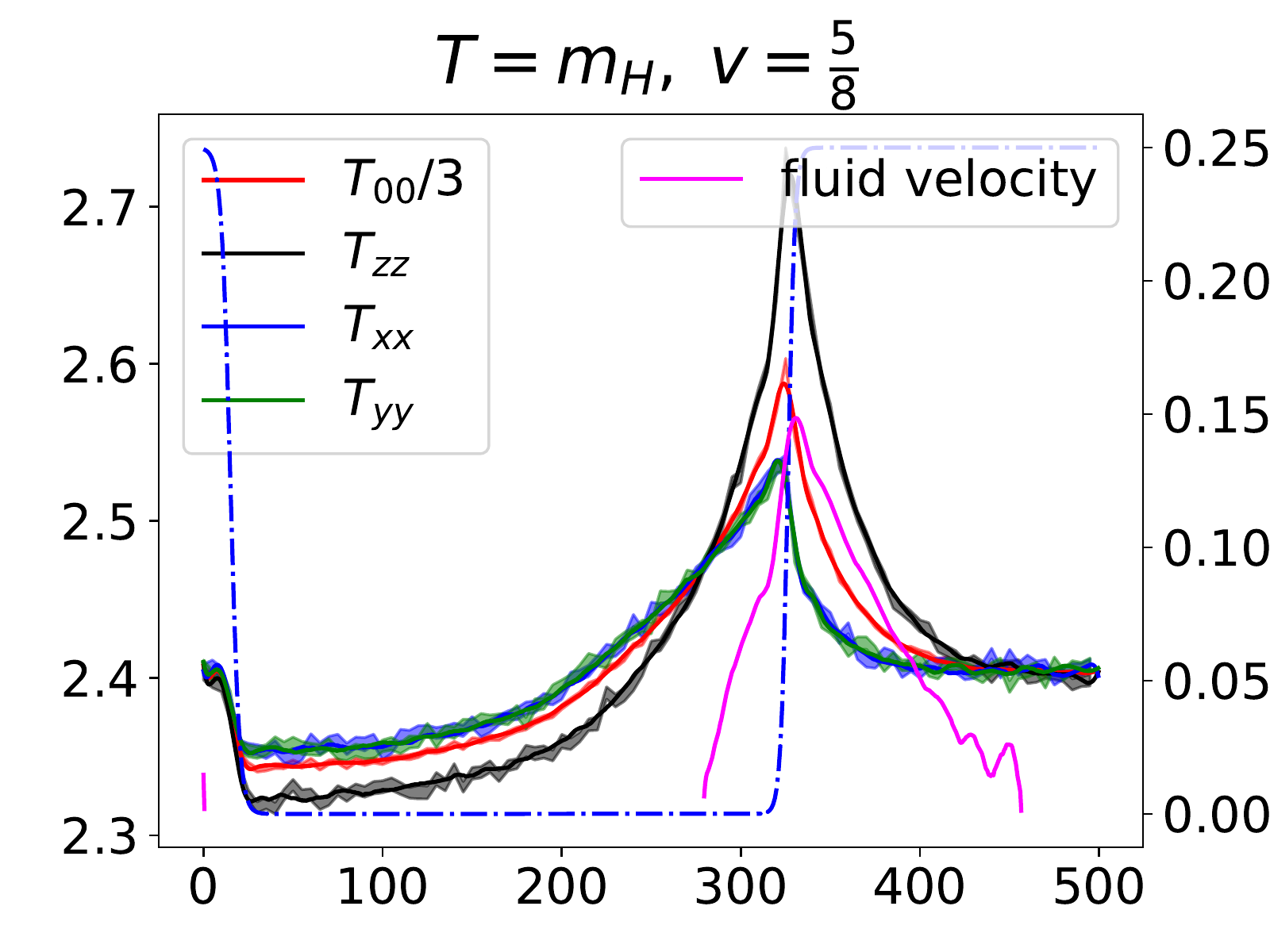} 
\\
\includegraphics[width=0.27\textwidth]{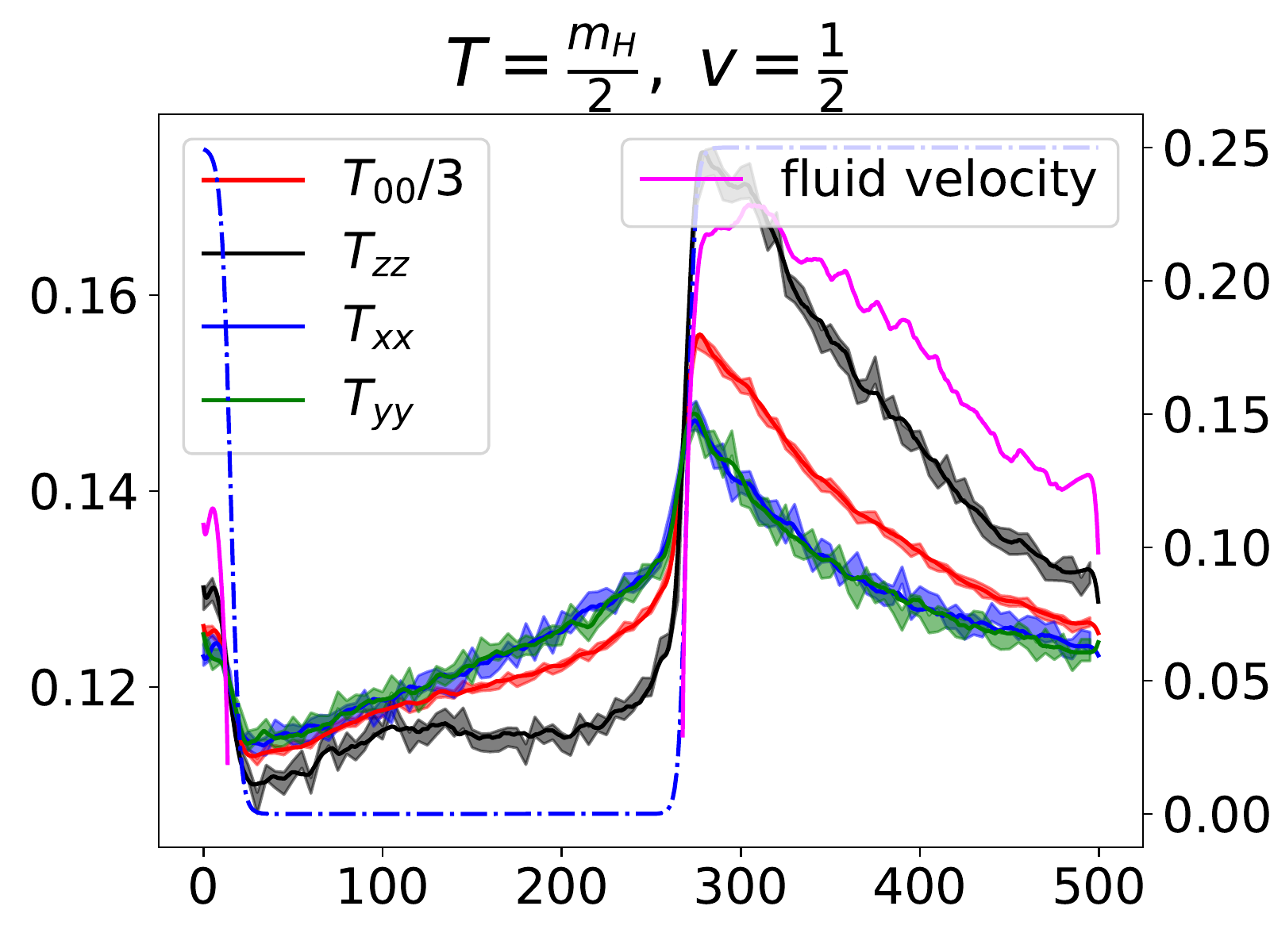} 
&
\includegraphics[width=0.27\textwidth]{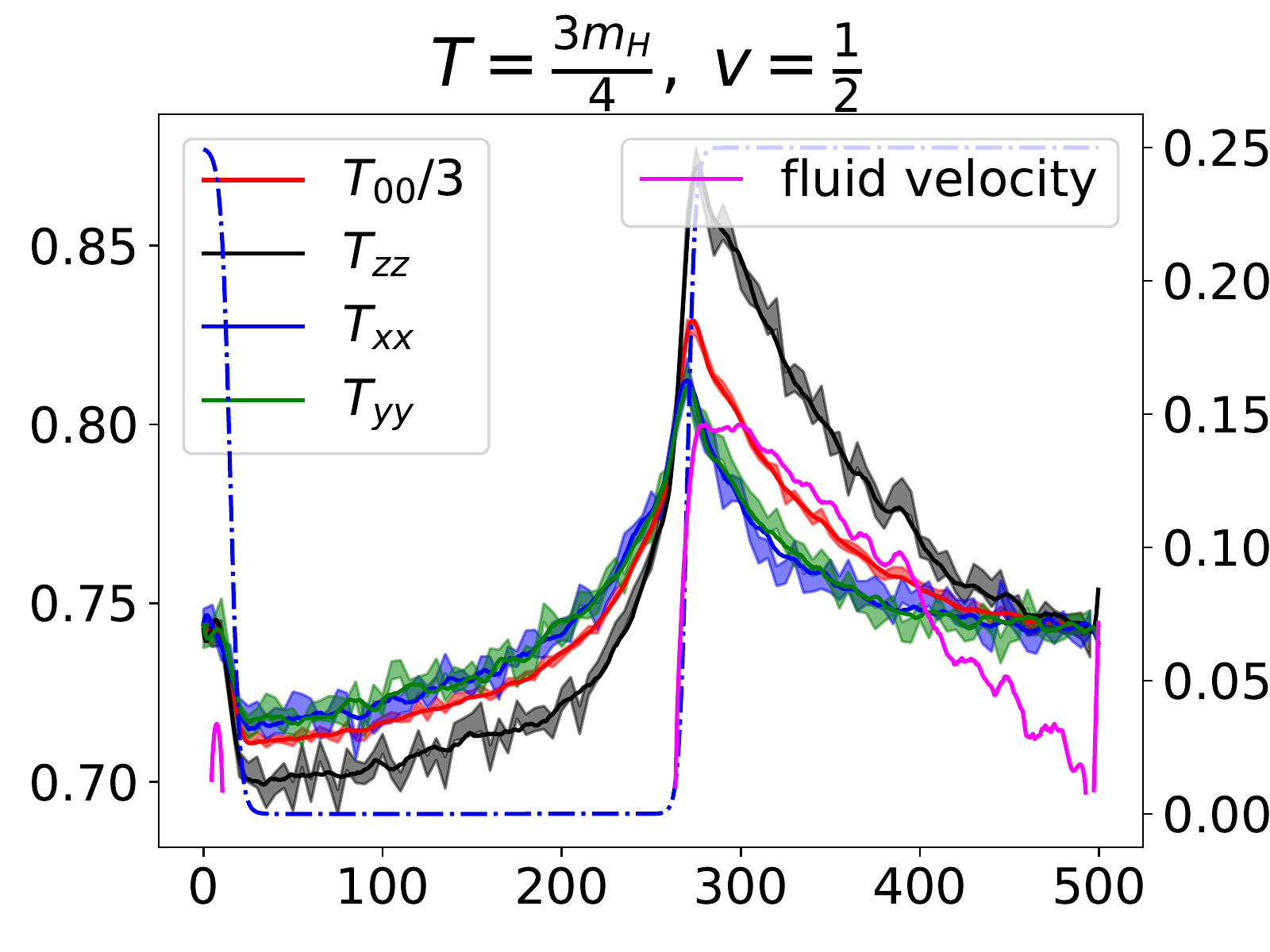} 
&
\includegraphics[width=0.27\textwidth]{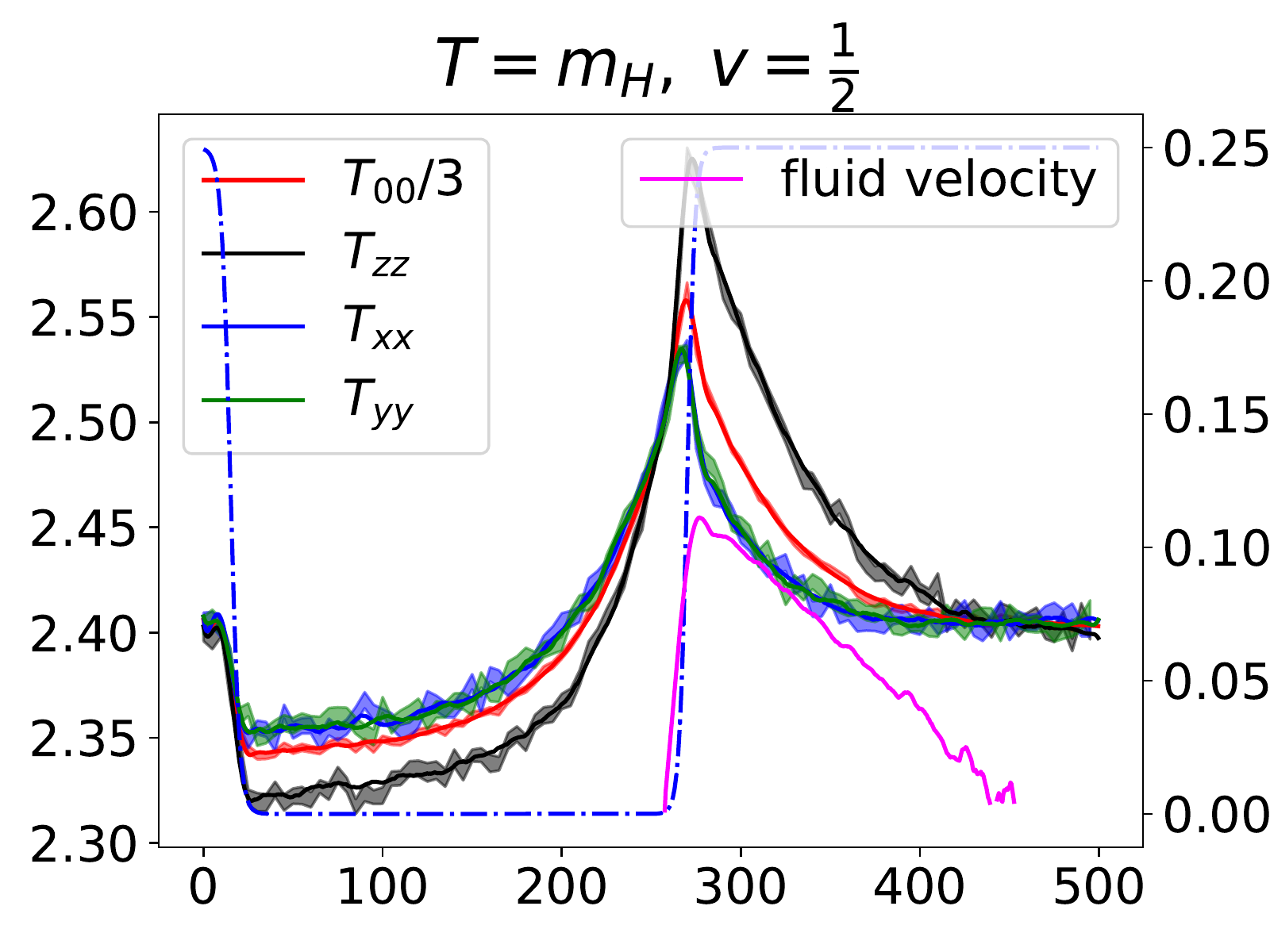} 
\\
\includegraphics[width=0.27\textwidth]{Placeholder.pdf} 
&
\includegraphics[width=0.27\textwidth]{Placeholder.pdf} 
&
\includegraphics[width=0.27\textwidth]{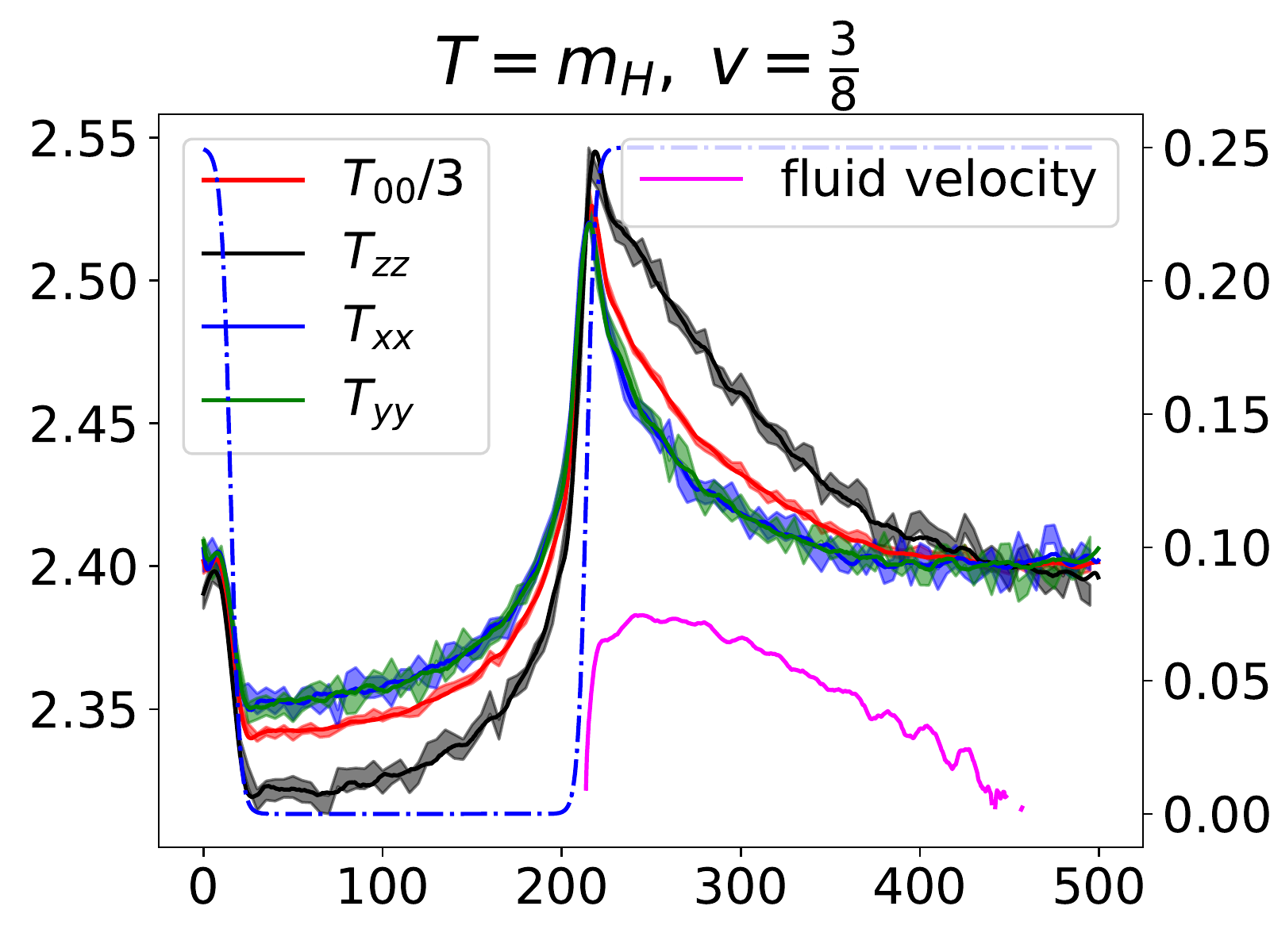} 
\\
\includegraphics[width=0.27\textwidth]{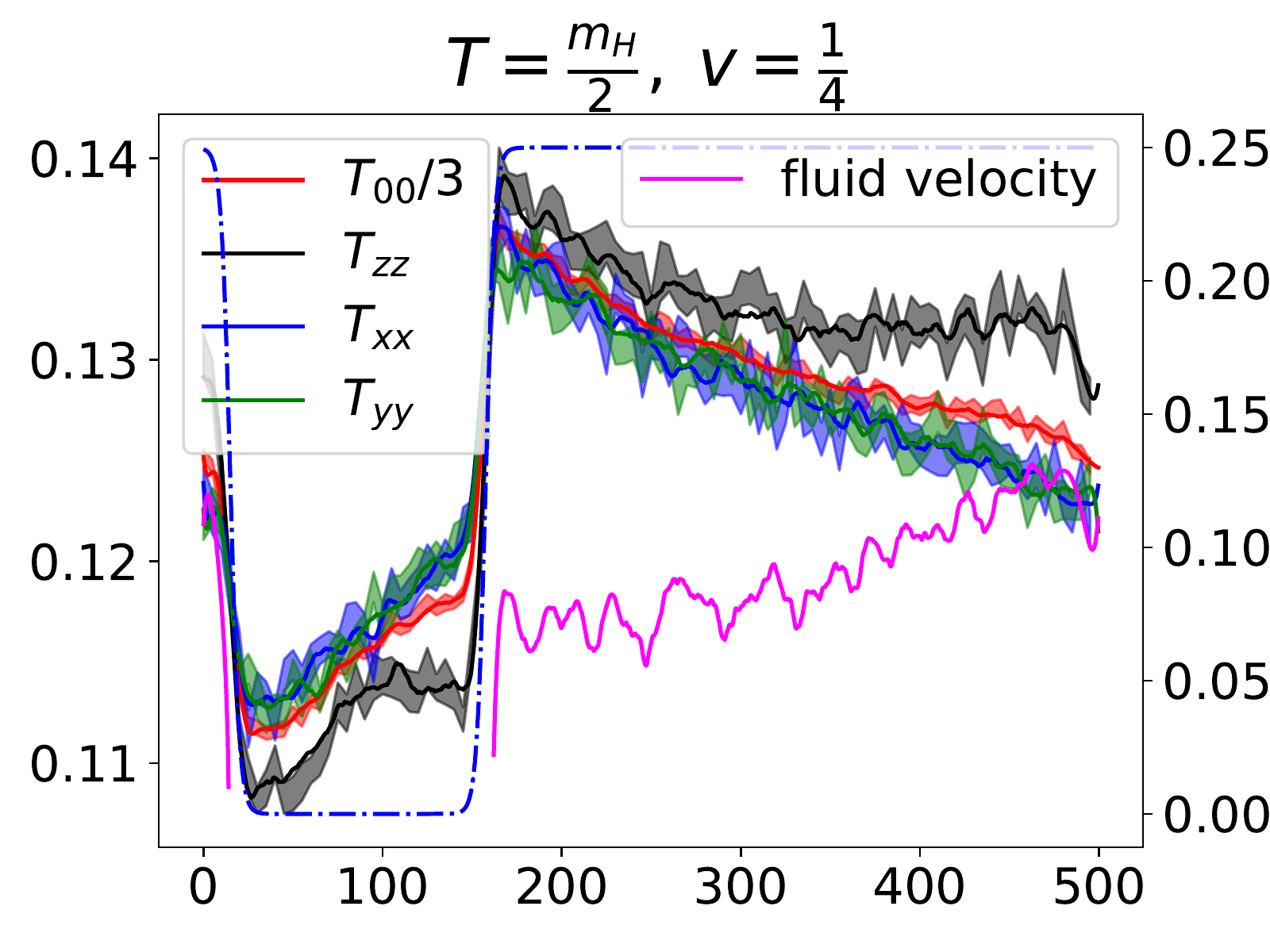} 
&
\includegraphics[width=0.27\textwidth]{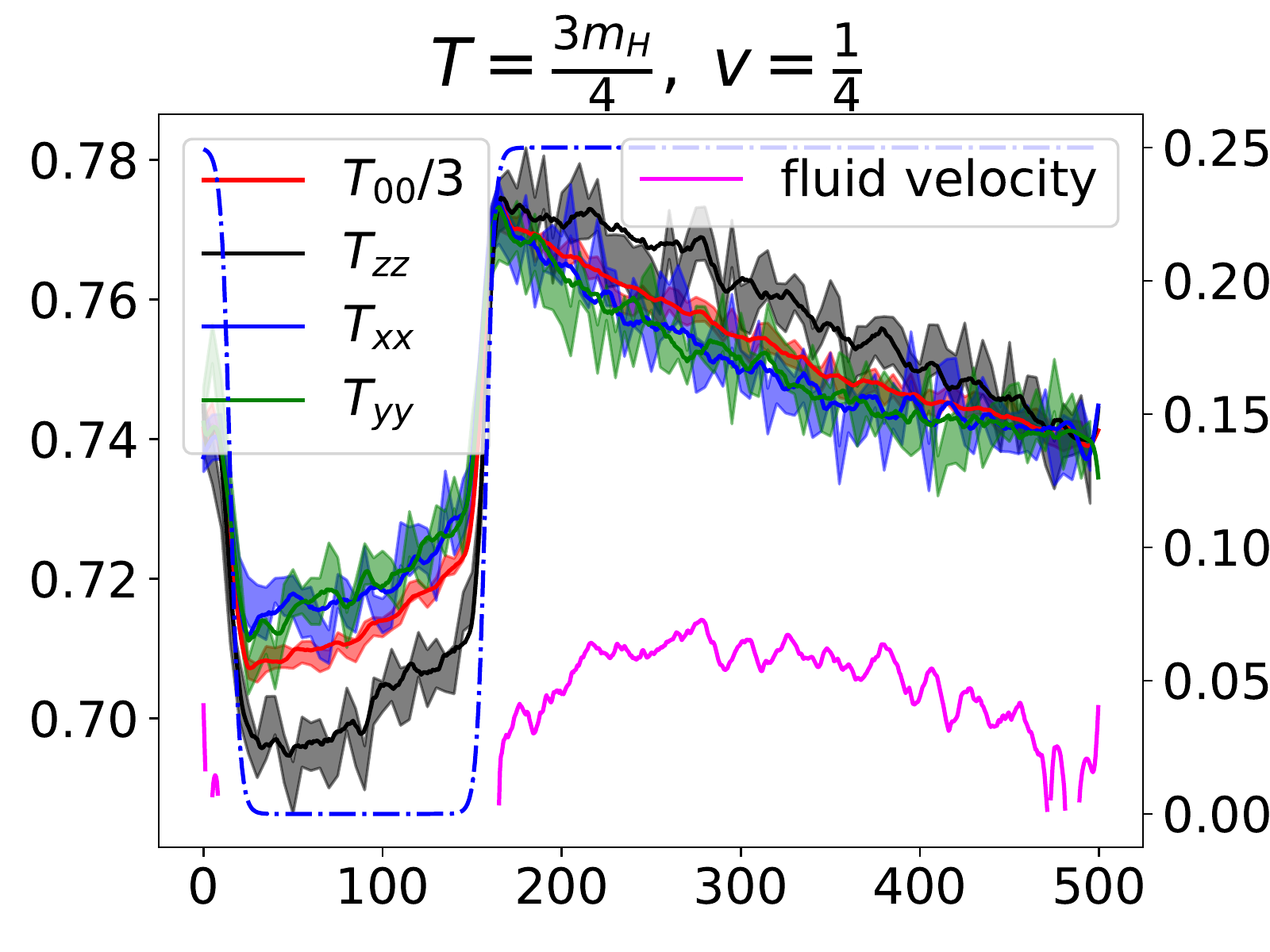} 
&
\includegraphics[width=0.27\textwidth]{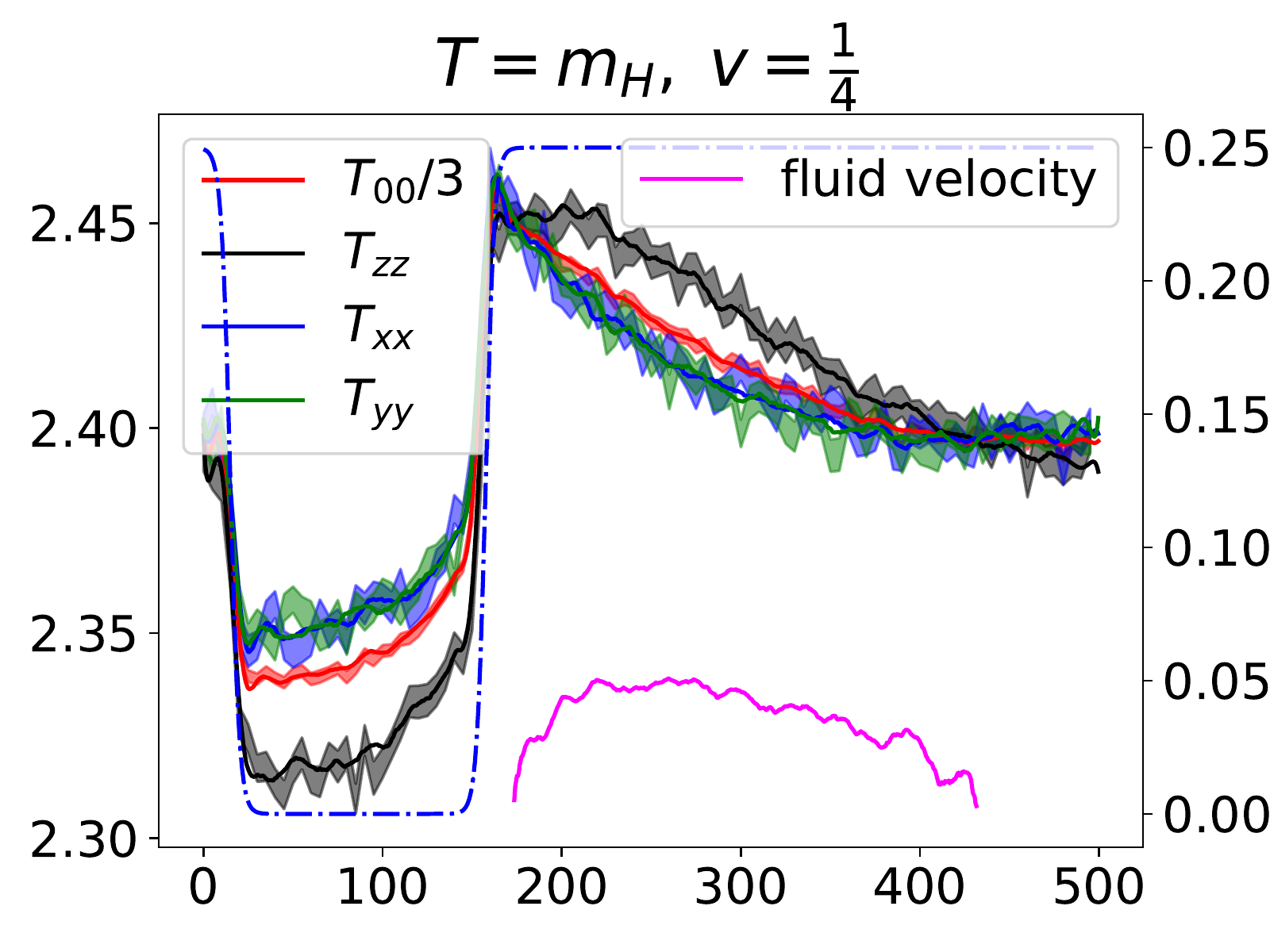} 
\\
\end{tabular}
\caption{ Slice-averages of all the diagonal components of the energy-momentum tensor for various initial temperatures, taken at time $m_Ht=600$ for  $v=0.25,\;0.375,\;0.5,\;0.625\;0.75$ and time $m_Ht=500$ for $v=0.99$.  Also plotted is the inferred fluid velocity (magenta). $T/m_H$ increases moving right in the plot table, increasing wall speed $v$ moving up. The data is averaged over 20 configurations, except for the $T/m_H=1$ ( the whole right-most column), which are averaged over 100. $am_H=0.5$, lattice size $64\times 64\times 1000$, $m_Hd=15$. The left-hand vertical axes of the individual plots refer to the stress tensor, while the right-hand vertical axes describe the fluid velocity and $R/4R_{\rm max}$ (blue dashed curve). The horizontal axes denote the $z$-coordinate.
}
\label{fig:thermal2}
\end{center}
\end{figure}

In addition to the energy density, we  also show the other diagonal components of the energy momentum tensor, Fig.~\ref{fig:thermal2}. All quantities have  been averaged over the $x$-$y$  plane, and are shown as a function of $z$ at particular times, $m_Ht=600$ for $v=0.25,\;0.375,\;0.5,\;0.625\;0.75$  and $m_Ht=500$ for $v=0.99$.  We also show the inferred fluid velocity, assuming a perfect fluid form of the energy-momentum tensor, as described in section \ref{sec:thermobs}. We must bear in mind that the effects of the periodic boundary conditions seem particularly pertinent for low temperatures, as can be seen from Fig.~\ref{fig:thermal1}, and so some care is needed when interpreting Fig.~\ref{fig:thermal2}. The blue dashed line is the external current  $R$ (divided by $4R_{\rm max}$ as measured on the right-hand vertical axes), and this describes the position of the wall as it moves outwards. 

Beginning again in the bottom left plot (low temperature, $0.5\,m_H$, low speed, 0.25), we see that all of the observables have a clear jump in magnitude, coinciding with the position of the wall. This reflects the observation in Fig. \ref{fig:thermal1} that at low speeds, the energy is deployed outside the bubble, ahead of the wall. The $z$-component is somewhat different from the $x$- and $y$-components, but the anisotropy is rather weak at these low speeds. This corresponds to a quite low fluid velocity around 0.1.

Picking up speed ($v=0.5$, upwards in the column of plots), we see that the picture persists, but with a larger overall amplitude, a much larger anisotropy and hence a much larger fluid velocity. All observables decay smoothly with the distance from the wall, as they interact with the ambient plasma. At a glance, the dumped energy-momentum reaches a distance of about $zm_H\sim150(m_Ht/600)\sim\frac{1}{4}m_Ht$ from the wall, where the factor of $m_Ht/600$ comes from the self-similar behaviour, allowing us to extrapolate to later times. The fluid velocity is still much smaller than the wall speed.

At a speed of $v=0.75$, the wall is able to overtake some of the thermodynamic envelope, before it has time to thermalize with the plasma. As a result, there is an energy peak, for which the maximum coincides with the position of the wall. The tails of the peak stretch both inside and outside the bubble. The peak is asymmetric, reaching a distance of about $zm_H\sim50(m_Ht/600)\sim\frac{1}{12}m_Ht$ outside the wall and as far as $zm_H\sim200(m_Ht/600)\sim\frac{1}{3}m_Ht$ inside the wall. The anisotropy is even more pronounced than before, because of the high wall speed. 

For asymptotically high speeds ($v=0.99$), the picture is even more extreme. The wall is now firmly ahead of most of the released energy, and no longer coincides with the maximum of the energy peak. This peak is now quite broad and stretches a distance of at least $zm_H\sim100(m_Ht/500)\sim\frac{1}{5}m_Ht$  inside the bubble. We also see an anisotropy between $z$ and $x,y$ components is a factor of 4, giving a large fluid velocity. Outside the bubble, in contrast, the plasma is fairly quiet.

Proceeding to larger temperatures (second and third column of plots), both the overall relative magnitudes of the observables increase substantially. The qualitative picture is the same, but quantitative differences exist. As an example, for the largest temperature $T/m_H=1$ and a speed of $v=0.75$, the energy peak reaches only about half as far into plasma outside the bubble. This likely means that a baryogenesis mechanism relying on out-of-equilibrium conditions and CP-bias emerging from the bubble wall will have less volume to act. 

Also, it seems that although the (maximum) fluid velocity is always smaller than the wall speed, the two are approximately proportional to each other, with the fluid velocity roughly inversely proportional to temperature (Table 1 ).

\begin{table}
\begin{center}
\begin{tabular}{|c|c|c|c|c|}
\hline
$v$\textbackslash $T$&0.5 $m_H$&0.75 $m_H$&$m_H$\\
\hline
0.0&0&0&0\\
\hline
0.25&0.12&0.07&0.05\\
\hline
0.375&&&0.07\\
\hline
0.5&0.22&0.15&0.11\\
\hline
0.625&&&0.15\\
\hline
0.75&0.43&0.27&0.19\\
\hline
0.99&0.48&0.24&0.14\\
\hline
\end{tabular}
\caption{Maximum fluid velocities for different wall speeds and temperatures.}
\end{center}
\label{tab:speed}
\end{table}

%%%%%%%%%%%%%%%%%%%%%%%%%%%%%%%%%%%%%%%%%%%%%%%%%%%%%
\section{Dynamics of topological observables}
\label{sec:topology}
%%%%%%%%%%%%%%%%%%%%%%%%%%%%%%%%%%%%%%%%%%%%%%%%%%%%%

%%%%%%%%%%%%%%%%%%%%%%%%%%%%%%%%%%%%%%%%%%%%%%%%%%%%%
\subsubsection*{Single configurations}
%%%%%%%%%%%%%%%%%%%%%%%%%%%%%%%%%%%%%%%%%%%%%%%%%%%%%

\begin{figure}[H]
%\vspace{-3cm}
\begin{tabular}{lr}
%\hspace{-2cm}
\includegraphics[width=0.5\textwidth]{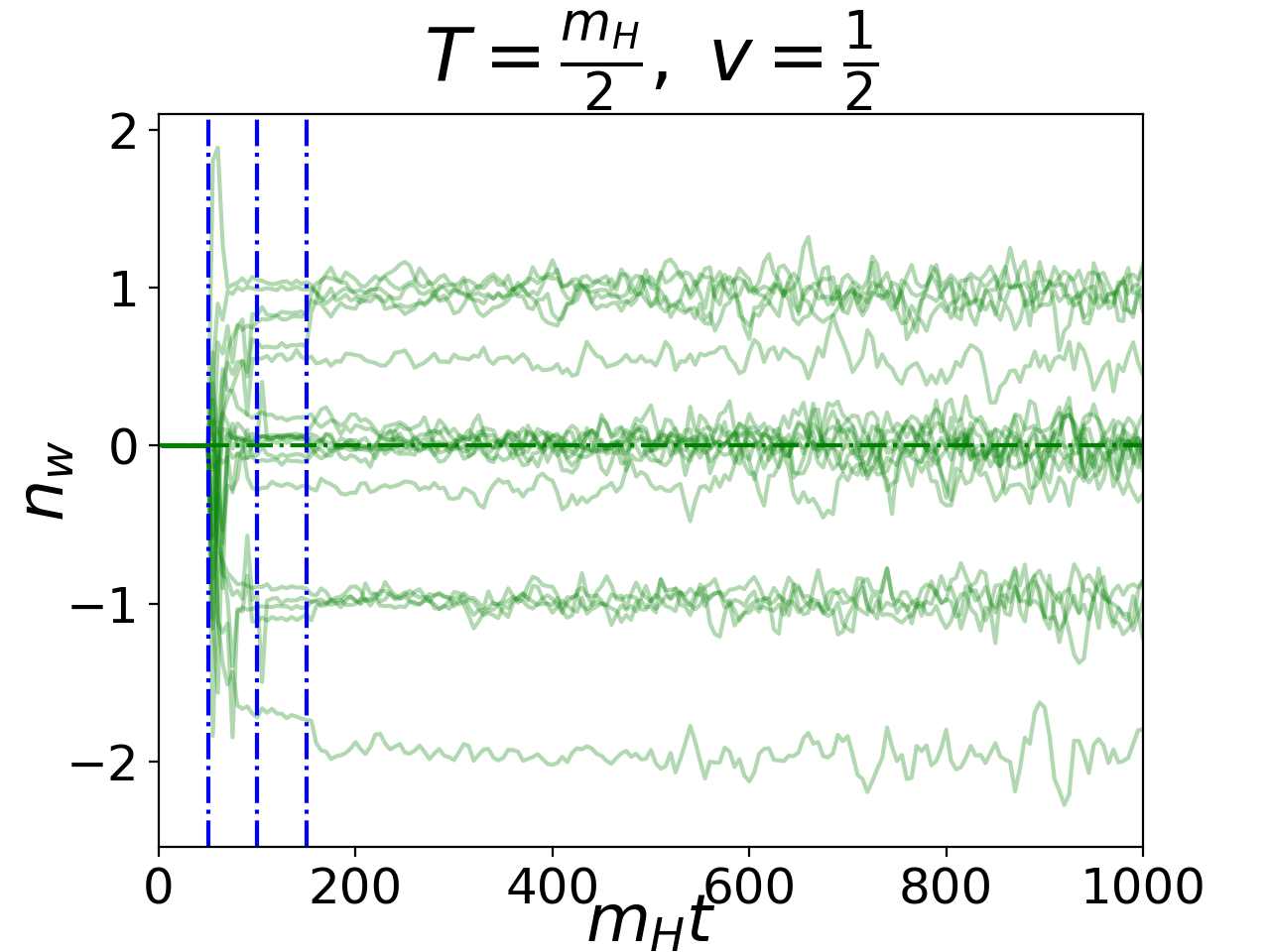} & \hspace{-0.5cm}
\includegraphics[width=0.5\textwidth]{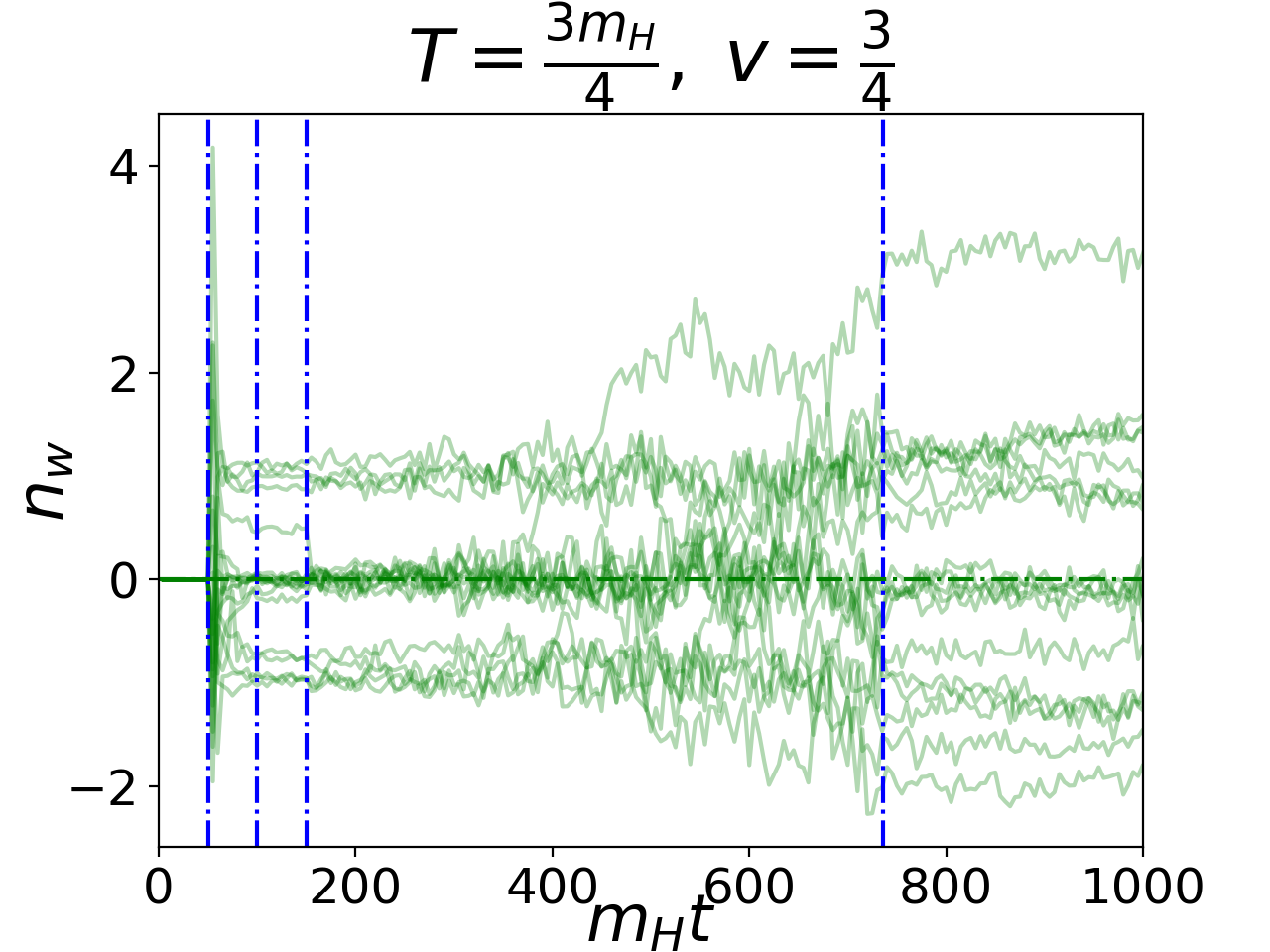} \\
\end{tabular}
\caption{The Higgs winding number integrated inside the bubble for a set of random realisations, with $v=0.5$, $T/m_H=0.5$ (left) and $v=0.75$, $T/m_H=0.75$ (right).
$am_H=0.5$, lattice size $64\times 64\times 1000$, $m_Hd=15$. The three dashed lines between $0<m_Ht<200$ correspond to $\tau_{\rm thermal}$, $\tau_{\rm wall}$ and $\tau_{\rm stable}$, while the fourth dashed line on the right-hand plot is when the current reaches the end of the box and stops moving.
}
\label{fig:winding}
\end{figure}

Electroweak baryogenesis relies on a CP-asymmetric fermion current biasing near-equilibrium sphaleron processes in the bulk. As the bubble wall collides with the plasma, CP-violating interactions produce a net left/righthanded fermion current through/reflected from the wall into the bubble/back into the symmetric phase plasma in front off the wall. If the sphalerons are sufficiently suppressed in the broken phase inside the bubble, baryon number is only generated outside the bubble in response to the CP-violating currents.

Since we do not include the fermion degrees of freedom, the role of a baryon number is played by Chern-Simons number and Higgs winding number. In order to potentially be able to react to a CP-bias, we need to have sufficiently high temperature for sphaleron processes (and more general processes changing Chern-Simons number and winding number) to readily take place, at least outside the bubble. 

In Fig.~\ref{fig:winding} we show  the winding number inside the bubble as a function of time for two different sets of realisations, with two different combinations of temperature and wall speed. The three blue dashed lines denote the set-up stages of the wall ($\tau_{\rm thermal}+\tau_{\rm wall}+\tau_{\rm stable}$). The last dashed line corresponds to the wall reaching the end of the lattice and stopping. 

The winding number is first recorded as the bubble wall comes up at around $m_Ht=50$, both because winding number is generated, but also since that is when an ``inside" of the bubble begins to exist. Once the bubble moves, more and more of the volume is included inside the bubble, and one would expect that the total winding number increases, assuming that there is a random distribution of winding number nuclei distributed throughout the volume. For the cold temperature (left-hand plot) this does not happen, and the integer winding number for each realisation remains at its initial value. This shows that the system is too  cold to spontaneously generate winding number nuclei.

For the larger temperature (right-hand plot) we see that there are instances of "flips", where the winding number increases as a symmetric  phase becomes subsumed by the bubble. The majority of realisations remain in their initial value, but a few add up to other integer values. At even higher temperature, the activity is even higher, but so is the noise on the observable.

%%%%%%%%%%%%%%%%%%%%%%%%%%%%%%%%%%%%%%%%%%%%%%%%%%%%%
\subsubsection*{Diffusion of winding number}
%%%%%%%%%%%%%%%%%%%%%%%%%%%%%%%%%%%%%%%%%%%%%%%%%%%%%

Another way to quantify the activity of winding number changing processes outside the bubble is to compute the average of winding number square,
\begin{eqnarray}
\bar{N_{\rm w}^2}=\langle(N_{\rm w}-\bar{N}_{\rm w})^2\rangle,
\end{eqnarray}
where the average is over the ensemble of realisations. On one hand, if $N_{\rm w}$ would evolve via diffusion in a fixed volume, one would expect this quantity to increase linearly with time as a random walk. The linear slope is the diffusion rate, well-known from the Sphaleron rate for Chern-Simons number \cite{DOnofrio:2014rug}. On the other hand, integrating a fixed set of configurations over an ever increasing bubble volume, if the winding nuclei are randomly  placed in space, we would again expect a linear increase as a function of volume, because bubble volume grows linearly in time with the speed of the wall. When combining the two  effects, a linearly increasing volume, spreading into a region where diffusion is taking place, we expect a time dependence of $\bar{N_{\rm w}^2}$ which is faster than linear.

\begin{figure}[H]
%\vspace{-3cm}
\begin{center}
\begin{tabular}{llll}
%\hspace{-2cm}
\includegraphics[width=0.27\textwidth]{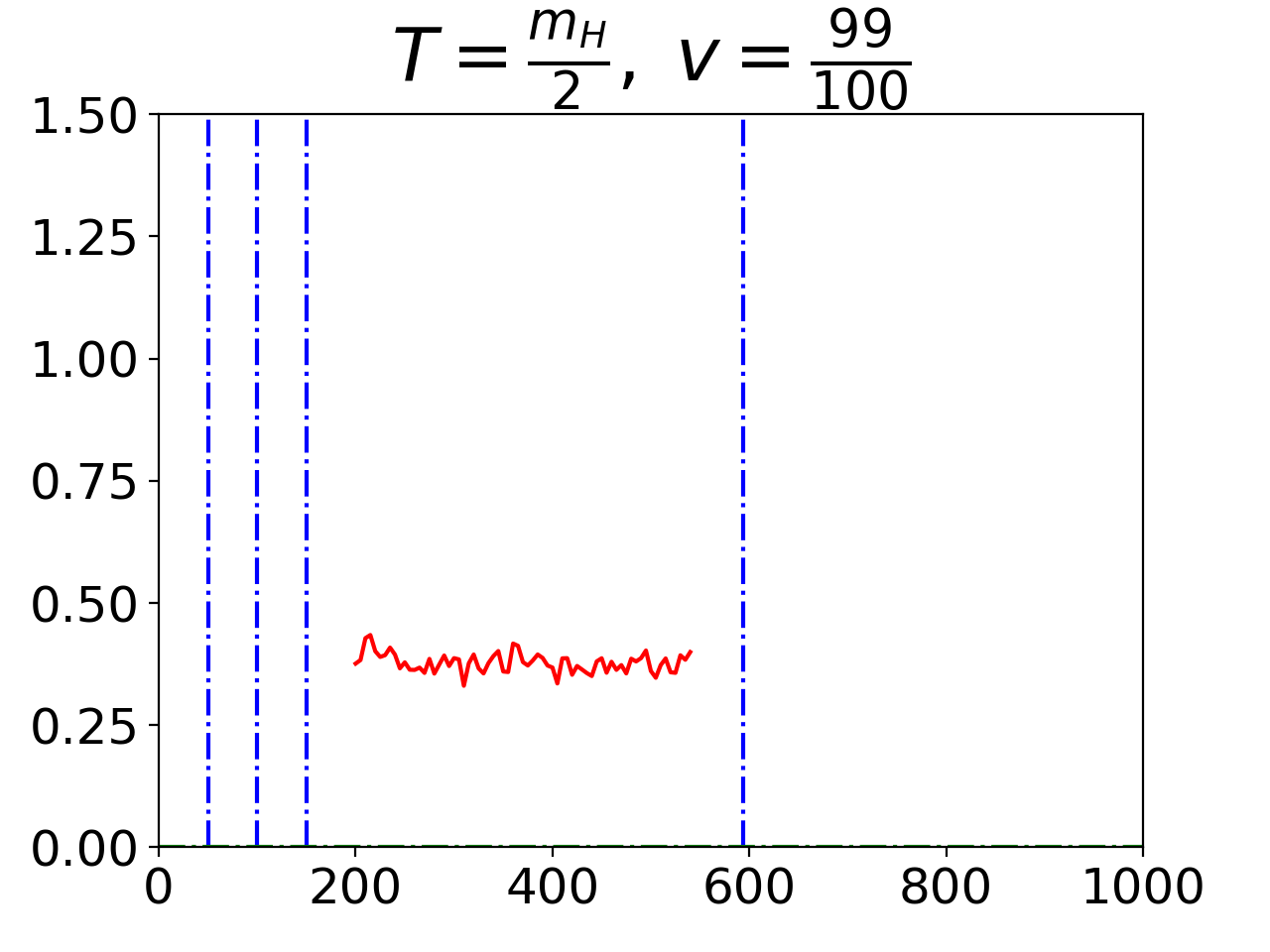} 
&
\includegraphics[width=0.27\textwidth]{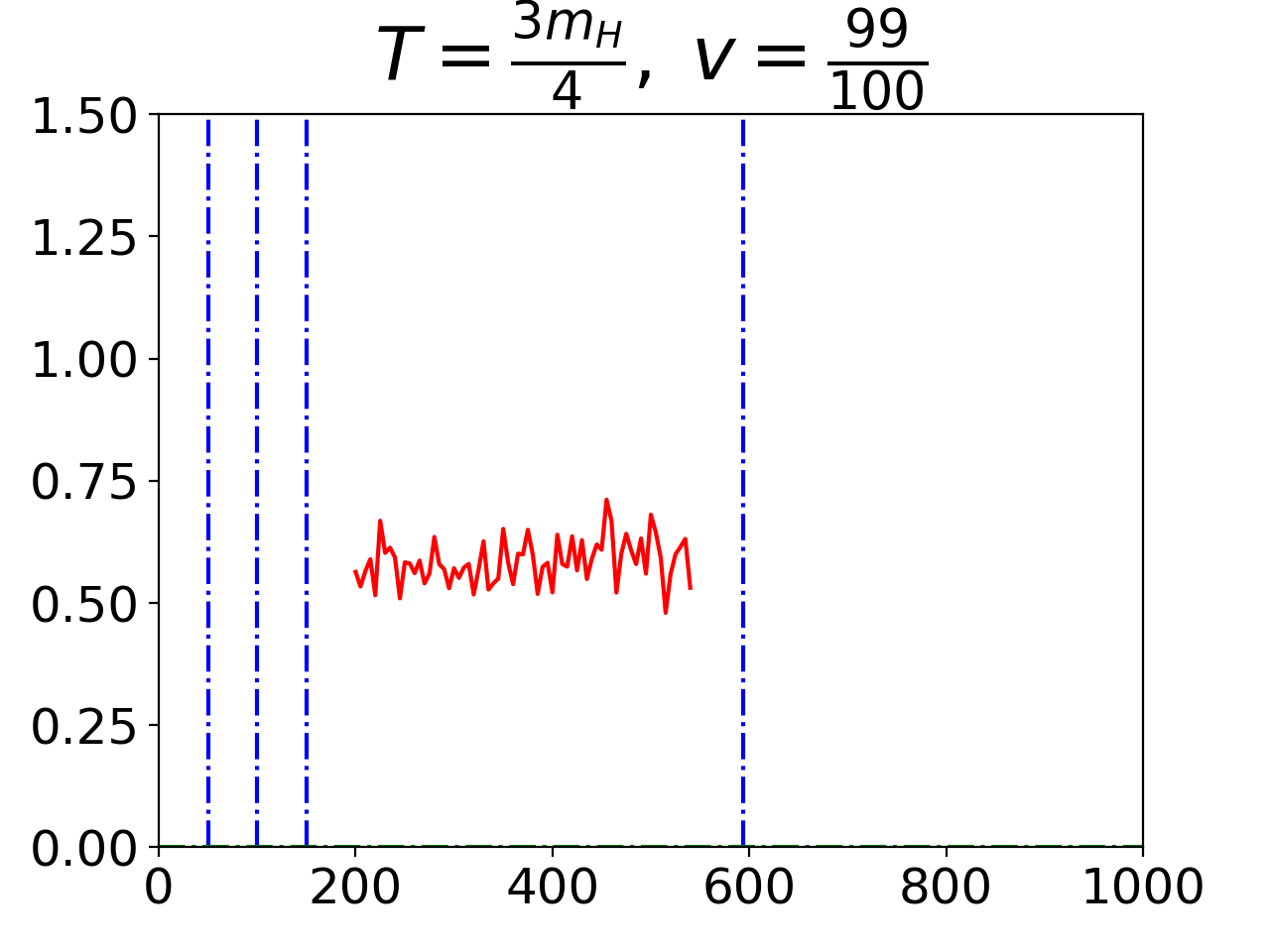} 
&
\includegraphics[width=0.27\textwidth]{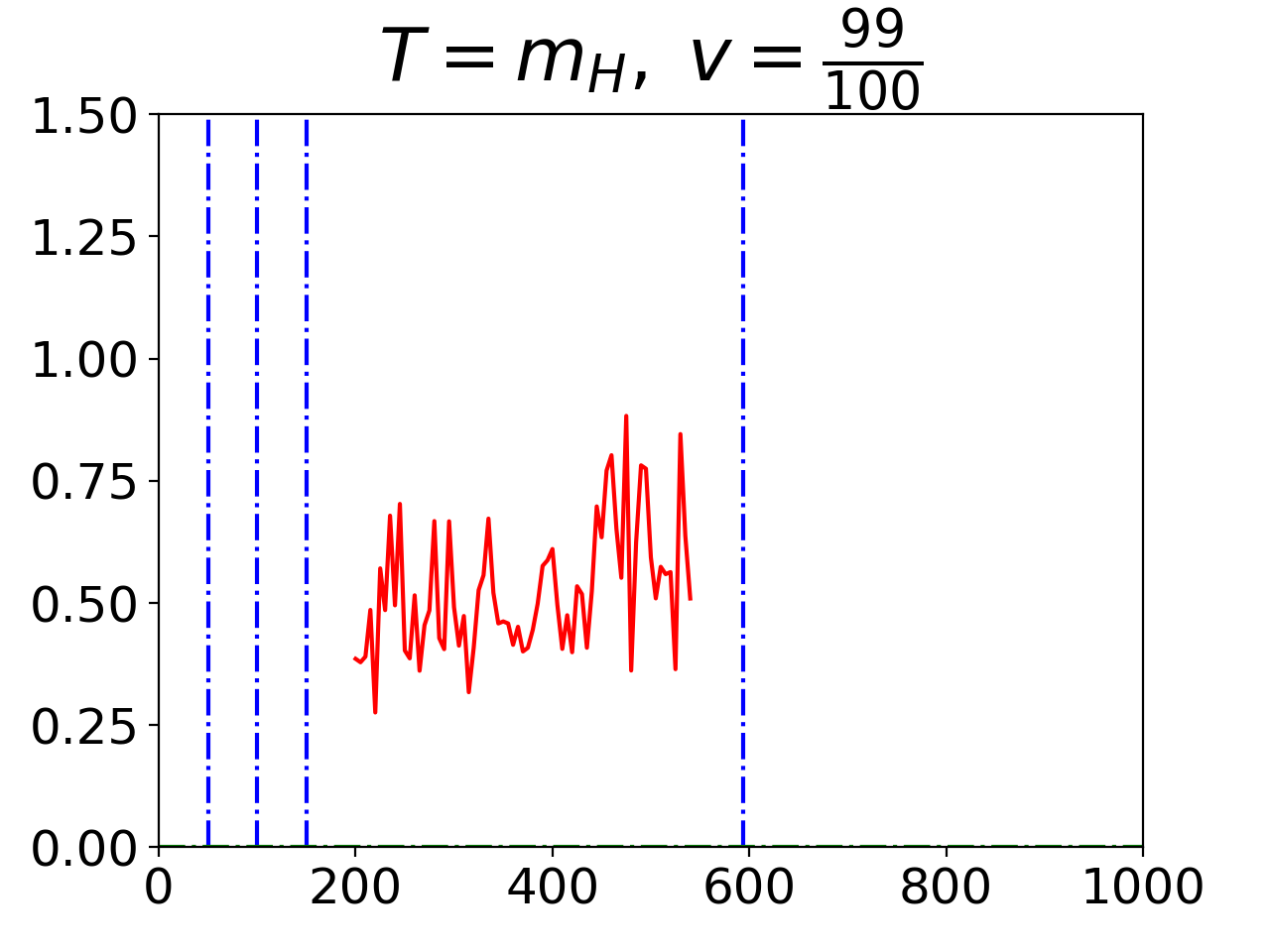} 
\\
\includegraphics[width=0.27\textwidth]{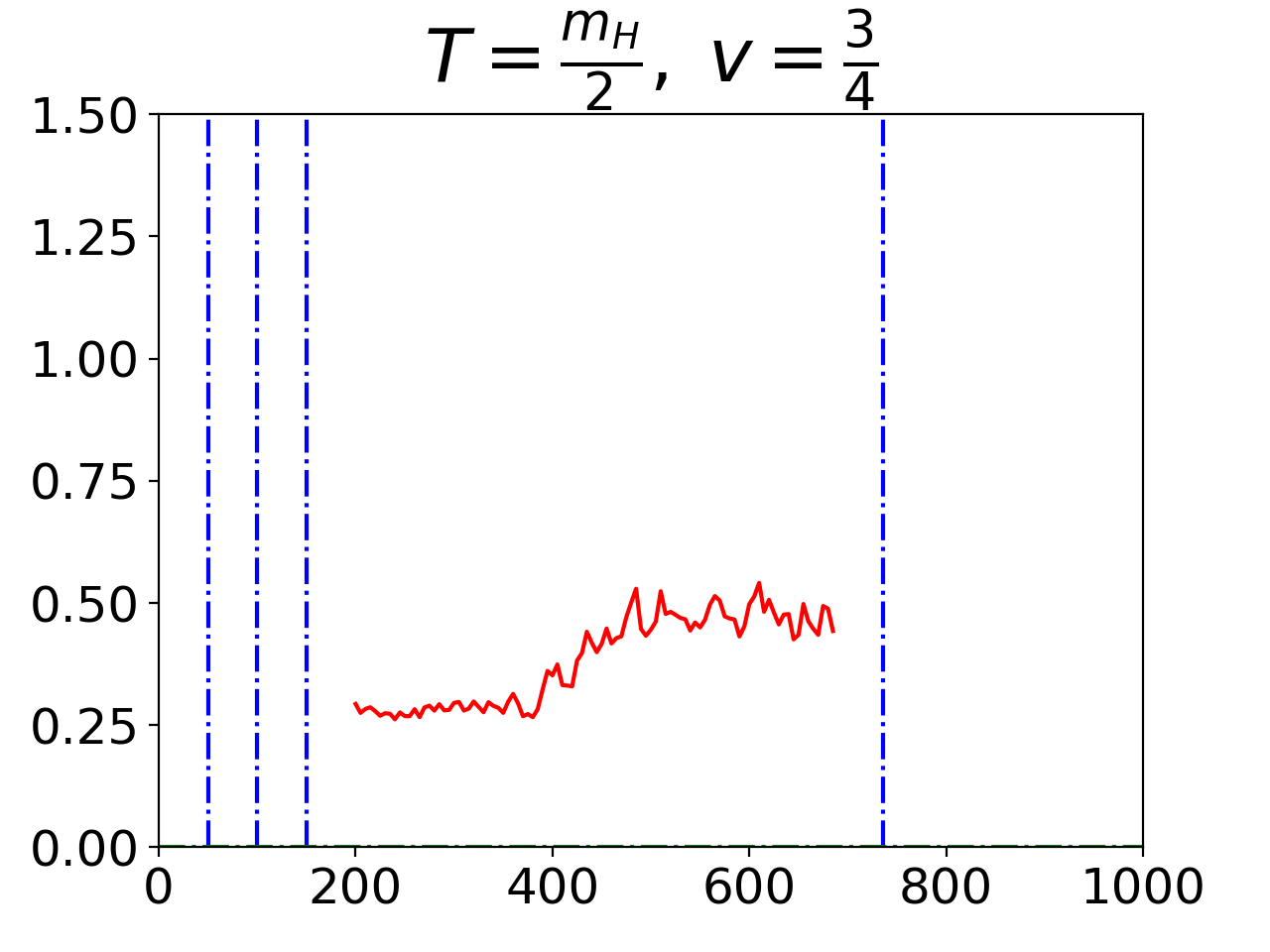} 
&
\includegraphics[width=0.27\textwidth]{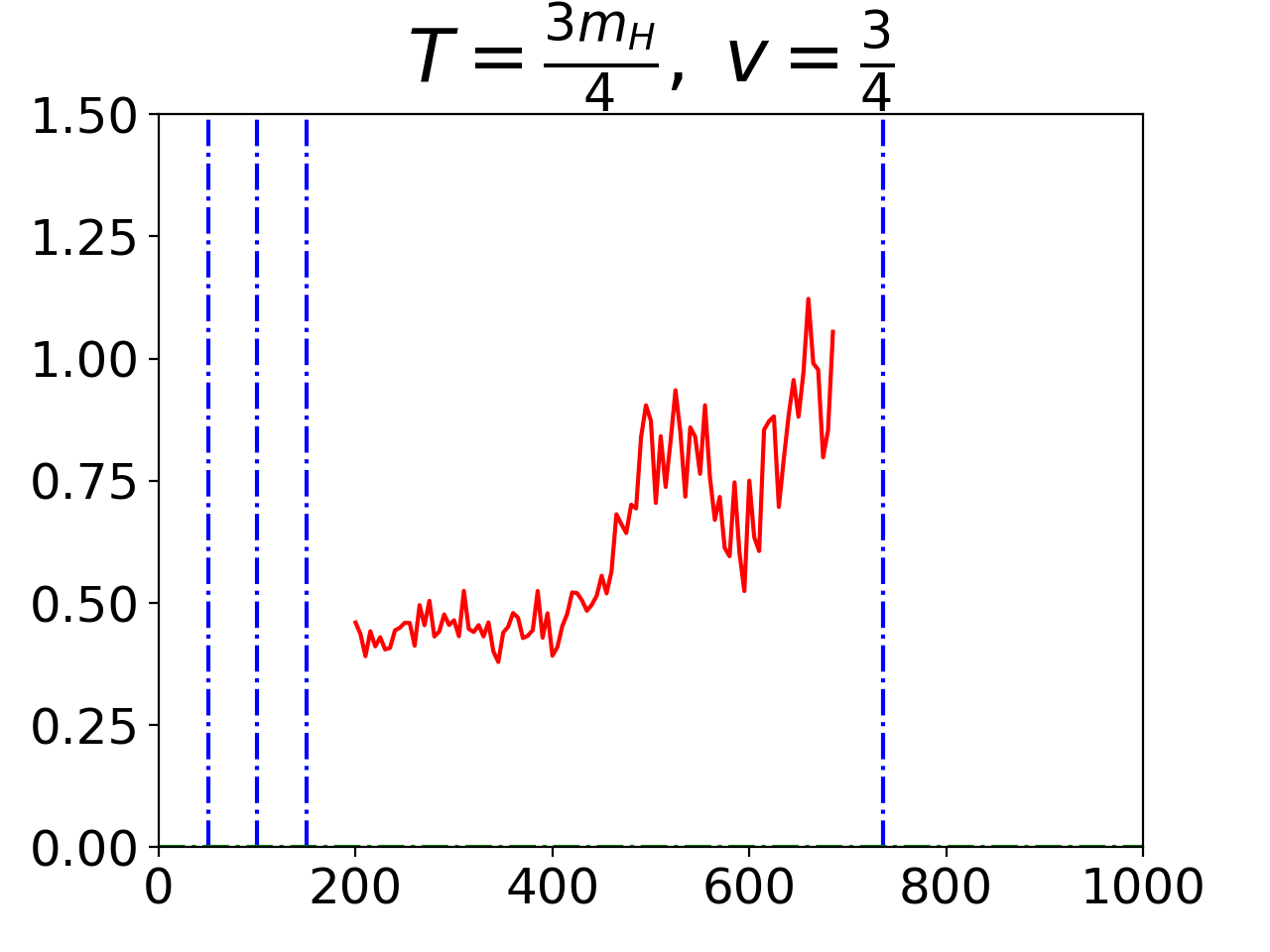} 
&
\includegraphics[width=0.27\textwidth]{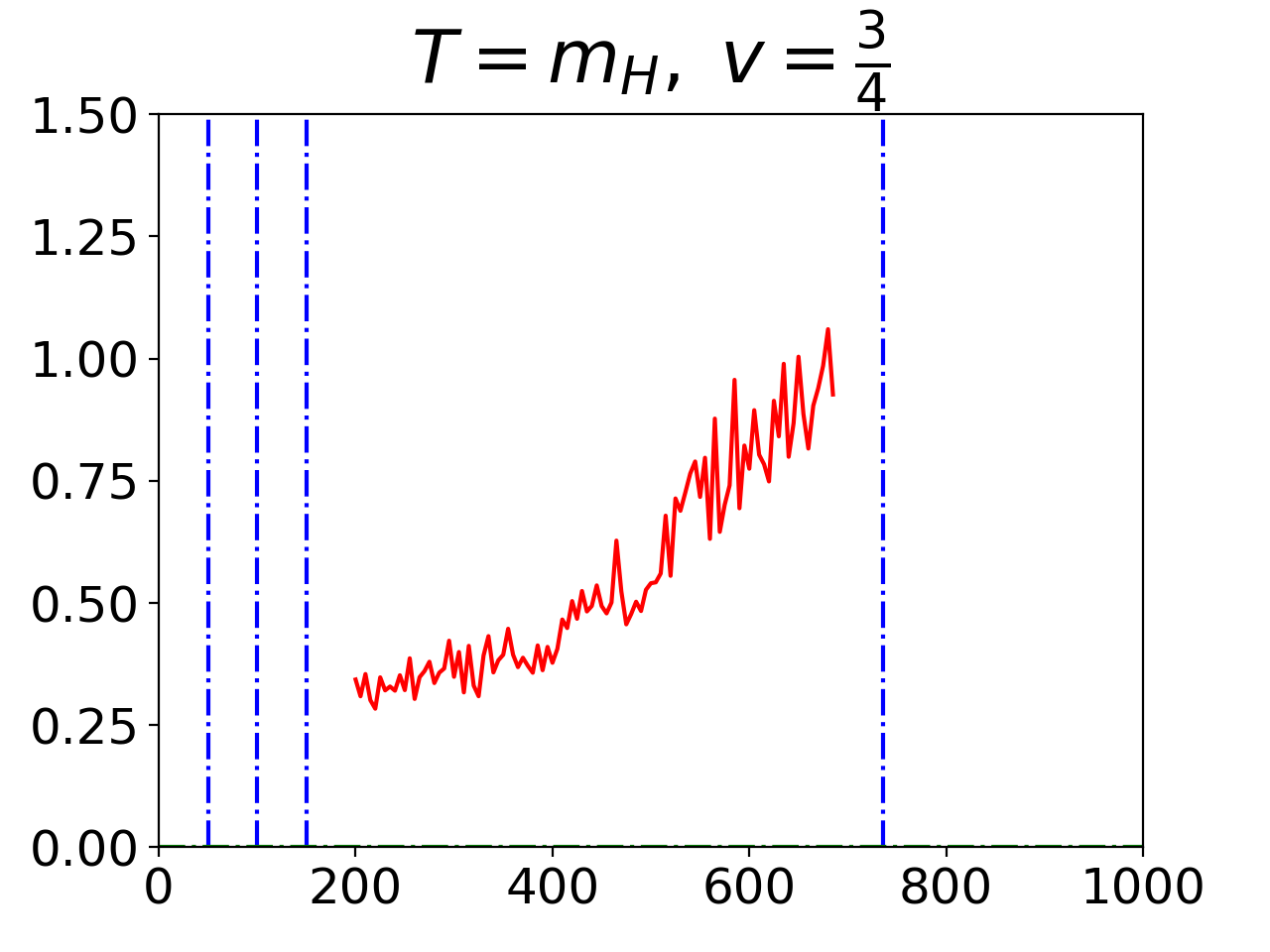} 
\\
\includegraphics[width=0.27\textwidth]{Placeholder.pdf} 
&
\includegraphics[width=0.27\textwidth]{Placeholder.pdf} 
&
\includegraphics[width=0.27\textwidth]{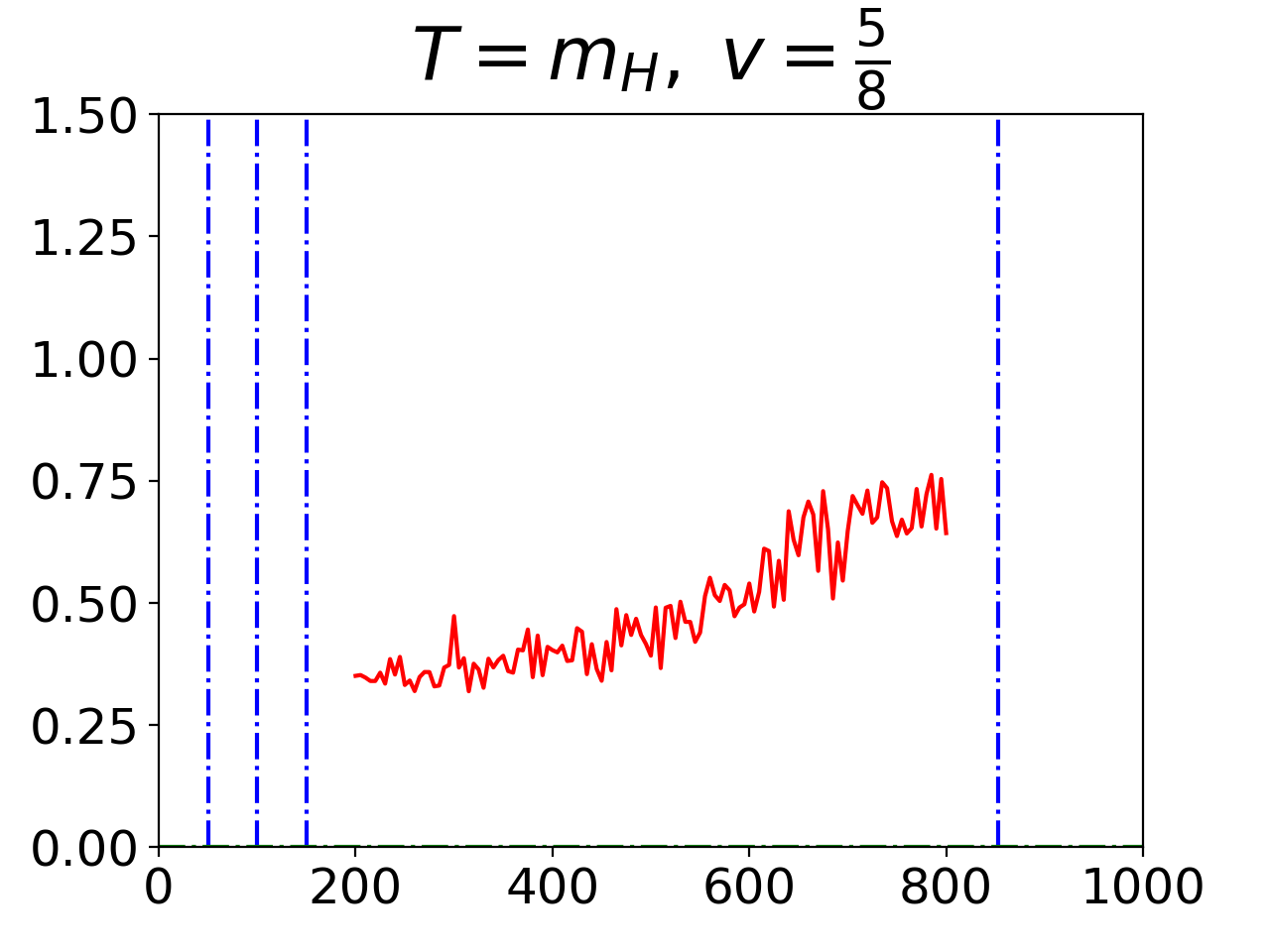} 
\\
\includegraphics[width=0.27\textwidth]{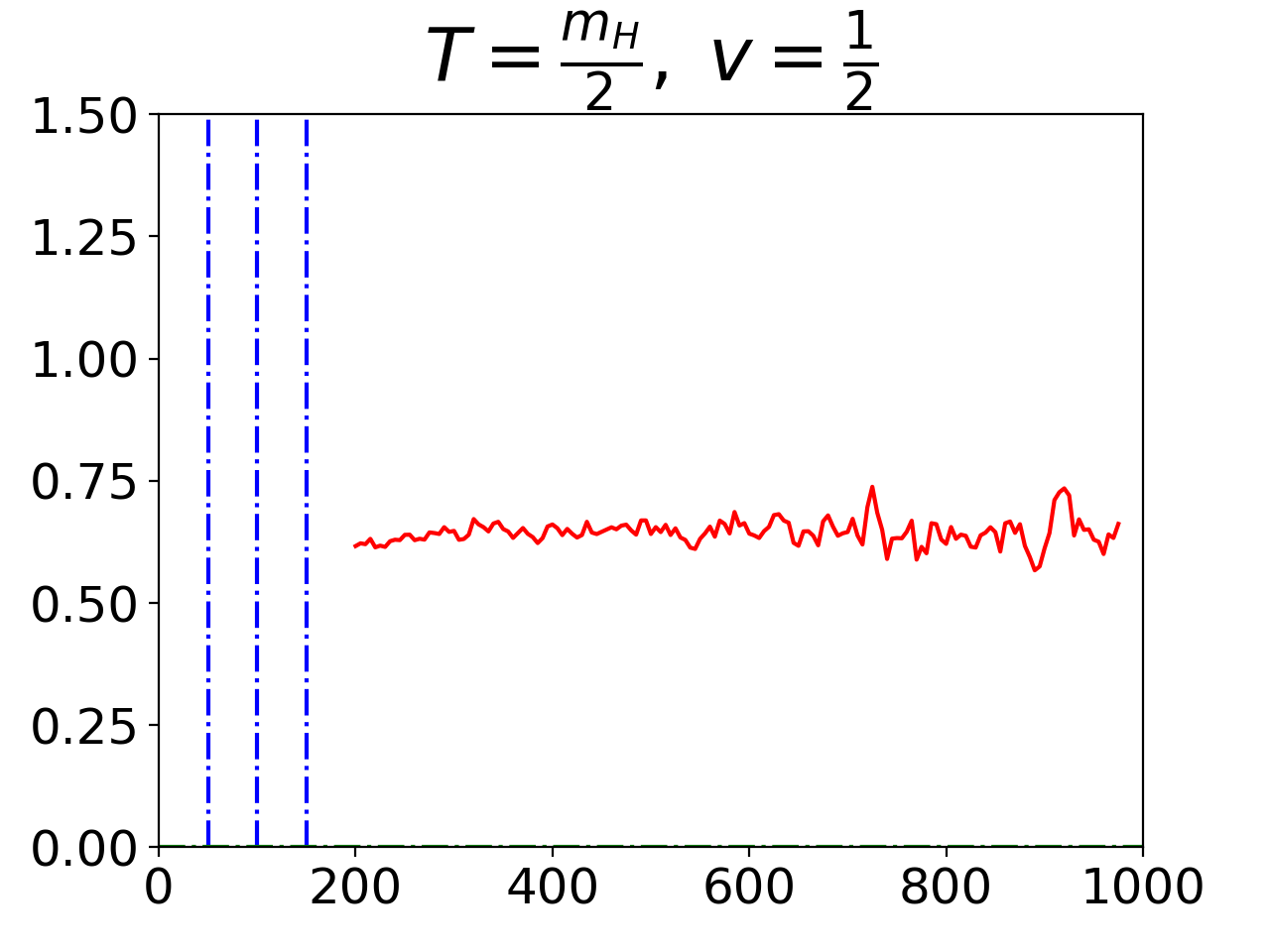} 
&
\includegraphics[width=0.27\textwidth]{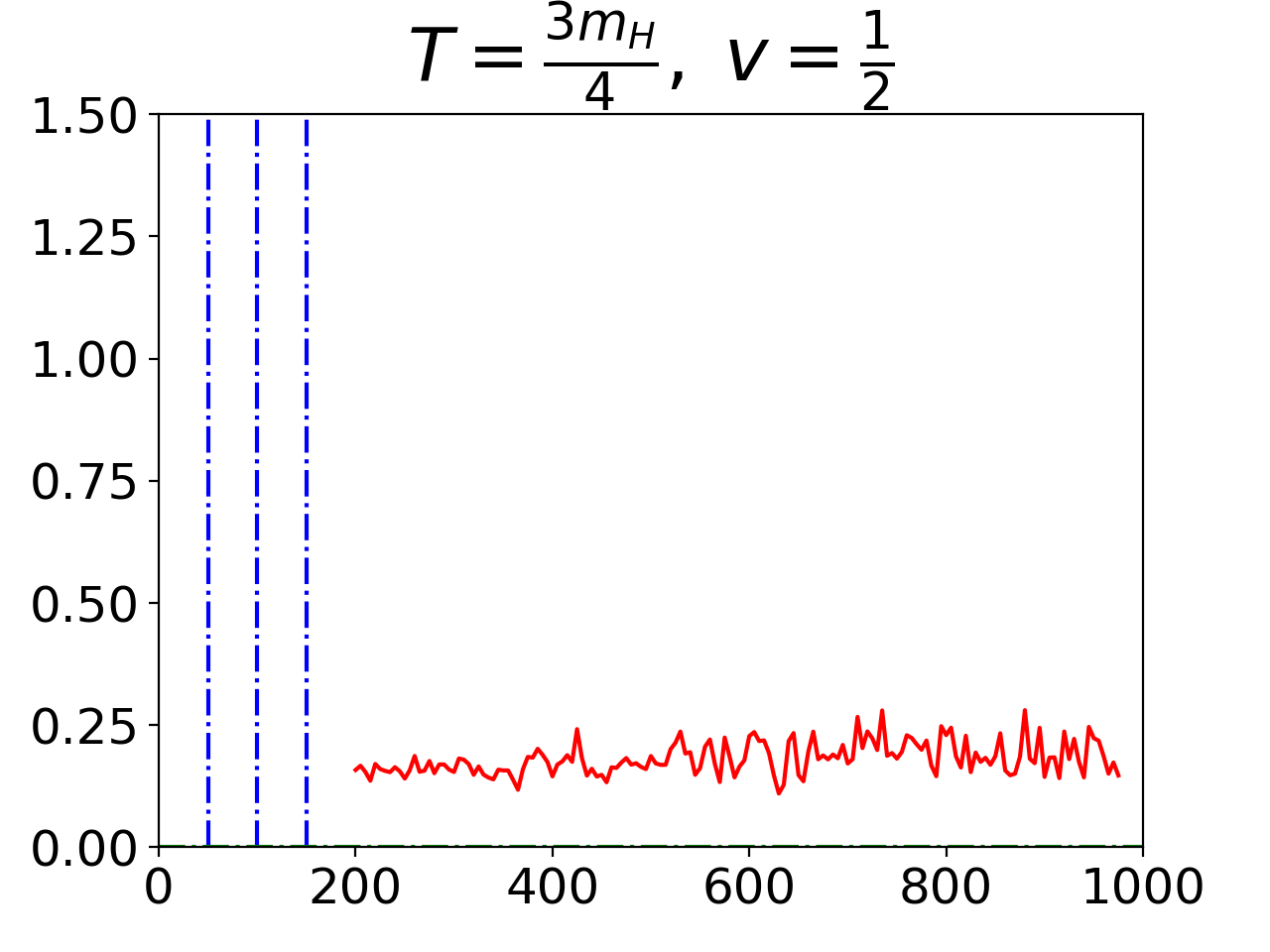} 
&
\includegraphics[width=0.27\textwidth]{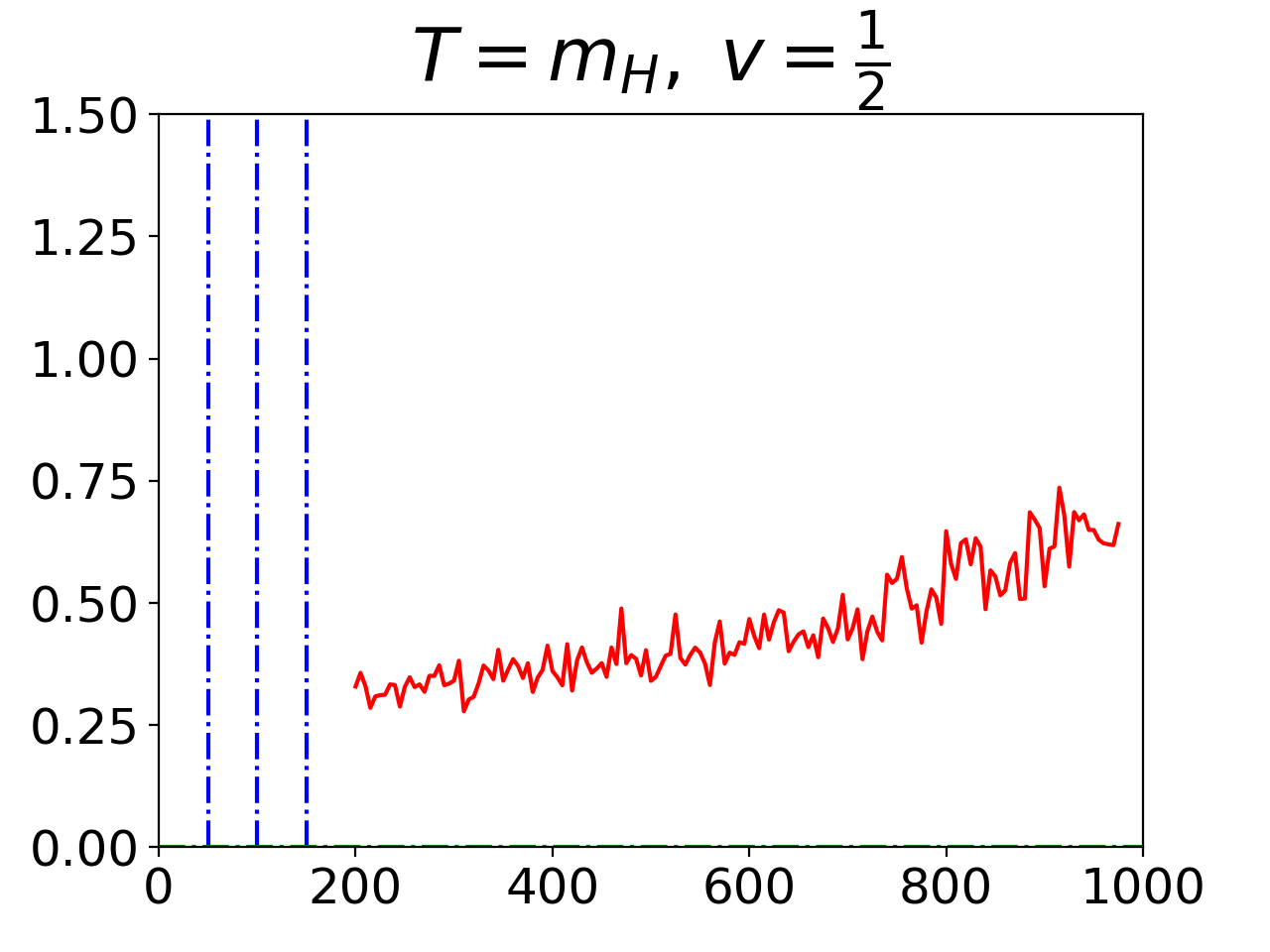} 
\\
\includegraphics[width=0.27\textwidth]{Placeholder.pdf} 
&
\includegraphics[width=0.27\textwidth]{Placeholder.pdf} 
&
\includegraphics[width=0.27\textwidth]{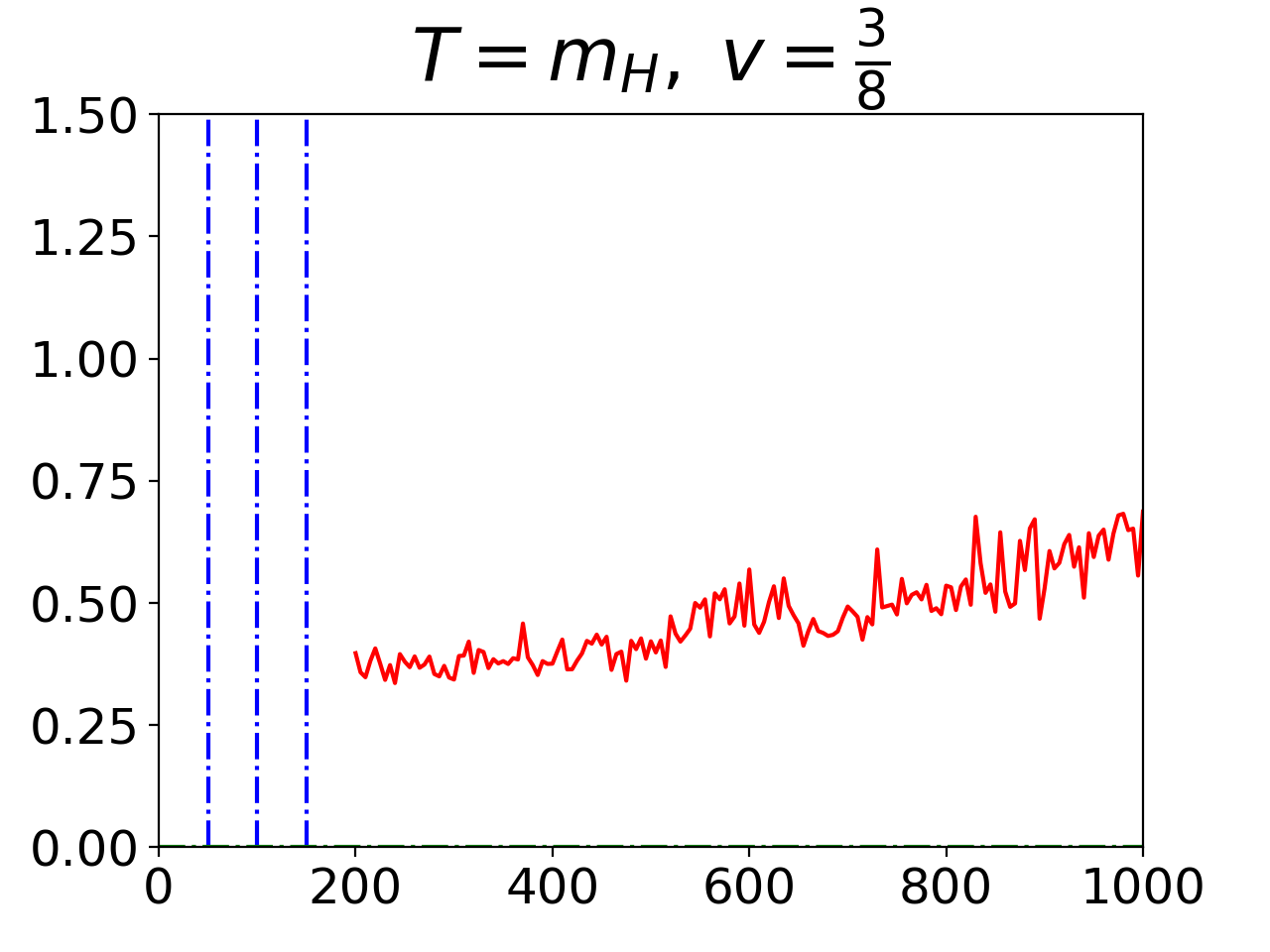} 
\\
\includegraphics[width=0.27\textwidth]{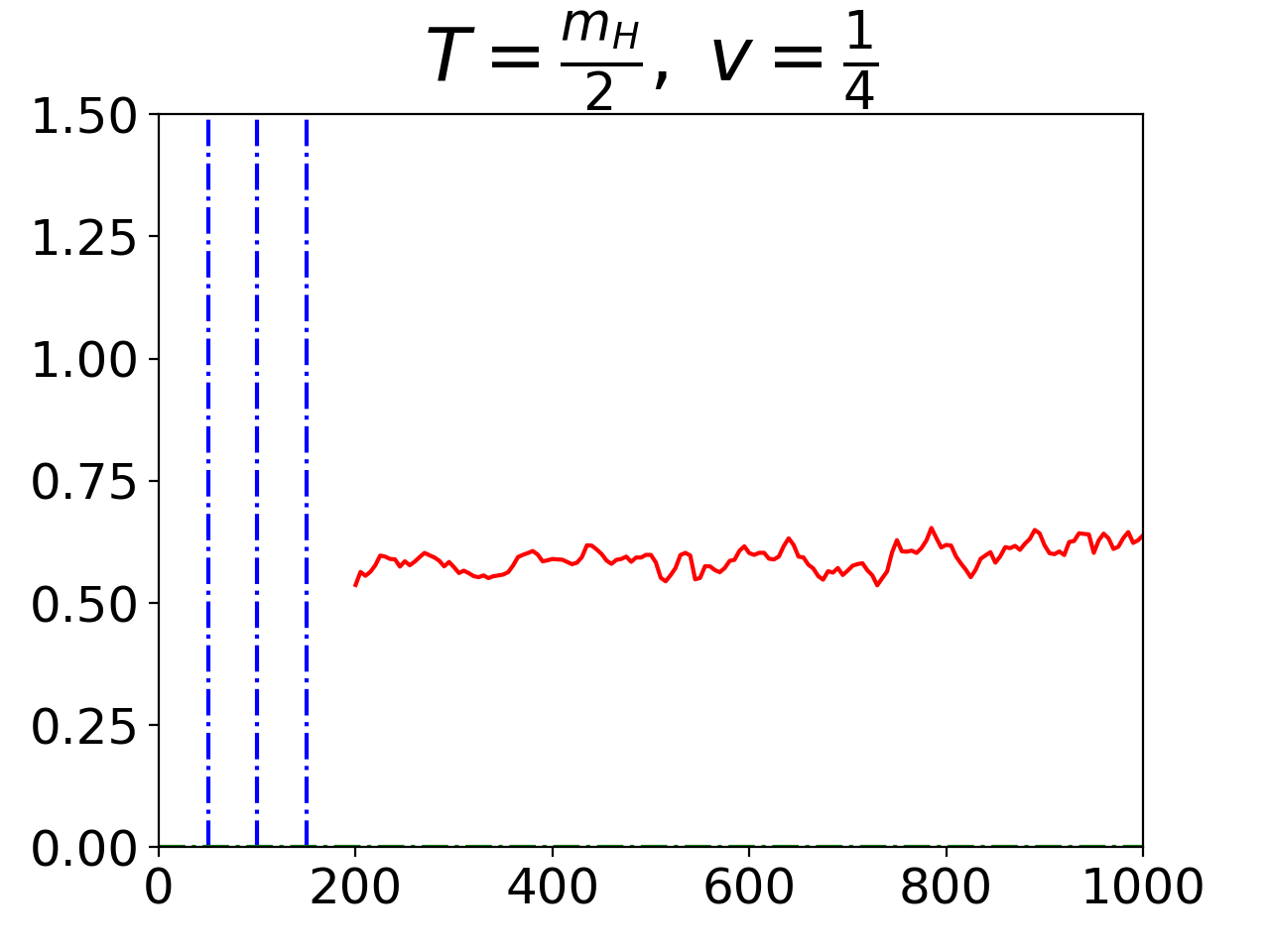} 
&
\includegraphics[width=0.27\textwidth]{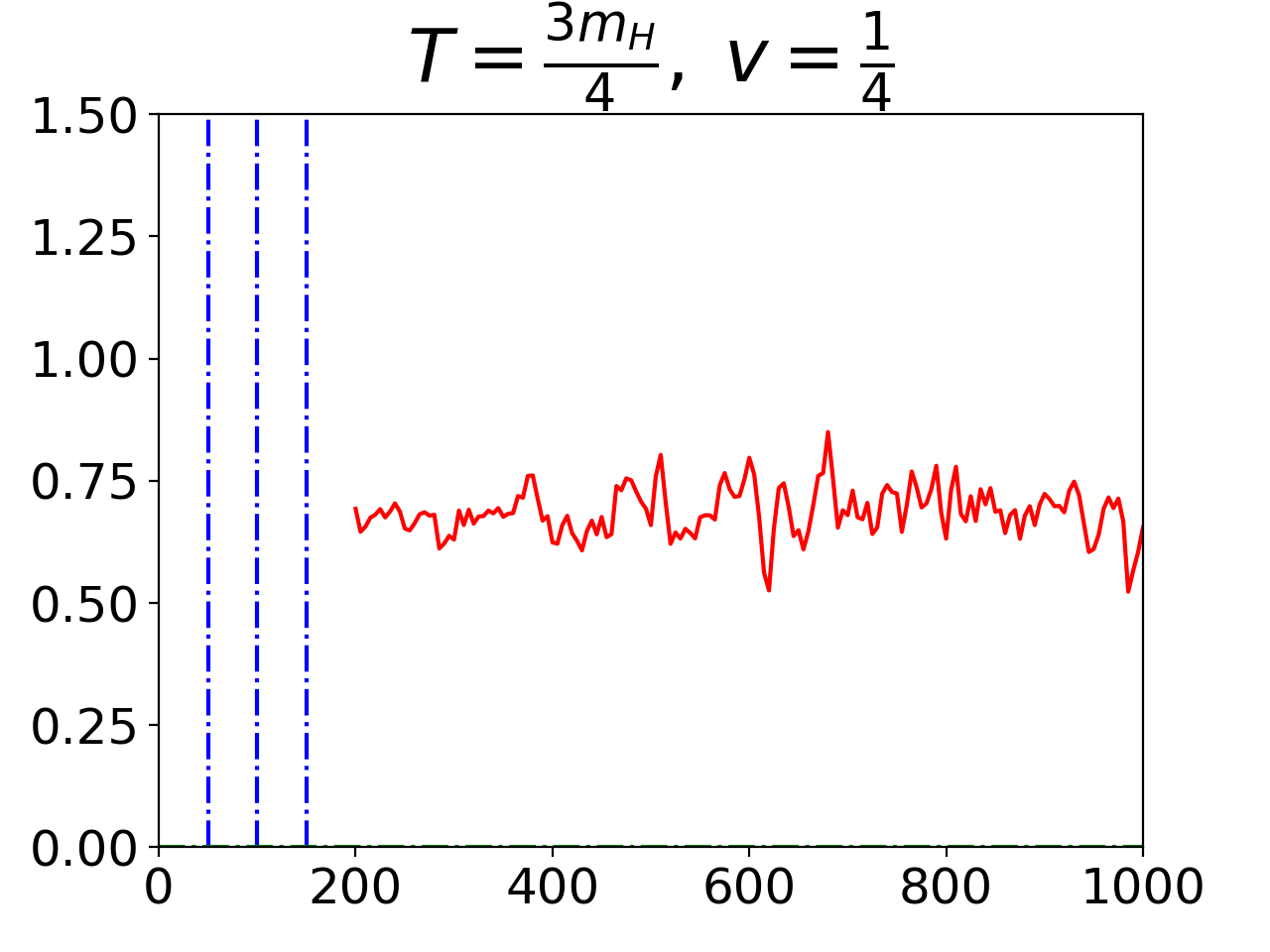} 
&
\includegraphics[width=0.27\textwidth]{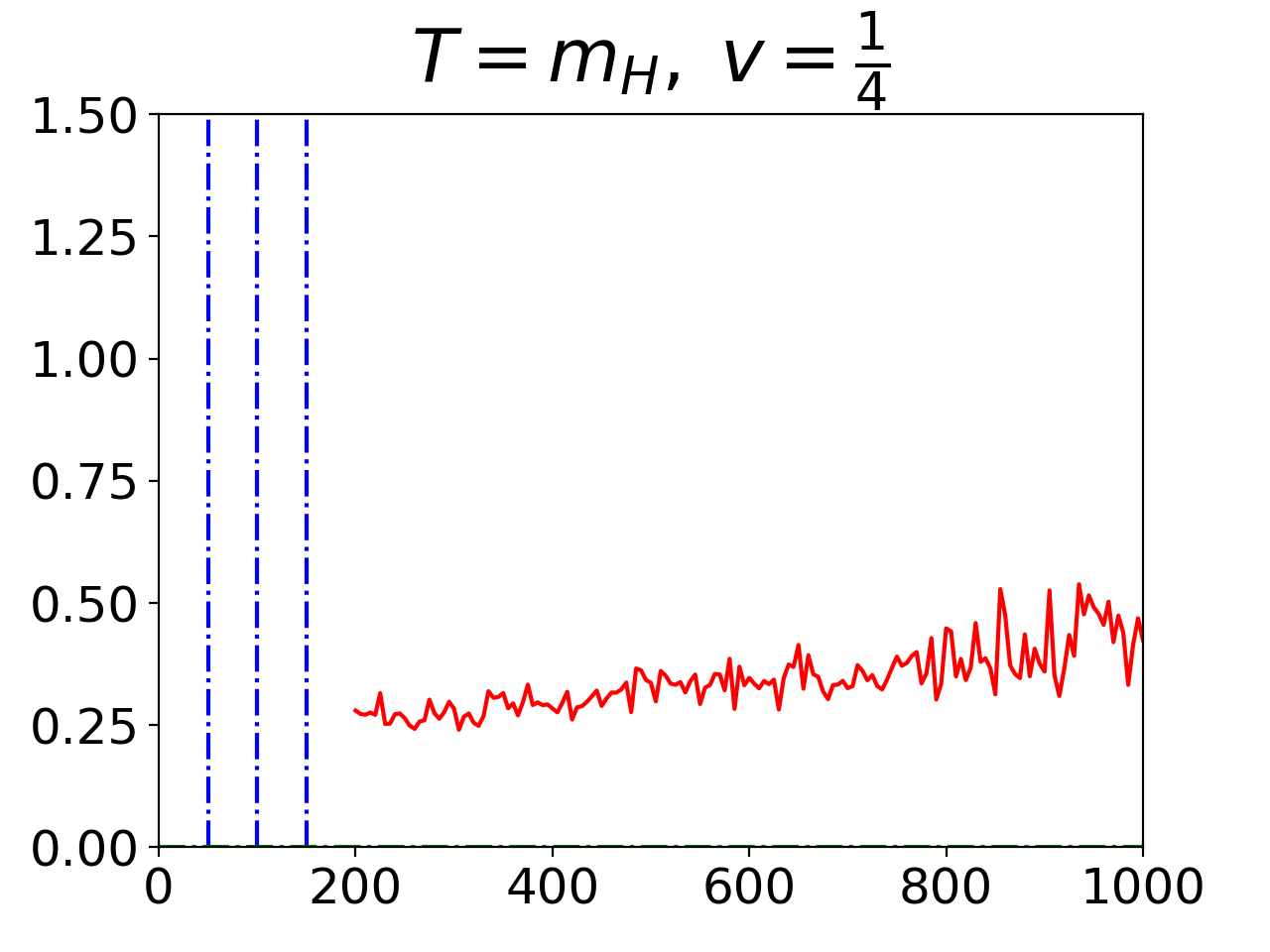} 
\\
\end{tabular}
\caption{ The width of the distribution of winding number inside the bubble, $\bar{N_{\rm w}^2}$ (vertical axis) as a function of time (horizontal axis). Results are shown for different combinations of speed $v$ and temperature $T/m_H$. $T/m_H$ increases moving right in the table of plots, increasing wall speed $v$ moving up. $am_H=0.5$, lattice size $64\times 64\times 1000$, $m_Hd=15$. The data is averaged over 20 configurations, except for the $T/m_H=1$ ( the whole right-most column), which are averaged over 100.
}
\label{fig:diffusion1}
\end{center}
\end{figure}

We can try and construct a simple model of this.  Let us assume that outside the bubble, there is a constant diffusion rate $\Gamma$. But as soon as a region is swallowed by the bubble, this diffusion shuts off. Also, the winding number density is assumed to be uncorrelated in space. Then we can then write
\begin{align}
\bar{N}^2_{\rm w} = \int dV\int_0^t d\tau \frac{d}{d\tau}n^2({\bf x},\tau) = \int dV\int_0^t d\tau \Gamma \theta(z-v\tau) = A\Gamma Z \frac{t}{2},
\end{align}
where $A$ is the cross-section area of the box ($x$-$y$ plane), and $Z$ is the distance covered by the wall up to time $t$, $Z=vt$. Then we have the simple expression for the winding number squared inside the bubble as a function of time:
\begin{align}
\bar{N}^2_{\rm w} =A\Gamma v \frac{t^2}{2},
\end{align}
proportional to the wall speed and to the diffusion rate, which is temperature-dependent, and notably quadratic in time. 

In Fig.~\ref{fig:diffusion1} we show this observable for the same set of speeds and temperatures as before. We see that for low temperatures, there is no activity at all. $\bar{N_{\rm w}^2}$ is constant (or marginally increasing) as a function of time. Only for larger temperatures  $T/m_H=1$ do we see activity in the winding number. It is indeed faster than linear in time, and it increases with wall speed. We remain wary of the results at the largest speed $0.99$, but for the next-to-fastest speed $v=0.75$ we also see some activity.

 As a complementary measurement to establish the presence of topological transitions, we may again consider winding number squared, but rather than integrating over the entire volume inside the bubble, we only integrate over a cubic volume ($z$-range $\Delta Z$ as large as the $x,y$-ranges), immediately on the inside of the wall. The upshot is, that as the bubble passes by, the winding number freezes in. And so we record a snapshot of the winding number diffusion process in a fixed-sized volume moving through space.  Modelling this again, we have
\begin{align}
\bar{N}^2_{\rm w} = \int dV\int_0^t d\tau \frac{d}{d\tau}n^2({\bf x},\tau) = \int dV\int_0^t d\tau \Gamma \theta(z-v\tau),
\end{align}
but now the volume integral is only over $A\Delta Z$, and we find
\begin{align}
\bar{N}^2_{\rm w} = A\Gamma\Delta Z t\left(1-\frac{\Delta Z}{2vt}\right),
\end{align}
which for large enough times is proportional to $\Gamma$, independent of speed $v$ and linear in time.

\begin{figure}[H]
%\vspace{-3cm}
\begin{center}
\begin{tabular}{llll}
%\hspace{-2cm}
\includegraphics[width=0.27\textwidth]{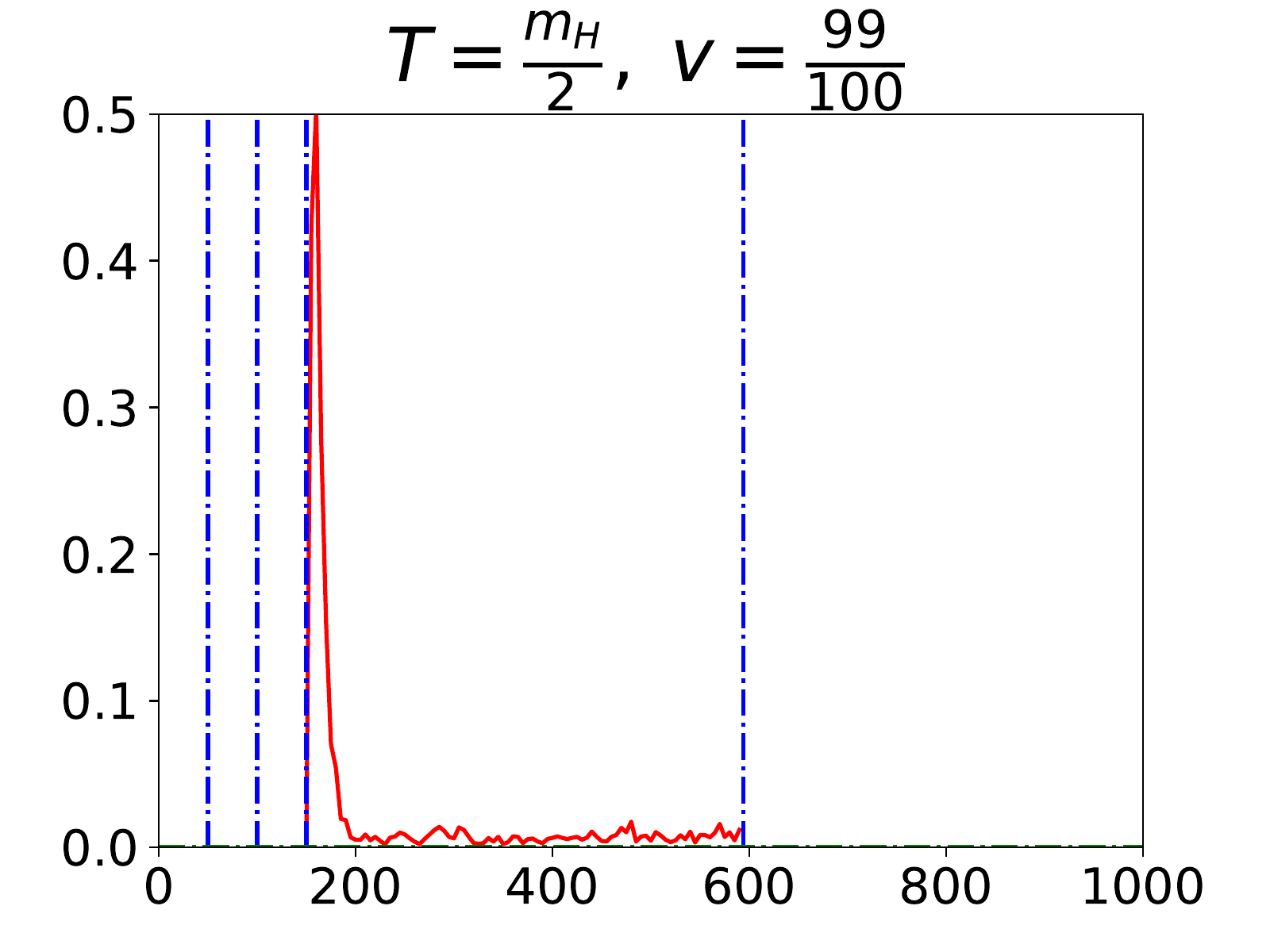} 
&
\includegraphics[width=0.27\textwidth]{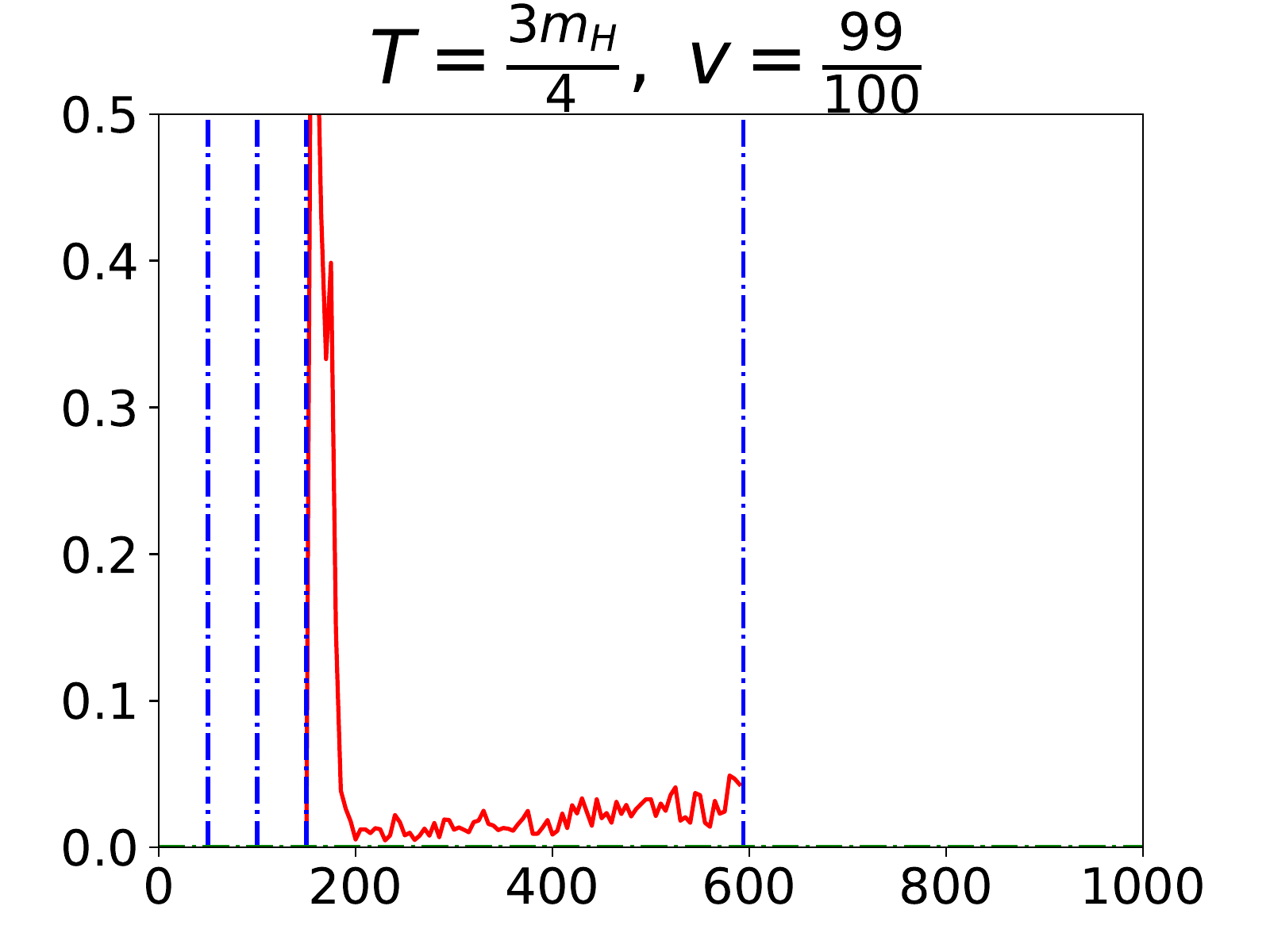} 
&
\includegraphics[width=0.27\textwidth]{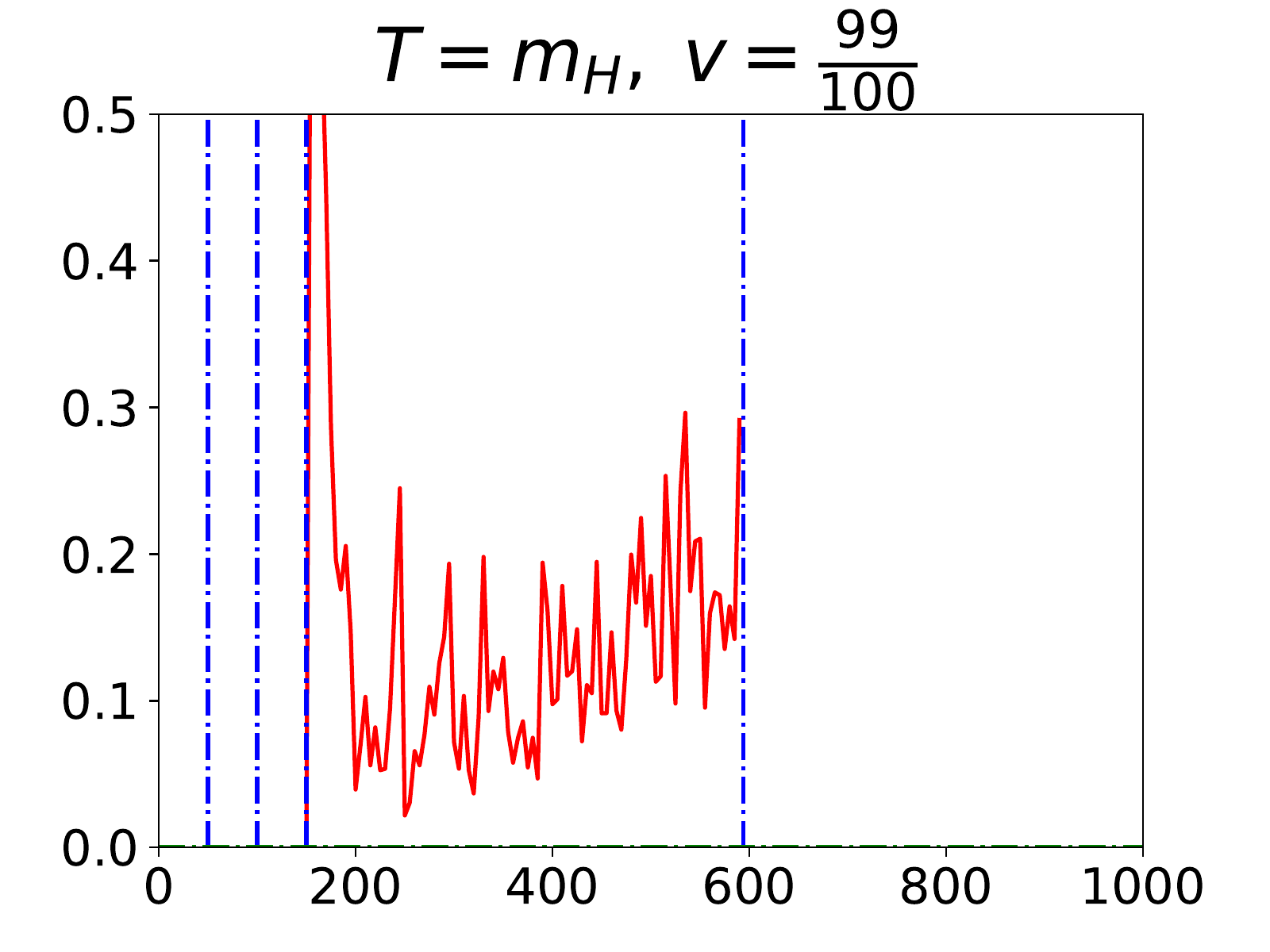} 
\\
\includegraphics[width=0.27\textwidth]{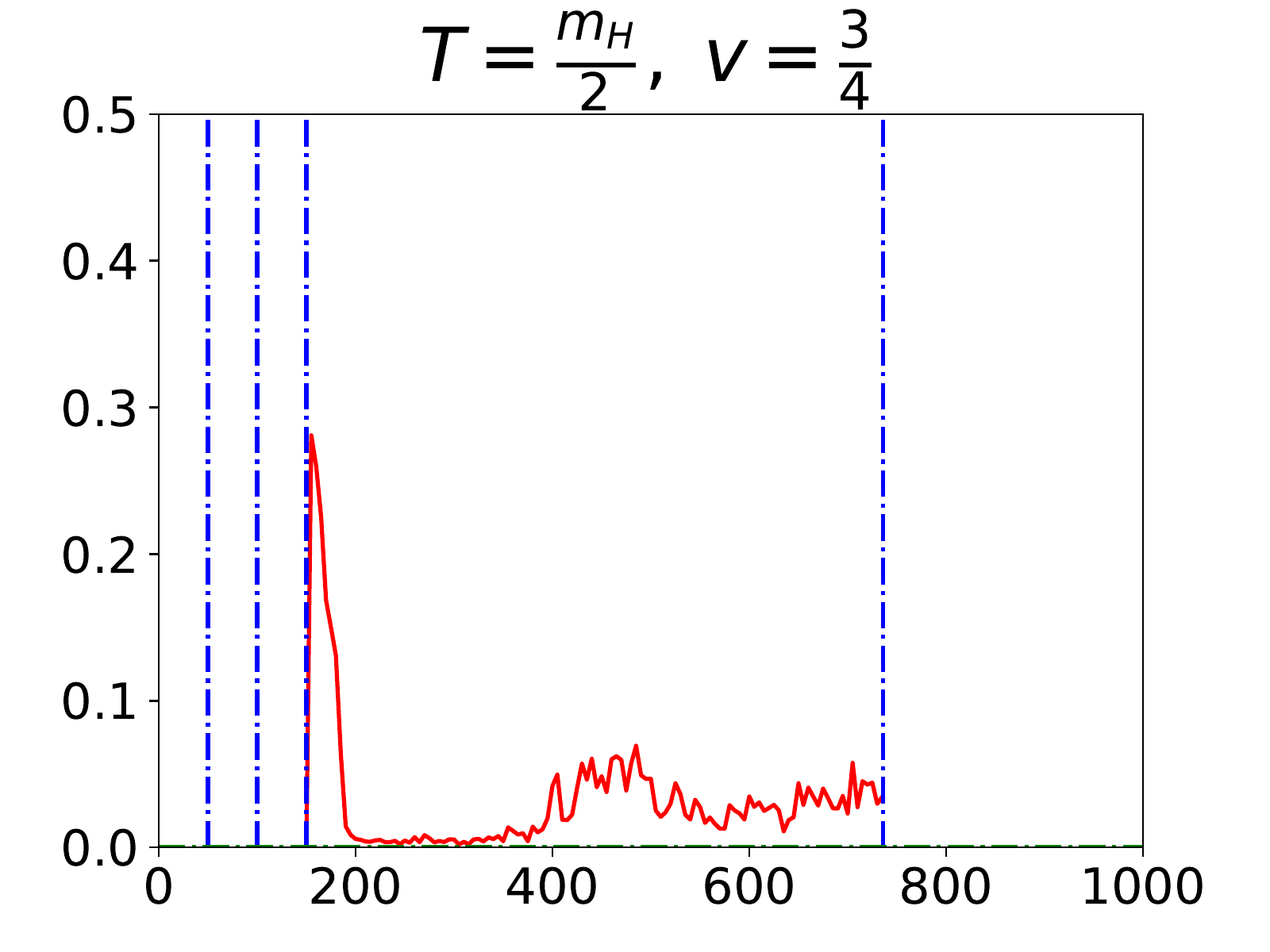} 
&
\includegraphics[width=0.27\textwidth]{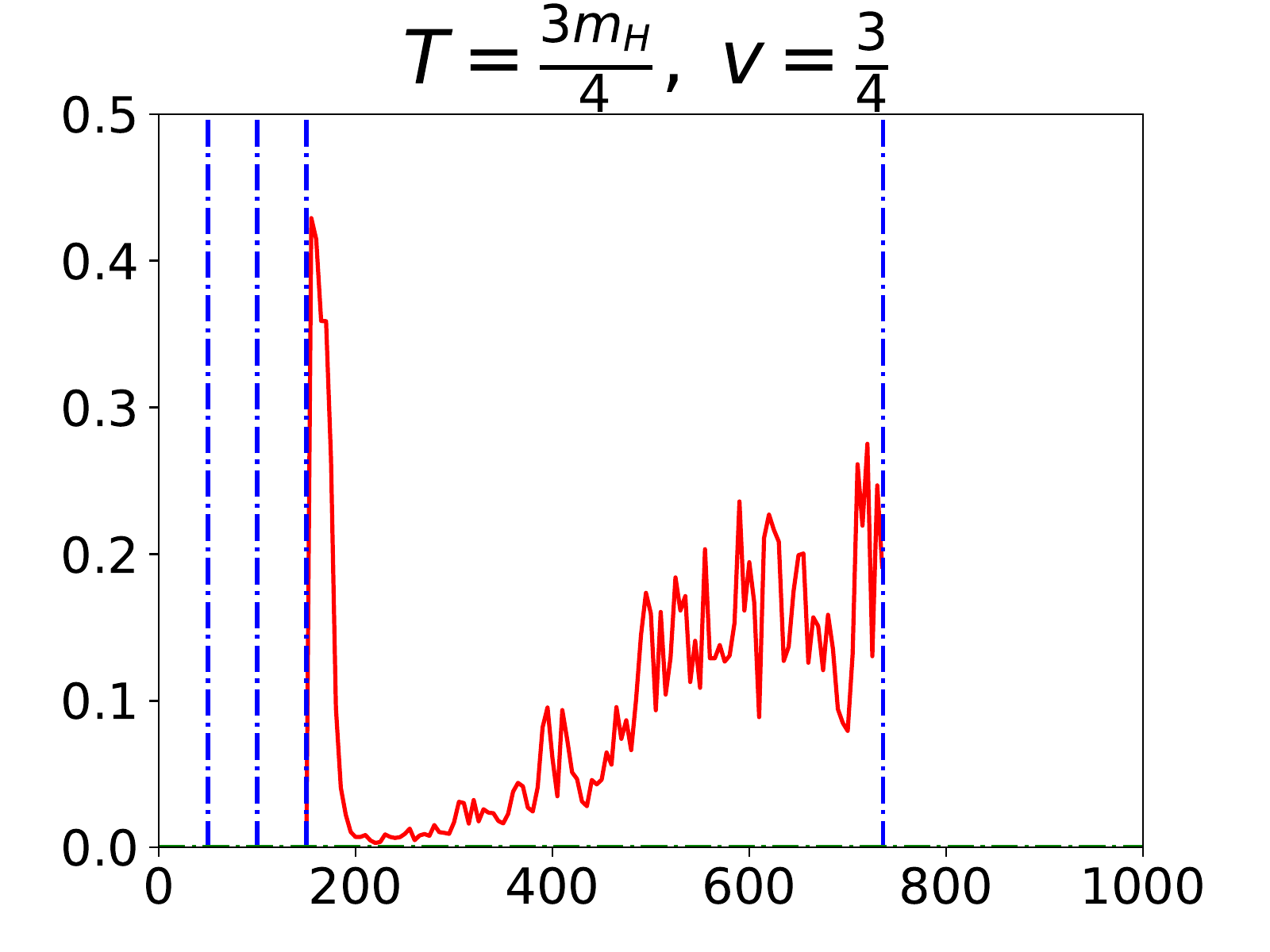} 
&
\includegraphics[width=0.27\textwidth]{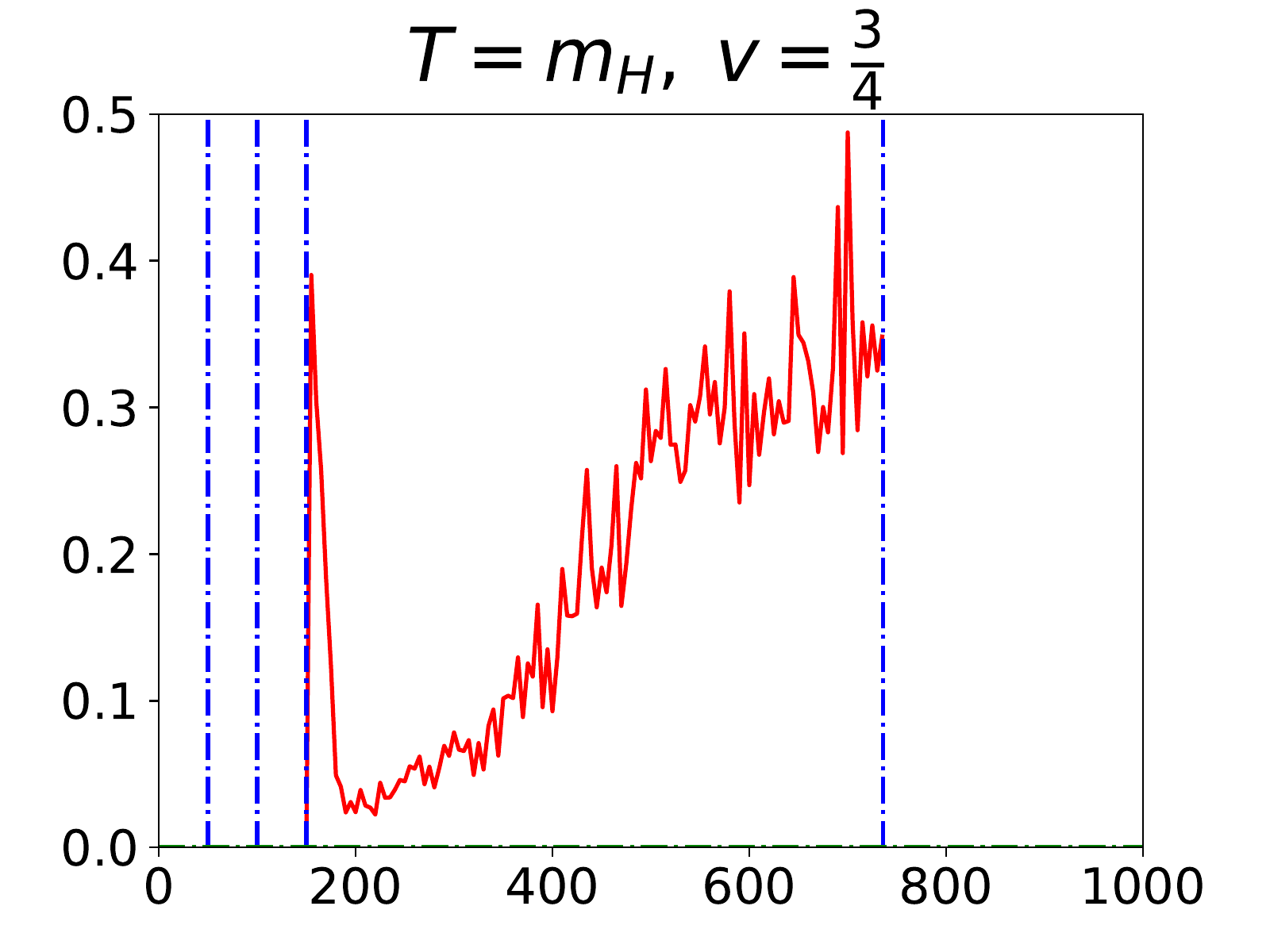} 
\\
\includegraphics[width=0.27\textwidth]{Placeholder.pdf} 
&
\includegraphics[width=0.27\textwidth]{Placeholder.pdf} 
&
\includegraphics[width=0.27\textwidth]{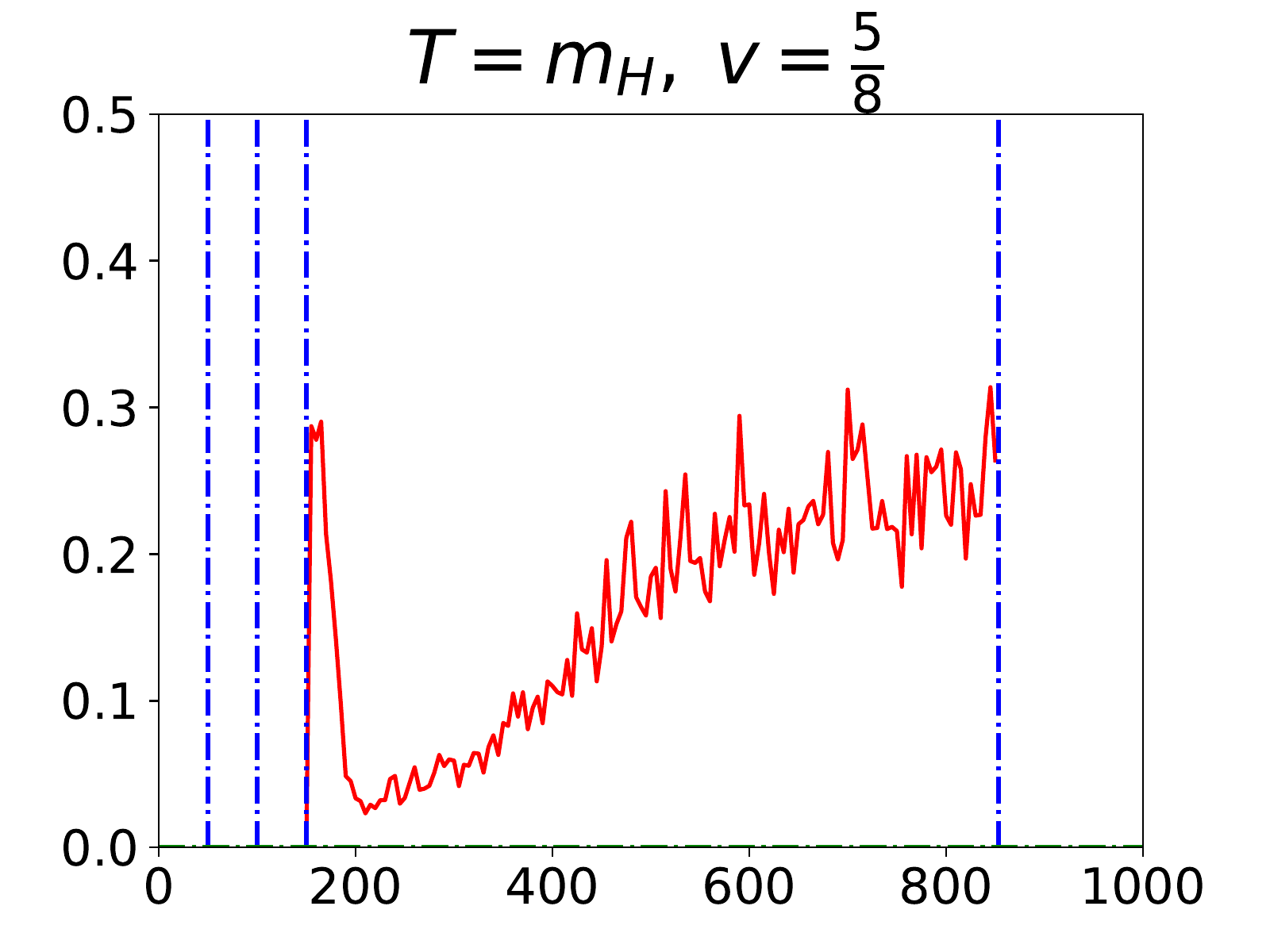} 
\\
\includegraphics[width=0.27\textwidth]{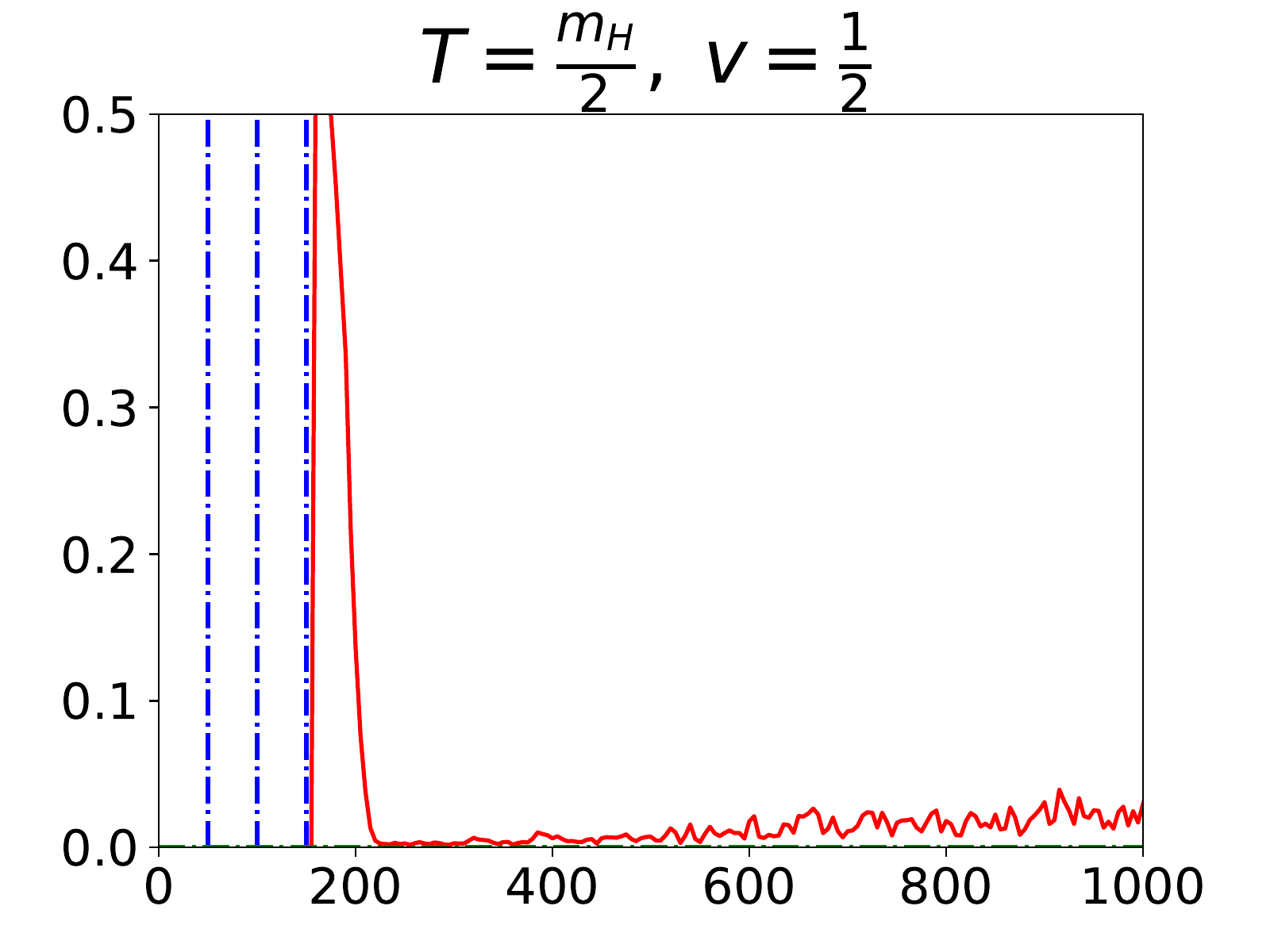} 
&
\includegraphics[width=0.27\textwidth]{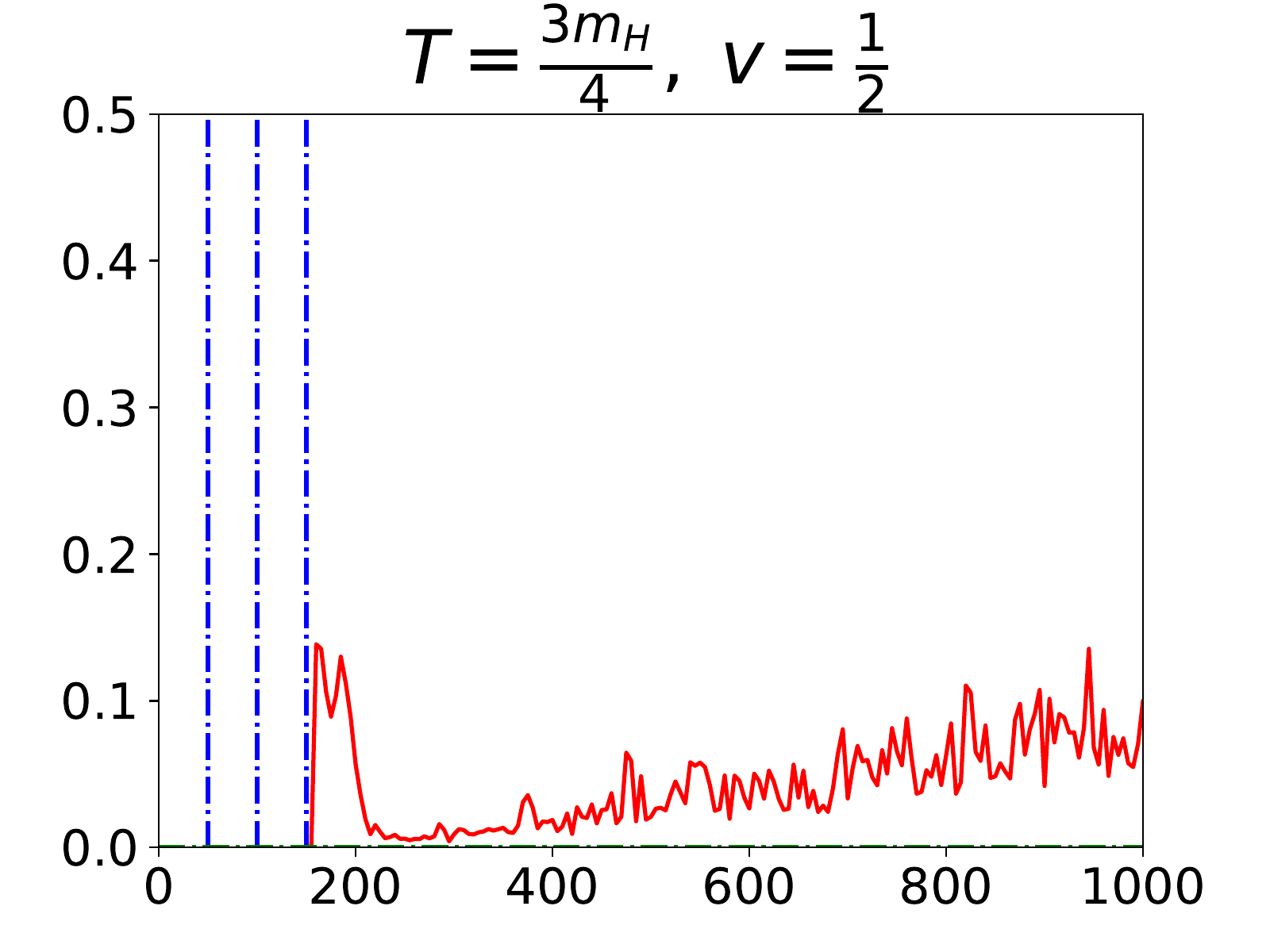} 
&
\includegraphics[width=0.27\textwidth]{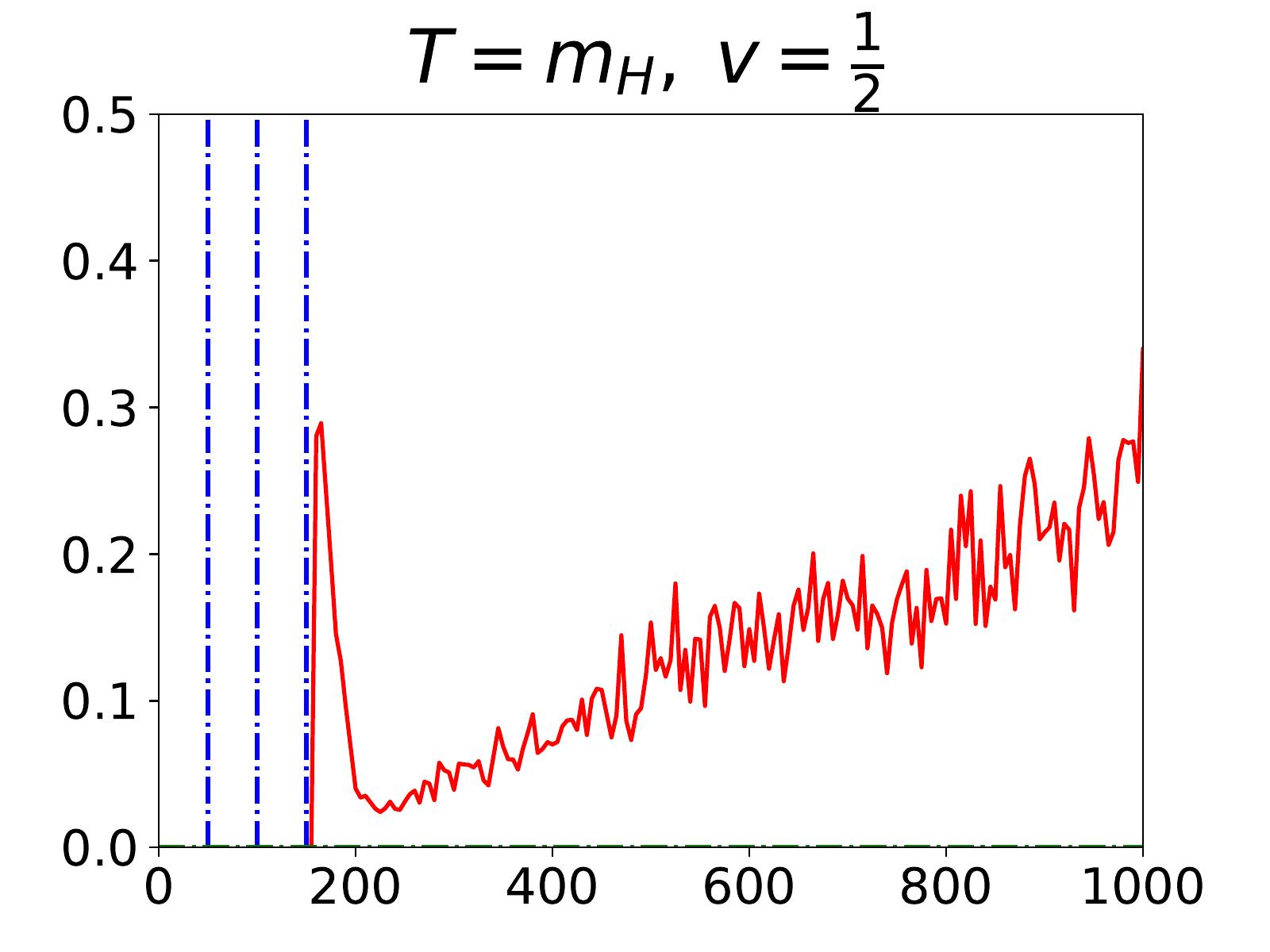} 
\\
\includegraphics[width=0.27\textwidth]{Placeholder.pdf} 
&
\includegraphics[width=0.27\textwidth]{Placeholder.pdf} 
&
\includegraphics[width=0.27\textwidth]{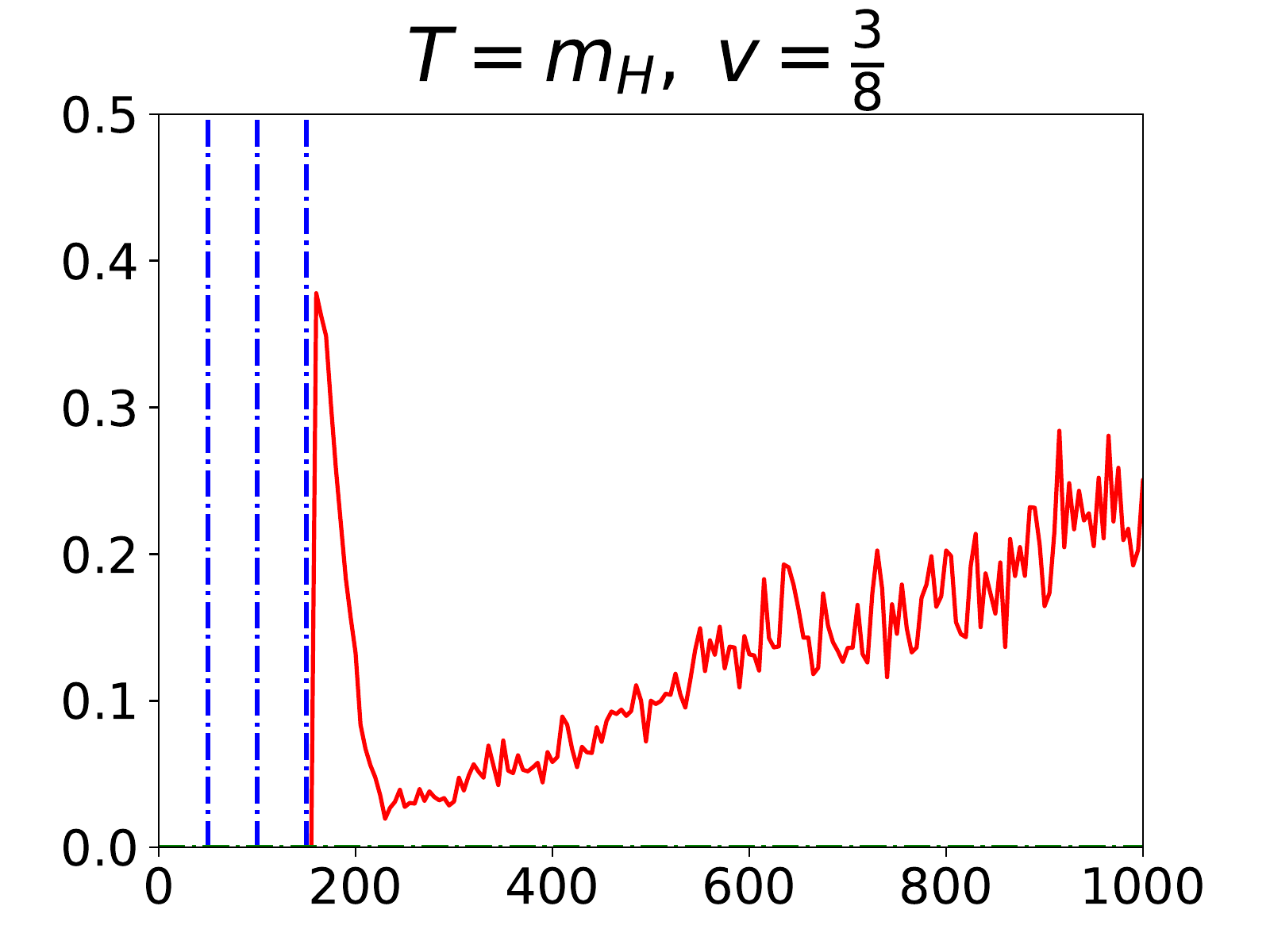} 
\\
\includegraphics[width=0.27\textwidth]{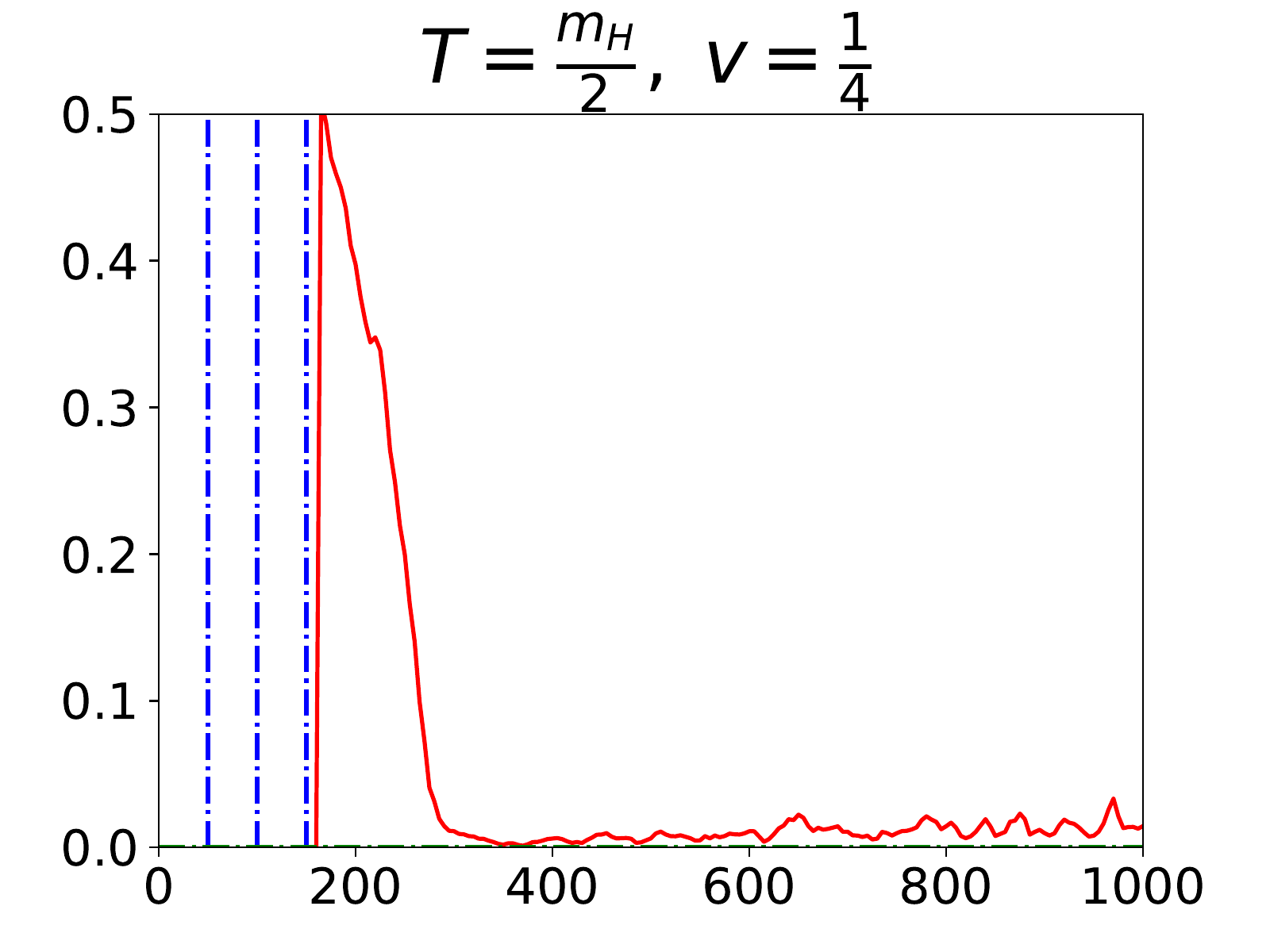} 
&
\includegraphics[width=0.27\textwidth]{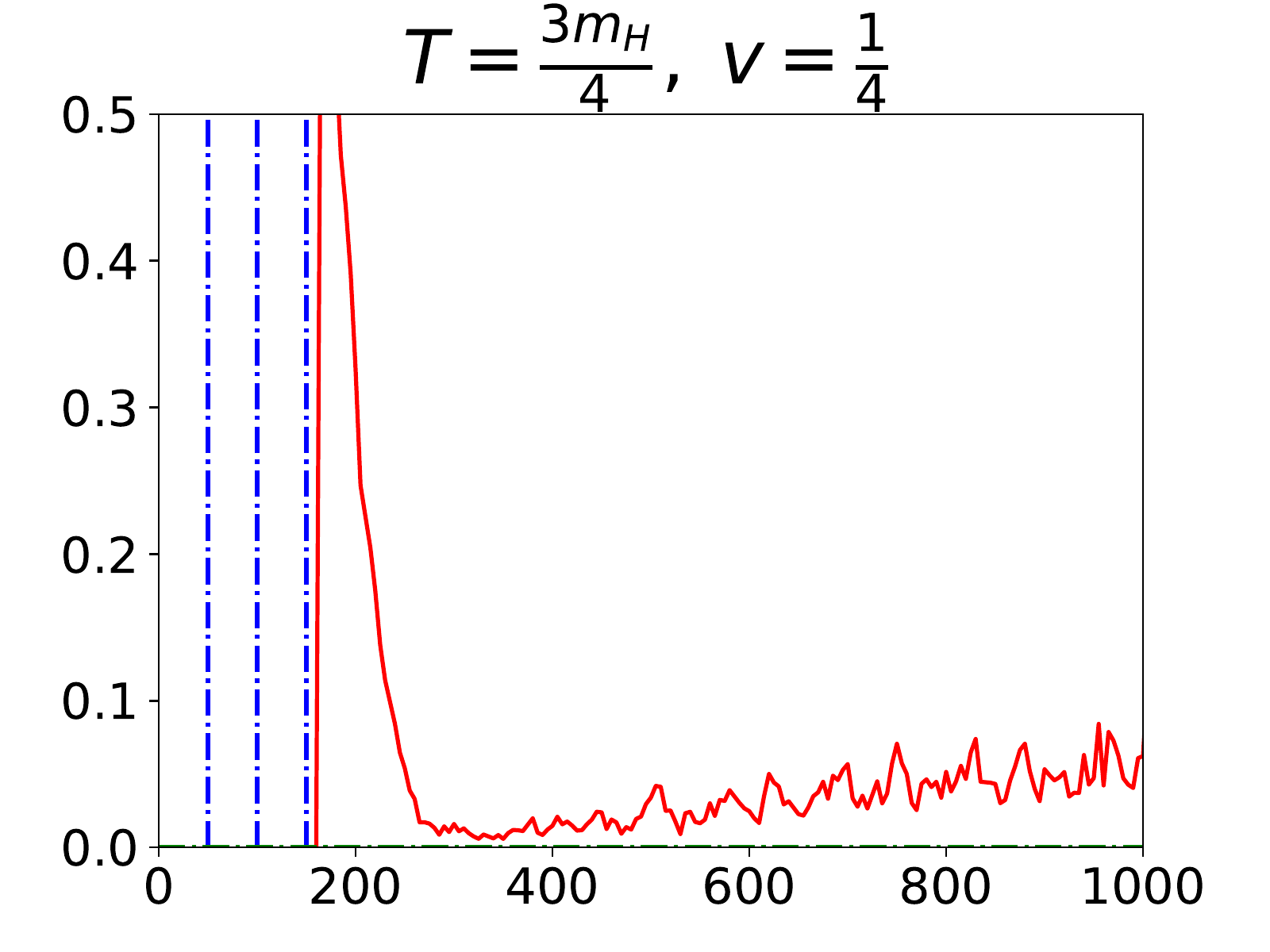} 
&
\includegraphics[width=0.27\textwidth]{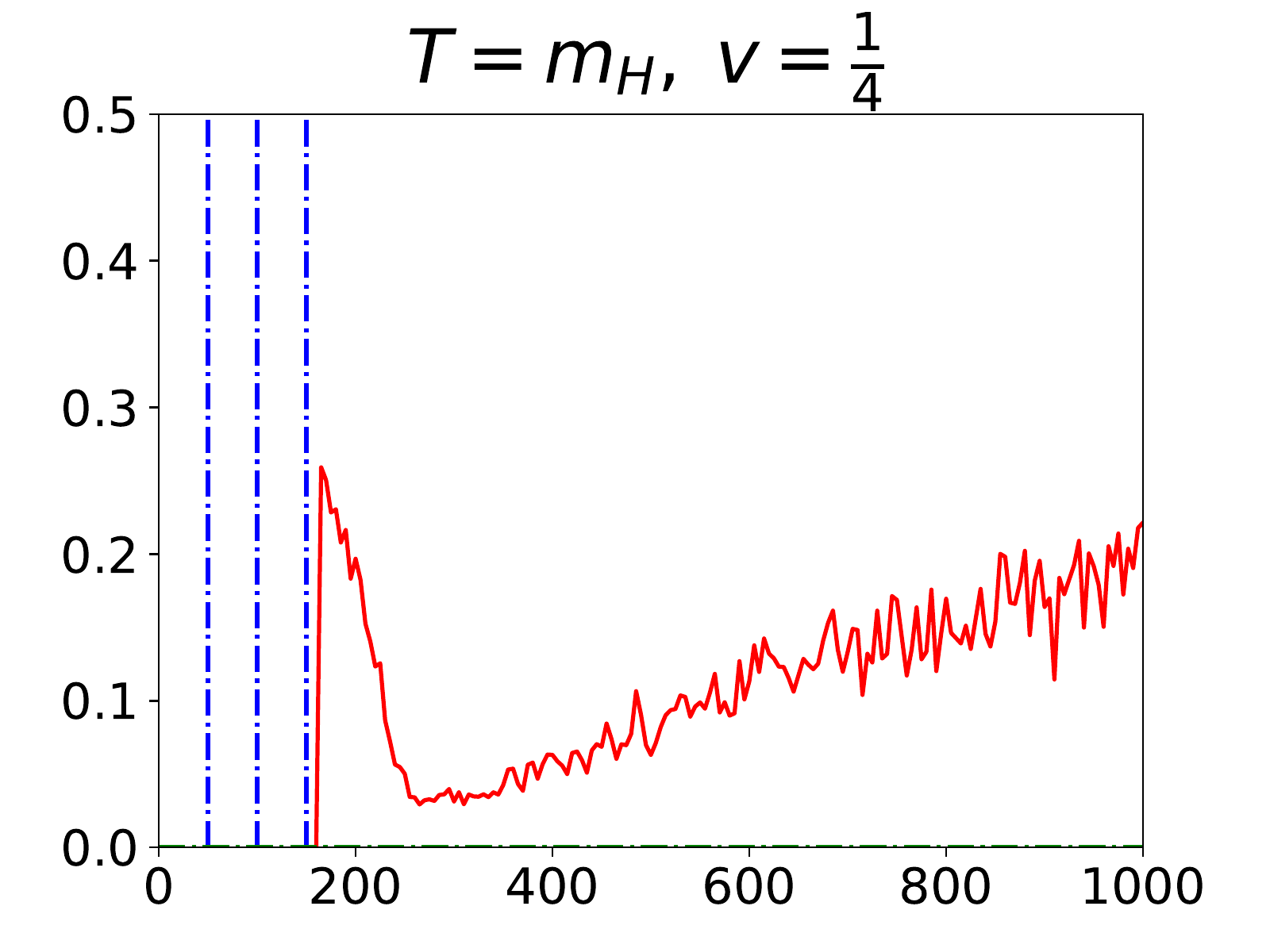} 
\\
\end{tabular}
\caption{ The width of the distribution of winding number, $\bar{N_{\rm w}^2}$, in a cubic volume just inside the bubble (vertical axis) as a function of time (horizontal axis). Results are shown for different combinations of speed $v$ and temperature $T/m_H$. $T/m_H$ increases moving right in the table of plots, increasing wall speed $v$ moving up.
$am_H=0.5$, lattice size $64\times 64\times 1000$, $m_Hd=15$. The data is averaged over 20 configurations, except for the $T/m_H=1$ ( the whole right-most column), which are averaged over 100.
}
\label{fig:diffusion2}
\end{center}
\end{figure}

We show this in Fig.~\ref{fig:diffusion2},  where we see that at least for small enough wall speeds, the time dependence is linear, independent of $v$ and increasing with increasing temperature. For speeds larger than $0.5$, our model breaks down, and the registered activity is larger than predicted. This may be because the passage of the wall itself and the heating up from the deployed latent heat influences the diffusion rate. 

%%%%%%%%%%%%%%%%%%%%%%%%%%%%%%%%%%%%%%%%%%%%%%%%%%%%%
\section{Comments, outlook and conclusion}
\label{sec:conc}
%%%%%%%%%%%%%%%%%%%%%%%%%%%%%%%%%%%%%%%%%%%%%%%%%%%%%

Electroweak baryogenesis relies on out-of-equilibrium, CP-violating interactions between an advancing Higgs bubble wall, fermions coupled to it, and the SU(2) and U(1) gauge fields together making up the finite-temperature plasma near the walls. 

Much work has gone into understanding the dynamics of this wall-plasma interaction and the transport of energy (latent heat) and fermion currents from the wall and into the plasma, where sphaleron processes generate the baryon asymmetry. This is extremely challenging, and a complete framework connecting all the different sub-processes has not been achieved.

It is appealing to attempt to simply simulate the whole thing numerically on the lattice. The dynamics of the bosonic fields is essentially classical, and the fermions are quantum mechanical but with linear equations of motion. One would hope and expect  that a steady state would establish itself, where the bubble wall is driven by the thermodynamics to sweep through the plasma at a constant speed, gradually generating the baryon asymmetry. This would require that we tune the temperature to precisely the nucleation temperature; that the lattice volume is large enough for the wall to run until it reaches the steady state; that the lattice is also large enough to hold both the bubble-dynamics region and the sphaleron transition region; that the entire set of fermion modes is included, as well as the UV-completion (Hard Thermal Loops) necessary to get the bosonic UV dynamics correct.

All these techniques exist on the market, although the combined numerical effort is daunting. In this work, we have focussed on how to set up the system and carry out the simulations, with the minimal set of dynamical degrees of freedom, and trying to bypass the difficult fine-tuning of the temperature. The bubble wall is driven by an external current, and we carefully investigated how to initialize the fields, the wall, the lattice implementation and what observables to consider. We studied the extent of the out-of-equilibrium region near the bubble wall and its dependence on the wall speed. We also did a number of tests of the bubble width and shape and optimizing lattice sizes and spacings, of which the main outcome is that the physical transverse lattice size should not be smaller than used here, and the discretization of the wall not coarser. There are some transient effects of turning on the current and establishing the wall, and one may consider whether careful use of damping and energy drains could improve on this. 

We have studied the (diagonal components of the) energy momentum tensor and how the energy released by the wall is transferred to the plasma both in front of the wall and inside the bubble. The fluid profile was confirmed to be self-similar and we observed, in particular, that how far outside and inside the bubble energy is projected depends strongly on the speed of the wall and the temperature of the plasma. In a simulation with fermions, it would be very interesting to similarly compute the distance that left- and righthanded currents are able to propagate before being thermalized by the plasma.

An obvious next step is making the connection to the fluid models of \cite{Hindmarsh:2013xza}, to see whether the first-principles field theory simulations map in a sensible way to a hydrodynamical approach. Most likely, this requires a substantial scaling up of the volumes, times and statistics achieved here. This is work in progress. 

We also studied the activity of winding number creation in the region near the wall, which is consistent with diffusion. The diffusion rate seems to depend somewhat on the wall speed, most likely through the heating of the plasma immediately in front of the wall. This sets the stage for including CP-violation. We did in fact implement a particular realisation of CP-violating dynamics in some test runs
\begin{align}
\Delta {\mathcal L_2}=&  -\frac{3\delta_{cp}}{m_W^2} \phi^\dagger\phi \frac{g_2^2}{64\pi^2}\epsilon^{\mu\nu\rho\sigma}W^a_{\mu\nu}.W^a_{\rho\sigma}.
\label{eq:CP}
\end{align}
It is the simplest CP-violating operator involving the fields included in our model, and is a stand-in for CP-violation generated by the fermion degrees of freedom, either through the CKM-matrix \cite{Brauner:2011vb}, or for a two-Higgs-doublet model when the relative complex phase of the Higgs doublets varies through the wall (see for instance \cite{Turok:1990zg}). 
This effective term is local in space, and only sizeable near the wall (where $\partial_t\phi$ is large). In contrast, the full mechanism of electroweak baryogenesis generates a handed current near the wall, which then propagates through and back away from the wall into the plasma, where sphaleron processes take place. It is hence quite non-local and so even if the effective term may be computed in the SM and applied to the dynamics, the resulting mechanism is very different as no propagating fermion currents enter. Including CP violation in this way is quite computing heavy (a factor of 4), and we were unable to see any net asymmetry created, given the volumes and statistics available to us. We intend to return to this problem in due course. 

The holy grail is to include fermions dynamically. At first as a probe, in the sense of computing the fermion currents in the background of the other fields, and studying the transport of the asymmetric currents into the bulk. The implementation is underway, but scaling to a lattice volume comparable to the one used here will require massive computing resources. Eventually, fermion current backreacting on the gauge fields will have to be included as well, in order to bias the sphaleron processes.

\vspace{0.2cm}

\noindent
{\bf Acknowledgments:}  The work of AT and ZGM was supported by a  UiS-ToppForsk grant from the University of Stavanger. PS acknowledges support by STFC Consolidated Grant No.
ST/P000703/1. The numerical work was performed in part
on the SAGA Cluster, owned by the University of Oslo and the Norwegian metacenter for High Performance Computing (NOTUR), and operated by the Department for Research Computing at USIT, the University of Oslo IT-department. This work also used the DIL managed by the University of Leicester on behalf of the STFC DiRAC HPC Facility (www.dirac.ac.uk). The equipment was funded by BEIS capital funding via STFC capital grants ST/K00042X/1, ST/P002293/1, ST/R002371/1 and ST/S002502/1, INSTITUTION and STFC operations grant ST/R000832/1. DiRAC is part of the National e-Infrastructure.
The authors would like to thank Mark Hindmarsh, Kari Rummukainen and David Weir for useful discussions.


\begin{thebibliography}{*}



%\cite{Hindmarsh:2013xza}
\bibitem{Hindmarsh:2013xza}
M.~Hindmarsh, S.~J.~Huber, K.~Rummukainen and D.~J.~Weir,
``Gravitational waves from the sound of a first order phase transition,''
Phys. Rev. Lett. \textbf{112} (2014), 041301
%doi:10.1103/PhysRevLett.112.041301
[arXiv:1304.2433 [hep-ph]].
%168 citations counted in INSPIRE as of 07 May 2020

%\cite{Caprini:2019egz}
\bibitem{Caprini:2019egz}
C.~Caprini, M.~Chala, G.~C.~Dorsch, M.~Hindmarsh, S.~J.~Huber, T.~Konstandin, J.~Kozaczuk, G.~Nardini, J.~M.~No, K.~Rummukainen, P.~Schwaller, G.~Servant, A.~Tranberg and D.~J.~Weir,
``Detecting gravitational waves from cosmological phase transitions with LISA: an update,''
JCAP \textbf{03} (2020), 024
%doi:10.1088/1475-7516/2020/03/024
[arXiv:1910.13125 [astro-ph.CO]].
%28 citations counted in INSPIRE as of 07 May 2020

%\cite{Kuzmin:1985mm}
\bibitem{Kuzmin:1985mm}
V.~Kuzmin, V.~Rubakov and M.~Shaposhnikov,
``On the Anomalous Electroweak Baryon Number Nonconservation in the Early Universe,''
Phys. Lett. B \textbf{155} (1985), 36
%doi:10.1016/0370-2693(85)91028-7
%2788 citations counted in INSPIRE as of 07 May 2020

%\cite{Cohen:1993nk}
\bibitem{Cohen:1993nk}
A.~G.~Cohen, D.~Kaplan and A.~Nelson,
``Progress in electroweak baryogenesis,''
Ann. Rev. Nucl. Part. Sci. \textbf{43} (1993), 27-70
%doi:10.1146/annurev.ns.43.120193.000331
[arXiv:hep-ph/9302210 [hep-ph]].
%881 citations counted in INSPIRE as of 07 May 2020

%

%\cite{Fromme:2006cm}
\bibitem{firstorder1}
L.~Fromme, S.~J.~Huber and M.~Seniuch,
``Baryogenesis in the two-Higgs doublet model,''
JHEP \textbf{11} (2006), 038
%doi:10.1088/1126-6708/2006/11/038
[arXiv:hep-ph/0605242 [hep-ph]].
%188 citations counted in INSPIRE as of 07 May 2020


%\cite{Huber:2006wf}
\bibitem{firstorder2}
S.~J.~Huber, T.~Konstandin, T.~Prokopec and M.~G.~Schmidt,
``Electroweak Phase Transition and Baryogenesis in the nMSSM,''
Nucl. Phys. B \textbf{757} (2006), 172-196
%doi:10.1016/j.nuclphysb.2006.09.003
[arXiv:hep-ph/0606298 [hep-ph]].
%120 citations counted in INSPIRE as of 07 May 2020

%\cite{Barger:2007im}
\bibitem{firstorder3}
V.~Barger, P.~Langacker, M.~McCaskey, M.~J.~Ramsey-Musolf and G.~Shaughnessy,
``LHC Phenomenology of an Extended Standard Model with a Real Scalar Singlet,''
Phys. Rev. D \textbf{77} (2008), 035005
%doi:10.1103/PhysRevD.77.035005
[arXiv:0706.4311 [hep-ph]].
%434 citations counted in INSPIRE as of 07 May 2020

%\cite{Barger:2008jx}
\bibitem{firstorder4}
V.~Barger, P.~Langacker, M.~McCaskey, M.~Ramsey-Musolf and G.~Shaughnessy,
``Complex Singlet Extension of the Standard Model,''
Phys. Rev. D \textbf{79} (2009), 015018
%doi:10.1103/PhysRevD.79.015018
[arXiv:0811.0393 [hep-ph]].
%237 citations counted in INSPIRE as of 07 May 2020

%\cite{Cline:2011mm}
\bibitem{firstorder5}
J.~M.~Cline, K.~Kainulainen and M.~Trott,
``Electroweak Baryogenesis in Two Higgs Doublet Models and B meson anomalies,''
JHEP \textbf{11} (2011), 089
%doi:10.1007/JHEP11(2011)089
[arXiv:1107.3559 [hep-ph]].
%114 citations counted in INSPIRE as of 07 May 2020

%\cite{Cline:2012hg}
\bibitem{firstorder6}
J.~M.~Cline and K.~Kainulainen,
``Electroweak baryogenesis and dark matter from a singlet Higgs,''
JCAP \textbf{01} (2013), 012
%doi:10.1088/1475-7516/2013/01/012
[arXiv:1210.4196 [hep-ph]].
%155 citations counted in INSPIRE as of 07 May 2020

%\cite{Damgaard:2013kva}
\bibitem{firstorder7}
P.~H.~Damgaard, D.~O'Connell, T.~C.~Petersen and A.~Tranberg,
``Constraints on New Physics from Baryogenesis and Large Hadron Collider Data,''
Phys. Rev. Lett. \textbf{111} (2013) no.22, 221804
%doi:10.1103/PhysRevLett.111.221804
[arXiv:1305.4362 [hep-ph]].
%28 citations counted in INSPIRE as of 07 May 2020

%\cite{Kozaczuk:2015owa}
\bibitem{firstorder8}
J.~Kozaczuk,
``Bubble Expansion and the Viability of Singlet-Driven Electroweak Baryogenesis,''
JHEP \textbf{10} (2015), 135
%doi:10.1007/JHEP10(2015)135
[arXiv:1506.04741 [hep-ph]].
%79 citations counted in INSPIRE as of 07 May 2020

%\cite{Dorsch:2016nrg}
\bibitem{firstorder9}
G.~Dorsch, S.~Huber, T.~Konstandin and J.~No,
``A Second Higgs Doublet in the Early Universe: Baryogenesis and Gravitational Waves,''
JCAP \textbf{05} (2017), 052
%doi:10.1088/1475-7516/2017/05/052
[arXiv:1611.05874 [hep-ph]].
%59 citations counted in INSPIRE as of 07 May 2020

%\cite{Alanne:2016wtx}
\bibitem{firstorder10}
T.~Alanne, K.~Kainulainen, K.~Tuominen and V.~Vaskonen,
``Baryogenesis in the two doublet and inert singlet extension of the Standard Model,''
JCAP \textbf{08} (2016), 057
%doi:10.1088/1475-7516/2016/08/057
[arXiv:1607.03303 [hep-ph]].
%26 citations counted in INSPIRE as of 07 May 2020

%\cite{Brauner:2016fla}
\bibitem{firstorder11}
  T.~Brauner, T.~V.~I.~Tenkanen, A.~Tranberg, A.~Vuorinen and D.~J.~Weir,
  %``Dimensional reduction of the Standard Model coupled to a new singlet scalar field,''
  JHEP {\bf 1703} (2017) 007
  doi:10.1007/JHEP03(2017)007
  [arXiv:1609.06230 [hep-ph]].
  %%CITATION = doi:10.1007/JHEP03(2017)007;%%
  %21 citations counted in INSPIRE as of 22 Jun 2020

%
%\cite{Andersen:2017ika}
\bibitem{firstorder12}
  J.~O.~Andersen, T.~Gorda, A.~Helset, L.~Niemi, T.~V.~I.~Tenkanen, A.~Tranberg, A.~Vuorinen and D.~J.~Weir,
  %``Nonperturbative Analysis of the Electroweak Phase Transition in the Two Higgs Doublet Model,''
  Phys.\ Rev.\ Lett.\  {\bf 121} (2018) no.19,  191802
  doi:10.1103/PhysRevLett.121.191802
  [arXiv:1711.09849 [hep-ph]].
  %%CITATION = doi:10.1103/PhysRevLett.121.191802;%%
  %15 citations counted in INSPIRE as of 22 Jun 2020

%

%\cite{Moore:2001vf}
\bibitem{bubsim1}
G.~D.~Moore, K.~Rummukainen and A.~Tranberg,
``Nonperturbative computation of the bubble nucleation rate in the cubic anisotropy model,''
JHEP \textbf{04} (2001), 017
%doi:10.1088/1126-6708/2001/04/017
[arXiv:hep-lat/0103036 [hep-lat]].
%12 citations counted in INSPIRE as of 07 May 2020

%\cite{Moore:2000jw}
\bibitem{bubsim2}
G.~D.~Moore and K.~Rummukainen,
``Electroweak bubble nucleation, nonperturbatively,''
Phys. Rev. D \textbf{63} (2001), 045002
%doi:10.1103/PhysRevD.63.045002
[arXiv:hep-ph/0009132 [hep-ph]].
%47 citations counted in INSPIRE as of 07 May 2020

%\cite{Hindmarsh:2015qta}
\bibitem{bubsim3}
M.~Hindmarsh, S.~J.~Huber, K.~Rummukainen and D.~J.~Weir,
``Numerical simulations of acoustically generated gravitational waves at a first order phase transition,''
Phys. Rev. D \textbf{92} (2015) no.12, 123009
%doi:10.1103/PhysRevD.92.123009
[arXiv:1504.03291 [astro-ph.CO]].
%152 citations counted in INSPIRE as of 07 May 2020

%\cite{Saffin:2011kc}
\bibitem{bubsim4}
P.~M.~Saffin and A.~Tranberg,
``Real-time Fermions for Baryogenesis Simulations,''
JHEP \textbf{07} (2011), 066
%doi:10.1007/JHEP07(2011)066
[arXiv:1105.5546 [hep-ph]].
%26 citations counted in INSPIRE as of 07 May 2020


%%%%

%\cite{Rubakov:1996vz}
\bibitem{EWBGrev2}
V.~Rubakov and M.~Shaposhnikov,
``Electroweak baryon number nonconservation in the early universe and in high-energy collisions,''
Usp. Fiz. Nauk \textbf{166} (1996), 493-537
%doi:10.1070/PU1996v039n05ABEH000145
[arXiv:hep-ph/9603208 [hep-ph]].
%798 citations counted in INSPIRE as of 07 May 2020

%\cite{Morrissey:2012db}
\bibitem{EWBGrev1}
D.~E.~Morrissey and M.~J.~Ramsey-Musolf,
``Electroweak baryogenesis,''
New J. Phys. \textbf{14} (2012), 125003
%doi:10.1088/1367-2630/14/12/125003
[arXiv:1206.2942 [hep-ph]].
%426 citations counted in INSPIRE as of 07 May 2020

%



%\cite{Celis:2013rcs}
\bibitem{Celis:2013rcs}
A.~Celis, V.~Ilisie and A.~Pich,
%``LHC constraints on two-Higgs doublet models,''
JHEP \textbf{07} (2013), 053
doi:10.1007/JHEP07(2013)053
[arXiv:1302.4022 [hep-ph]].
%148 citations counted in INSPIRE as of 18 May 2020

%\cite{Tanabashi:2018oca}
\bibitem{Tanabashi:2018oca}
M.~Tanabashi \textit{et al.} [Particle Data Group],
%``Review of Particle Physics,''
Phys. Rev. D \textbf{98} (2018) no.3, 030001
doi:10.1103/PhysRevD.98.030001
%4597 citations counted in INSPIRE as of 18 May 2020

%

%\cite{Saffin:2011kn}
\bibitem{Saffin:2011kn}
P.~M.~Saffin and A.~Tranberg,
``Dynamical simulations of electroweak baryogenesis with fermions,''
JHEP \textbf{02}, 102 (2012)
%doi:10.1007/JHEP02(2012)102
[arXiv:1111.7136 [hep-ph]].
%28 citations counted in INSPIRE as of 13 May 2020

%%
%\cite{Tranberg:2003gi}
\bibitem{Tranberg:2003gi}
A.~Tranberg and J.~Smit,
``Baryon asymmetry from electroweak tachyonic preheating,''
JHEP \textbf{11} (2003), 016
%doi:10.1088/1126-6708/2003/11/016
[arXiv:hep-ph/0310342 [hep-ph]].
%98 citations counted in INSPIRE as of 07 May 2020

%\cite{Carrington:1991hz}
\bibitem{Carrington:1991hz}
M.~Carrington,
``The Effective potential at finite temperature in the Standard Model,''
Phys. Rev. D \textbf{45} (1992), 2933-2944
%doi:10.1103/PhysRevD.45.2933
%415 citations counted in INSPIRE as of 15 May 2020

%\cite{Arnold:1992rz}
\bibitem{Arnold:1992rz}
P.~B.~Arnold and O.~Espinosa,
``The Effective potential and first order phase transitions: Beyond leading-order,''
Phys. Rev. D \textbf{47} (1993), 3546
%doi:10.1103/PhysRevD.47.3546
[arXiv:hep-ph/9212235 [hep-ph]].
%449 citations counted in INSPIRE as of 15 May 2020

%
%\cite{Mou:2017zwe}
\bibitem{Mou:2017zwe}
  Z.~G.~Mou, P.~M.~Saffin and A.~Tranberg,
  ``Simulations of Cold Electroweak Baryogenesis: Hypercharge U(1) and the creation of helical magnetic fields,''
  JHEP {\bf 1706} (2017) 075
  %doi:10.1007/JHEP06(2017)075
  [arXiv:1704.08888 [hep-ph]].
  %%CITATION = doi:10.1007/JHEP06(2017)075;%%
  %4 citations counted in INSPIRE as of 22 Nov 2018

%\cite{Rezzolla-Zanotti}
\bibitem{Rezzolla-Zanotti}
L.~Rezzolla and O.~Zanotti,
``Relativistic Hydrodynamics,'' (Oxford University Press, Oxford, UK, 2013).

%\cite{DOnofrio:2014rug}
\bibitem{DOnofrio:2014rug}
M.~D'Onofrio, K.~Rummukainen and A.~Tranberg,
``Sphaleron Rate in the Minimal Standard Model,''
Phys. Rev. Lett. \textbf{113} (2014) no.14, 141602
%doi:10.1103/PhysRevLett.113.141602
[arXiv:1404.3565 [hep-ph]].
%149 citations counted in INSPIRE as of 07 May 2020

%\cite{Brauner:2011vb}
\bibitem{Brauner:2011vb}
T.~Brauner, O.~Taanila, A.~Tranberg and A.~Vuorinen,
``Temperature Dependence of Standard Model CP Violation,''
Phys. Rev. Lett. \textbf{108} (2012), 041601
%doi:10.1103/PhysRevLett.108.041601
[arXiv:1110.6818 [hep-ph]].
%23 citations counted in INSPIRE as of 07 May 2020

%\cite{Turok:1990zg}
\bibitem{Turok:1990zg}
N.~Turok and J.~Zadrozny,
``Electroweak baryogenesis in the two doublet model,''
Nucl. Phys. B \textbf{358} (1991), 471-493
%doi:10.1016/0550-3213(91)90356-3
%265 citations counted in INSPIRE as of 07 May 2020


\end{thebibliography}
\end{document}